\documentclass[12pt,a4paper]{article}
\usepackage{fullpage}
\usepackage{float}
\usepackage{cite}
\usepackage{floatflt}
\usepackage[centertags]{amsmath}
\allowdisplaybreaks[1]
\numberwithin{equation}{section}
\usepackage{amsbsy}
\usepackage{amsfonts}
\usepackage{amssymb}
\usepackage[dvips]{graphicx}
\usepackage{url}
\newcommand{\be}{\begin{equation}}
\newcommand{\ee}{\end{equation}}
\newcommand{\eV}{\mathrm{eV}}
\begin{document}

\begin{flushright}
% \text{\sf \today}
\text{\sf 27 March 2003}
\\
UAB--FT--533
\\
KIAS--P02070
\\
hep-ph/0211462
\end{flushright}

%\vspace{1cm}

\begin{center}
\large
\textbf{Absolute Values of Neutrino Masses: Status and Prospects}
\normalsize
\\[0.5cm]
\large
S. M. Bilenky${}^{1,2,3}$,
C. Giunti${}^{2,4}$,
J. A. Grifols${}^{1}$
and
E. Mass\'o${}^{1}$
\normalsize
\\[0.5cm]
${}^{1}$
Department of Physics and IFAE, Universitat Aut\`onoma de Barcelona,
\\
08193 Bellaterra, Barcelona, Spain
\\[0.5cm]
${}^{2}$
INFN, Sezione di Torino, and Dipartimento di Fisica Teorica,
\\
Universit\`a di Torino,
Via P. Giuria 1, I--10125 Torino, Italy
\\[0.5cm]
${}^{3}$
Joint Institute
for Nuclear Research, Dubna, R-141980, Russia
\\[0.5cm]
${}^{4}$
Korea Institute for Advanced Study,
\\
207-43 Cheongryangri-dong, Dongdaemun-gu Seoul 130-012, Korea
\\[0.5cm]
\begin{minipage}[t]{0.8\textwidth}
\begin{center}
\textbf{Abstract}
\end{center}
Compelling evidences in favor of neutrino masses and mixing
obtained in the last years in
Super-Kamiokande, SNO, KamLAND
and other
% atmospheric, solar and long-baseline
neutrino experiments
made the physics of massive and mixed neutrinos a frontier field of research in
particle physics and astrophysics.
There are many open problems in this new field. In this review we 
consider the problem of the absolute values of neutrino masses, which apparently
is the most difficult one
from the experimental point of view.
We discuss the present limits and the future 
prospects of $\beta$-decay neutrino mass measurements
and
neutrinoless double-$\beta$ decay.
We consider the important problem
of the calculation of nuclear matrix elements of neutrinoless double-$\beta$ decay
and discuss the possibility
to check the results of different model 
calculations of the nuclear matrix elements through their comparison
with the experimental data.
We discuss the upper bound of the total mass of neutrinos that was
obtained recently from the data of the 2dF Galaxy Redshift Survey  and other
cosmological data and we discuss future prospects of the
cosmological measurements of the total mass of neutrinos.   
We discuss also the possibility to obtain information on neutrino masses from
the observation of the ultra high-energy cosmic rays
(beyond the GZK cutoff).
Finally, we review the main aspects of the physics of core-collapse supernovae,
the limits on the
absolute values of neutrino masses from the observation of SN1987A neutrinos
and the future prospects of supernova neutrino detection.
\end{minipage}
\end{center}

%\begin{flushleft}
%PACS Numbers: 14.60.Pq, 14.60.Lm
%\\
%Keywords: Neutrino Mass, Neutrino Mixing
%\end{flushleft}

\newpage
{\small\tableofcontents}
\newpage

\section{Introduction}

Compelling evidences in favor of neutrino oscillations,
driven by small neutrino masses and neutrino mixing, were obtained
in the
Super-Kamiokande
\cite{Fukuda:1998mi,%
Fukuda:1998ah,%
Fukuda:2000np,%
Fukuda:2001nj,%
Fukuda:2002pe},
SNO
\cite{Ahmad:2001an,%
Ahmad:2002jz,%
Ahmad:2002ka},
KamLAND
\cite{hep-ex/0212021}
and other
atmospheric
\cite{Allison:1999ms,%
Ambrosio:2000qy},
solar
\cite{Cleveland:1998nv,%
Hampel:1998xg,%
Altmann:2000ft,%
astro-ph/0204245}
and long-baseline
\cite{Ahn:2002up}
neutrino experiments.
These findings have brought
\emph{the physics of massive and mixed neutrinos}
in the front line of the research in particle physics and astrophysics\footnote{
See Ref.~\cite{Neutrino-Unbound-note}
for an
extensive bibliography on neutrino physics and astrophysics.
}.

From all the existing terrestrial and astrophysical data
it follows that neutrino masses are smaller
than the masses of other fundamental fermions (lepton and quarks)
by many orders of magnitude.
There is a general consensus that the smallness of neutrino masses is 
due to 
New Physics beyond the Standard Model. 
In the most attractive see-saw mechanism of neutrino mass generation
\cite{see-saw},
the smallness of neutrino masses is due to
the violation of the total lepton number on a scale which is much larger than
the electroweak scale.

There are many open problems in the physics of massive and mixed neutrinos:

\begin{itemize}

\item
How many light neutrinos with definite mass exist in nature?

The minimal number of massive neutrinos is equal to the number of
active (flavor) neutrinos (three).
If, however, sterile neutrinos exist, the number of massive neutrinos is larger than three
(see Refs.~\cite{BGG-review-98,hep-ph/0202058}). 
The data of all the existing neutrino oscillation experiments
(solar
\cite{Cleveland:1998nv,%
Hampel:1998xg,%
Altmann:2000ft,%
astro-ph/0204245,%
Fukuda:2002pe,%
Ahmad:2001an,%
Ahmad:2002jz,%
Ahmad:2002ka},
atmospheric
\cite{Fukuda:1998mi,%
Fukuda:1998ah,%
Fukuda:2000np,%
Allison:1999ms,%
Ambrosio:2000qy}
and LSND
\cite{LSND})
require the existence of (at least) four massive neutrinos.
LSND is the single accelerator experiment in which the
transition $\bar \nu_{\mu}\to \bar \nu_{e}$ has been observed.
The check of the LSND claim is an urgent problem. This will be done
by the MiniBooNE experiment
\cite{hep-ex/0211039}, which started recently.

\item
What is the nature of the neutrinos with definite mass: are they
purely neutral Majorana particles, or Dirac particles, possessing a
conserved total lepton number?

The answer to this fundamental question can be obtained through
the investigation of processes in which the total lepton number is not conserved.
The most promising process is neutrinoless double $\beta$-decay
of some even-even nuclei.
There are many new experiments on the search for 
neutrinoless double $\beta$-decay now in preparation
(see Ref.~\cite{Cremonesi:2002is}),
which will push the experimental sensitivity at a level
that is about two orders of magnitude better than today's sensitivity.
We will discuss neutrinoless double $\beta$-decay in this review.

\item
What are the absolute values of the neutrino masses?

Neutrino oscillations are due to differences of phases which
different massive components 
of the initial flavor neutrino states pick up during their evolution.
As a result,
neutrino oscillation experiments allow to obtain information only
on neutrino mass-squared differences
(see Refs.~\cite{CWKim-book-93,Mohapatra-book-98,BGG-review-98,hep-ph/0202058}).
It is very important that neutrino oscillation experiments 
are sensitive to tiny neutrino mass-squared differences,
because of
the possibility to explore very large distances and small energies.
The measurement of the absolute values of neutrino masses 
at a level of a few eV is a
challenging problem. 
This review is dedicated to the discussion of this problem
(see also Ref.~\cite{Pas:2001nd}).

\end{itemize}

Let us mention also the very important problems of 
the precise determination of the values of the neutrino oscillation parameters
and the search for CP violation in neutrino oscillations.
These problems will be investigated in experiments at the future Super-Beam facilities
and Neutrino factories
(see Refs.~\cite{Lindner:2002vt,Aoki:2002ks,Nakaya-Nu2002,hep-ph/0210113}).
We will not discuss them here.

We will start
in Section~\ref{Status of neutrino oscillations}
with a short review of the present status of neutrino oscillations. 
We will consider neutrino oscillations in
solar, atmospheric and long-baseline neutrino experiments
in the framework of mixing of three neutrinos.
The importance for neutrino mixing
of the results of the long-baseline reactor experiments CHOOZ and Palo Verde,
in which no
indication in favor of neutrino oscillations was found, will be stressed.

In Section~\ref{Neutrino mass from beta-decay experiments}
we will consider the Mainz
\cite{Weinheimer:1999tn,Bonn:2002jw,hep-ex/0210050}
and Troitsk
\cite{Lobashev:1999tp,Lobashev:2001uu}
experiments on the measurement of the neutrino mass through the detailed investigation of the end-point 
part of the $\beta$-spectrum of tritium.
We will discuss also the future KATRIN tritium experiment
\cite{Osipowicz:2001sq}.

In Section~\ref{Muon and tau neutrino mass measurements}
we briefly review the most recent results of
the experiments on the measurement of the effect of neutrino masses in pion and tau
decays.

Section~\ref{Neutrinoless double-beta decay}
is dedicated to neutrinoless double $\beta$-decay.
Even though in this review we are mainly interested in the possibilities to obtain information
about the absolute values of the neutrino masses from the investigation of this process,
some aspects of the theory of the process will be also presented.

%%%%%%%%%%%%%%%%%%%

The role of neutrinos in Astrophysics and Cosmology has been under the
scrutiny of physicists with ever increasing intensity over the last few
decades
(see Refs.~\cite{Pas:2001nd,Dolgov:2002wy,Dolgov:2002ad,Kainulainen:2002pu,%
Raffelt:2002ed,Raffelt:2002tu,hep-ex/0202043,Olive:2002qg,hep-ph/0210089}).
Actually, the intimate relationship of neutrinos and astrophysics
goes even further back in time when Bethe and others realized that the
inner workings of the Sun proceed via the thermonuclear reactions that
burn hydrogen into helium and release neutrinos. Since then there has
been a steady increase in interest and involvement in the study of neutrinos
in astrophysical environments. Not only the interest in solar neutrinos
has been particularly intense over the last years, where such epochal
events as their detection on Earth in a variety of underground experiments
have been milestones of late 20th century high energy physics, but also
the involvement of neutrinos in stellar core collapse has been theoretically
analyzed and observationally established in the momentous detection of neutrinos
of SN1987A and, furthermore, the influence and role of neutrinos in the
cosmic evolution has been a major area of research in contemporary high energy
physics.

Seminal work on neutrinos and Cosmology was pursued in the late
sixties and early seventies when neutrinos appeared as ideal candidates
to contribute substantially to the matter density of the Universe
(see Ref.~\cite{Dolgov:1981hv}). In fact,
hot dark matter models were popular for quite some time as they seemed to
render a satisfactory model for structure formation. Of course, the interest
in neutrinos worked then and still works now both ways; from the cosmological arena, neutrinos
were welcomed in the new cosmological Paradigm but also from the Particle
Physics side, Cosmology/Astrophysics was used and is used even more so today to constrain and sharpen
the still not well known properties of neutrinos.

The main focus of this review is the mass of neutrinos and especially their
absolute mass. So we have selected the issues in Cosmology/Astrophysics
that have relevant impact on the extraction of information concerning
neutrino mass. They are contained in Sections~\ref{Cosmology},
\ref{Cosmic Rays} and \ref{Supernova Neutrinos}.
We start in Subsection~\ref{The Gerstein-Zeldovich limit on neutrino masses} by
introducing the famous Gerstein-Zeldovich upper bound on the total sum of neutrino
masses that can contribute to the matter density of the Universe
\cite{Gershtein:1966gg,Cowsik:1972gh}.
Subsection~\ref{Microwave Background Anisotropies}
is dedicated to an overview of the temperature fluctuations of the
Cosmic Microwave Background
Radiation (CMB) with a special attention to the characteristics of the peak structure
of the angular power spectrum
(see Ref.~\cite{Hu:2001bc}).
There, the influence of neutrino mass on the anisotropy
spectrum is explicitly discussed. Although it is shown that this influence is
not as significant as the role of other cosmological parameters that enter
the angular spectrum of temperature fluctuations,
Subsection~\ref{Microwave Background Anisotropies} is relatively
long as compared to the other Subsections in Section~\ref{Cosmology} because in any analysis
of cosmological import the CMB is of pivotal importance. 
Subsection~\ref{Galaxy Redshift Surveys} is devoted
to Galaxy Redshift Surveys. Neutrino mass has a remarkable effect on the
power spectrum of matter distribution and this effect is observable in the
large samples of data compiled in present galaxy distribution surveys or to be
collected in future surveys. The final astrophysical source of information
discussed in this review, namely Lyman $\alpha$ forests studies, is dealt with in
Subsection~\ref{Lyman alpha forests}. The last Subsection
in Section~\ref{Cosmology}, Subsection~\ref{Neutrino mass bounds}, contains the summary of all
relevant neutrino mass limits obtained in the actual analysis by different groups and
by different authors of the astrophysical/cosmological sources that have been
discussed in the foregoing Subsections. It contains also a brief report on the prospects
for neutrino mass in this rapidly changing field of Cosmology and Astrophysics.

Another topic that we cover in our review concerns cosmic rays. A probe
of neutrino properties could come from the observation
of cosmic rays
with energies exceeding the Greisen-Zapsepin-Kuzmin cutoff
\cite{Greisen:1966jv,Zatsepin:1966jv}. 
A possible explanation could be the so-called $Z$-burst scenario
\cite{Fargion:1997ft,Weiler:1997sh},
where a flux of ultra high energy neutrinos interacts with relic
cosmological neutrinos, producing cosmic rays through the $Z$-resonance.
The resonance condition involves the masses of neutrinos and we
review the status of this mechanism in Section~\ref{Cosmic Rays}
(see also Ref.~\cite{Pas:2001nd}).

In 1987 the observation of neutrinos coming from supernova 1987A
in the Large Magellanic Cloud
marked the beginning of extra solar system neutrino astronomy
and allowed to get information on the supernova mechanism
and neutrino properties
(see Refs.~\cite{Trimble-RMP60-859-1988,Wheeler:2002gw}).
In particular,
the values of the neutrino masses are limited
by the lack of spread of the observed neutrino signal,
which would be caused by energy-dependent
velocities of sufficiently massive neutrinos
\cite{Zatsepin:1968,Pakvasa:1972gz,Cabibbo:1980,Piran:1981zz,Fargion:1981xc}.
In Section~\ref{Supernova Neutrinos}
we review
the classification and rate of supernovae
(Section~\ref{Supernova Types and Rates}),
the current theory of core-collapse supernova dynamics
(Section~\ref{Core-Collapse Supernova Dynamics}),
the observation of SN1987A neutrinos
(Section~\ref{SN1987A}),
the inferred limits on neutrino masses
(Section~\ref{Neutrino Mass}),
and the future prospects for supernova neutrino detection
(Section~\ref{Future}).

\section{Status of neutrino oscillations}
\label{Status of neutrino oscillations}

Strong evidences in favor of neutrino oscillations 
were obtained recently in
Super-Kamiokande
\cite{Fukuda:1998mi,%
Fukuda:1998ah,%
Fukuda:2000np,%
Fukuda:2001nj,%
Fukuda:2002pe},
SNO
\cite{Ahmad:2001an,%
Ahmad:2002jz,%
Ahmad:2002ka},
KamLAND
\cite{hep-ex/0212021}
and other
atmospheric
\cite{Allison:1999ms,%
Ambrosio:2000qy},
solar
\cite{Cleveland:1998nv,%
Hampel:1998xg,%
Altmann:2000ft,%
astro-ph/0204245}
and long-baseline
\cite{Ahn:2002up}
neutrino experiments.
These findings gave us the first evidence that neutrino masses
are different from zero and
the fields of neutrinos with definite mass
$\nu_{i}$ enter into
the standard charged current (CC) and neutral current (NC)
\begin{equation}
j^{\mathrm{CC}}_{\alpha} = \sum_{l=e,\mu,\tau} \bar\nu_{lL} \gamma_{\alpha}l_{L}
\,,
\qquad
j^{\mathrm{NC}}_{\alpha} =\sum_{l=e,\mu,\tau} \bar\nu_{lL}\gamma_{\alpha}\nu_{lL}
\label{001}
\end{equation}
in the \emph{mixed form}
\begin{equation}
\nu_{lL} = \sum_{i} U_{li} \nu_{iL}
\qquad
(l=e,\mu,\tau)
\,,
\label{002}
\end{equation}
where $U$ is the unitary mixing matrix.
The minimal number of massive neutrinos $\nu_{i}$ is equal to the 
number of active (flavor) neutrinos (three). The number of massive neutrinos
can be larger than three (see Ref.~\cite{BGG-review-98}). 
In this case, in addition to Eq.~(\ref{002}) we have
\begin{equation}
\nu_{sL} = \sum_{i} U_{si} \nu_{iL}
\,,
\label{003}
\end{equation}
where $\nu_{sL}$ ($s=s_{1},s_{2},\ldots$)
are the fields of sterile neutrinos\footnote{
The fields
$\nu_{sL}$ do not enter into the standard CC and NC
in Eq.~(\ref{001}).
They could be right-handed neutrino fields, SUSY fields, etc..
}.

The most plausible mechanism of neutrino mass generation is
the see-saw mechanism \cite{see-saw}.
In order to explain this mechanism,
let us consider the simplest case of one generation and assume that the
standard Higgs mechanism with one Higgs doublet generates
the Dirac mass term
\begin{equation}
\mathcal{L}^{\mathrm{D}} = - m_{\mathrm{D}} \, \bar\nu_{R} \nu_{L} + \mathrm{h.c.}
\,. 
\label{004}
\end{equation}
It is natural to expect that $m_{\mathrm{D}}$ is of the same order of magnitude as
the mass of the charged lepton or quarks in the same generation.
We know, however,
from experimental data that
neutrino masses are much smaller than the masses of
charged leptons and quarks.
In order to ``suppress'' the neutrino mass let us assume that there is
a lepton-number violating
mechanism beyond the Standard Model
which generates
the right-handed Majorana mass term\footnote{
Notice that,
since $\nu_R$ is a SU(2) singlet
and has zero hypercharge,
the Majorana mass term $\mathcal{L}^{\mathrm{Mj}}_{R}$
is allowed by the electroweak gauge symmetries.
}
\begin{equation}
\mathcal{L}^{\mathrm{Mj}}_{R}
=
- \frac{1}{2} \, M_{R} \, \bar\nu_{R} \, (\nu_{R})^{c} + \mathrm{h.c.}
\,, 
\label{005}
\end{equation}
with $M_{R} \gg m_{\mathrm{D}}$ (usually it is assumed that $M_{R} \simeq M_{\mathrm{GUT}}\sim
10^{15}\, \mathrm{GeV}$).
Here
$ (\nu_{R})^{c} = \mathcal{C} \overline{\nu_R}^T $,
where
$\mathcal{C}$
is the charge conjugation matrix.

After the diagonalization of the total neutrino mass term,
for the light neutrino mass
we obtain
\begin{equation}
m
=
- \frac{1}{2}\,M_{R}+\frac{1}{2}\,\sqrt{M_{R}^{2}+ 4\,m_{\mathrm{D}}^{2}}
\simeq
\frac{m_{\mathrm{D}}^{2}}{M_{R}}\ll m_{\mathrm{D}}
\label{006}
\end{equation}

In the case of three generations,
the see-saw mechanism leads to
a spectrum of masses of Majorana particles
with three light neutrinos with masses $m_i$ ($i=1,2,3$)
much smaller than the quark and charged-lepton masses,
and three very heavy masses
of the order of the scale of violation of the total lepton number
(for recent reviews see \cite{hep-ph/0206077,King:2002gx}).

Let us stress that, if the neutrino masses have a standard see-saw origin\footnote{
In non-standard see-saw models
neutrinos could be Dirac particles
\cite{Ecker:1987ib,Branco:1989ex}.
},
neutrinos with definite masses are \emph{Majorana particles}.
In some models which implement the see-saw mechanism
(see, for example, Ref.~\cite{King:2002gx}),
the neutrino masses naturally satisfy the hierarchy
\begin{equation}
m_{1} \ll m_{2} \ll m_{3}
\,.
\label{008}
\end{equation}

If there is neutrino mixing, the state of a neutrino (active or sterile)
with momentum $ \vec p $ is given by the
\emph{coherent superposition of the states of neutrinos with
definite masses}
\begin{equation}
|\nu_{\alpha}\rangle
= \sum_{i} U_{\alpha i}^* \, |\nu_i\rangle \,,
\label{009}
\end{equation}
where $|\nu_i\rangle $ is the state of neutrinos with momentum $\vec p $,
mass $m_{i}$ and energy
\begin{equation}
E_i = \sqrt{p^2 + m_i^2 } \simeq p + \frac{ m_i^2 }{ 2 p }
\qquad
( p^2 \gg m_i^2 )
\,.
\label{010}
\end{equation}
From Eq.~(\ref{009})
it follows that if at time $t=0$
a neutrino $\nu_{\alpha}$ ($\alpha =e, \mu, \tau$) is produced,
the probability amplitude to find $\nu_{\alpha'}$ at time $t$
is given by
\begin{equation}
A(\nu_\alpha\to\nu_{\alpha'}) = 
\langle \nu_{\alpha'}\,|e^{-iH_{0}\,t}\,|\nu_{\alpha}\rangle =
\sum_i U_{\alpha' i} \, e^{-iE_it} \, U_{\alpha i}^*
\,.
\label{011}
\end{equation}
Thus, the probability of the transition $\nu_{\alpha}\to\nu_{\alpha'}$ 
in vacuum is given by
\begin{equation}
\mathrm{P}(\nu_\alpha\to\nu_{\alpha'})
=
\left|
\delta_{{\alpha'}\alpha}
+
\sum_{i \geq 2} U_{\alpha' i}  U_{\alpha i}^*
\left( e^{- i \Delta{m}^2_{i 1} \frac{L}{2E}} -1 \right)
\right|^2
\,.
\label{012}
\end{equation}
Here 
$\Delta{m}^2_{i 1}= m^2_{i}-m^2_{1}$,
$L\simeq t$ is the distance between the neutrino source and the neutrino 
detector, and $E$
is the neutrino energy.

In the simplest case of transitions between two types of neutrinos
($\nu_{\mu}\to\nu_{\tau}$ or  $\nu_{\mu}\to\nu_{e}$, etc.), 
the index $i$ in Eq.~(\ref{012}) takes only one value $i=2$ and for the 
transition probability we obtain the standard expression
\begin{equation}
\mathrm{P}(\nu_\alpha\to\nu_{\alpha'})
=
\frac{1}{2} \, \sin^{2}2\vartheta
\left( 1 - \cos \Delta{m}^{2} \frac{L}{2E} \right)
\qquad
(\alpha'\neq\alpha)
\,,
\label{013}
\end{equation}
where $\Delta{m}^{2}= m^{2}_{2}-m^{2}_{1}$,
$|U_{\alpha 2}|^{2} = \cos^{2}\vartheta$,
$|U_{\alpha' 2}|^{2}=1-|U_{\alpha 2}|^{2} = \sin^{2}\vartheta $
($\vartheta$ is the mixing angle).
For the probability of $\nu_{\alpha}$
to survive we have
\begin{equation}
\mathrm{P}(\nu_\alpha\to\nu_\alpha) 
=
1-\mathrm{P}(\nu_\alpha\to\nu_{\alpha'})
=
1 - \frac{1}{2} \, \sin^{2} 2\vartheta
\left( 1 - \cos \Delta{m}^{2} \frac{L}{2E} \right)
\,.
\label{014}
\end{equation}
The expressions (\ref{013})
and (\ref{014}) describe periodical transitions between two types of neutrinos
(neutrino oscillations).
They are widely used in the 
analysis of 
experimental data.

\begin{figure}[t]
\begin{center}
\includegraphics*[bb=45.1211 300.4404 549.3418 799.3911, width=0.7\textwidth]{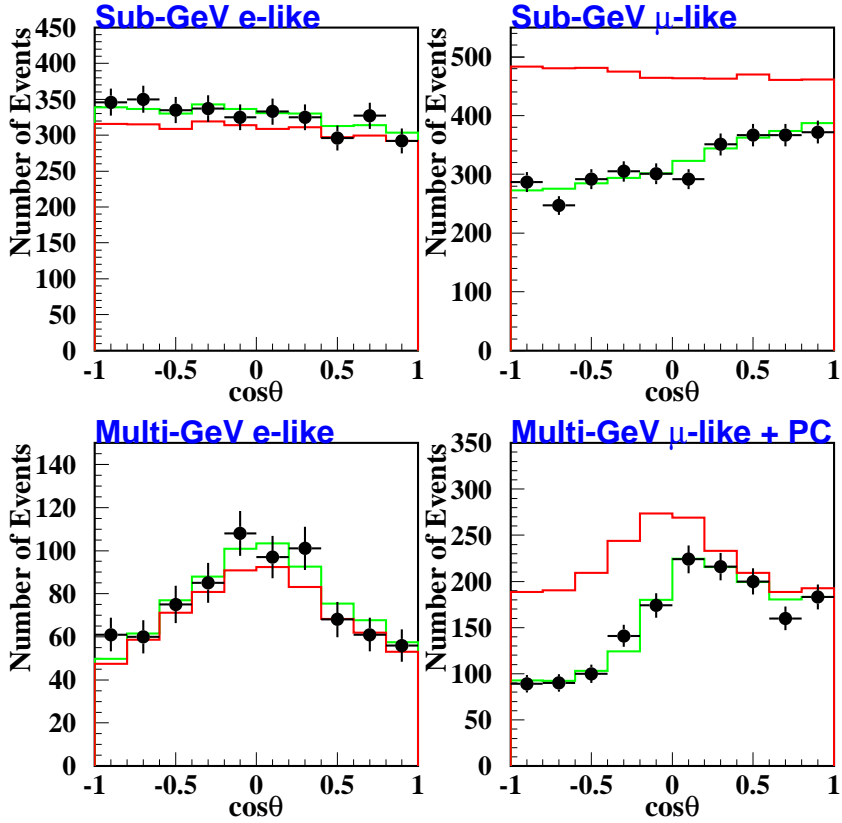}
\end{center}
\caption{ \label{sk-nakaya-0209036-f1}
Zenith angle distribution of Super-Kamiokande sub-GeV
single ring $e$-like events, $\mu$-like events, multi-GeV single ring $e$-like
events, and $\mu$-like events + partially contained (PC) events.
The black histogram shows the Monte Carlo prediction
and
the gray histogram is the best fit for
$\nu_\mu\to\nu_\tau$
oscillations with
$\Delta m^2 = 2.5 \times 10^{-3} \, \mathrm{eV}^2$
and
$\sin^22\vartheta=1.0$.
Figure taken from Ref.~\cite{Nakaya:2002ki}.
}
\end{figure}

\begin{figure}[t]
\begin{center}
\includegraphics*[bb=0 18 531 550, height=7cm]{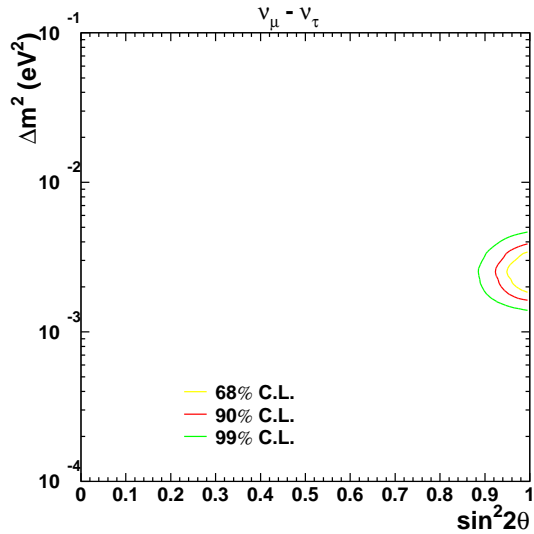}
\caption{ \label{sk-wilkes-0212035-f4}
Allowed region of the oscillation parameters
$\mathsf{\Delta{m}^{2}} = \Delta{m}^{2}_{\mathrm{atm}}$
and
$\mathsf{\sin^{2}2\theta} = \sin^{2}2 \vartheta_{\mathrm{atm}}$
from the analysis
of Super-Kamiokande atmospheric neutrino data. 
Figure taken from Ref.~\cite{hep-ex/0212035}.
}
\end{center}
\end{figure}

\subsection{Atmospheric neutrinos}
\label{Atmospheric neutrinos}

The first model independent
evidence in favor of neutrino oscillations was obtained in the 
atmospheric Super-Kamiokande (S-K) experiment \cite{Fukuda:1998mi,Fukuda:1998ah,Fukuda:2000np}. 
In this experiment a significant zenith angle $\theta_{z}$ asymmetry of the
high energy muon events was observed. At high energies the zenith angle
$\theta_{z}$ is determined by the distance
$L$, which neutrinos pass from the production region in the atmosphere
to the detector.
If there are no neutrino oscillations,
the number of detected
multi-GeV ($E \geq 1.3 $ GeV)
electron (muon) events satisfy the symmetry relation
\begin{equation}
N_{l}(\cos\theta_{z})= N_{l}( -\cos \theta_{z})
\qquad
(l=e,\mu)
\label{015}
\end{equation}
As one can see from Fig.~\ref{sk-nakaya-0209036-f1},
the number of multi-GeV electron events observed in the S-K experiment is in
good agreement with this relation.
On the other hand,
Fig.~\ref{sk-nakaya-0209036-f1}
shows that
the multi-GeV muon events
observed in the S-K experiment strongly violate the relation (\ref{015}).
Let us define the ratio $U/D$, where 
$U$ is the number of up-going muons
($ -1 \leq \cos\theta_{z} \leq -0.2$,
$500 \, \mathrm{km} \lesssim L \lesssim 13000 \, \mathrm{km}$)
and $D$ is the number of down-going
muons ($0.2 \leq \cos\theta_{z} \leq 1$,
$20 \, \mathrm{km} \lesssim L \lesssim 500 \, \mathrm{km}$).
For the multi-GeV events in the S-K experiment it was found
\cite{Kajita:2000mr}
\begin{equation}
\frac{(U/D)_{\mathrm{meas}}}{(U/D)_{\mathrm{MC}}} = 0.54 \pm 0.04\, \mbox{(stat.)}
 \pm 0.01\, \mbox{(syst.)}\,,
\label{016}
\end{equation}
where $(U/D)_{\mathrm{MC}}$ is the
ratio predicted by Monte Carlo
under the assumption that there are no neutrino oscillations.
If there are no neutrino oscillations, the ratio
of ratios
${(U/D)_{meas}}/{(U/D)_{\mathrm{MC}}}$
must be equal to one.
The S-K value
(\ref{016})
differs from one by $11\,\sigma$.

The data of the S-K \cite{Fukuda:1998mi,Fukuda:1998ah,Fukuda:2000np} and other atmospheric neutrino experiments
(SOUDAN 2 \cite{Allison:1999ms}, MACRO \cite{Ambrosio:2000qy}) are well described 
assuming that two-neutrino oscillations $\nu_{\mu}\to
\nu_{\tau}$ take place.
The allowed region of
the neutrino oscillation parameters
$\Delta{m}^{2}_{\mathrm{atm}}$ and $\sin^{2}2 \vartheta_{\mathrm{atm}}$
from the analysis of the S-K data is shown in Fig.~\ref{sk-wilkes-0212035-f4}.
At 90\% $\mathrm{C.L.}$
the oscillation parameters
are in the ranges
\begin{equation}
1.6\times 10^{-3} \leq
\Delta{m}^{2}_{\mathrm{atm}}
\leq 3.9 \times 10^{-3} \, \mathrm{eV}^{2}
\,,
\qquad
\sin^{2}2 \vartheta_{\mathrm{atm}} > 0.92
\,.
\label{017}
\end{equation}
The best-fit values of the parameters are
\begin{equation}
\Delta{m}^{2}_{\mathrm{atm}}=2.5\times 10^{-3}\mathrm{eV}^{2}
\,,
\qquad
\sin^{2}2 \vartheta_{\mathrm{atm}}=1.0 
\qquad
(\chi^{2}_{\mathrm{min}}= 163.2/ 170\,\mathrm{d.o.f.})
\,.
\label{018}
\end{equation}

\begin{figure}[t]
\begin{center}
\includegraphics*[bb=0 0 544 514, height=7cm]{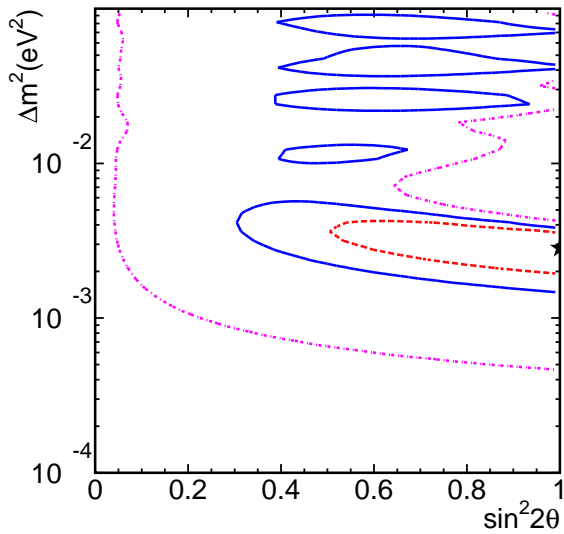}
\caption{ \label{k2k-0212007-f3}
Allowed regions of the oscillation parameters
$\mathsf{\Delta{m}^{2}} = \Delta{m}^{2}_{\mathrm{K2K}}$
and
$\mathsf{\sin^{2}2\theta} = \sin^{2}2 \vartheta_{\mathrm{K2K}}$
obtained from the results of the K2K experiment. 
The dashed, solid and dot-dashed lines are, respectively, 68.4\%, 90\% and
99\% C.L. contours.
The best fit point is indicated by the star.
Figure taken from Ref.~\cite{Ahn:2002up}.
}
\end{center}
\end{figure}

Recently the results of the first long-baseline accelerator experiment K2K
have been published
\cite{Ahn:2002up}.
In this experiment neutrino oscillations in the atmospheric range of
$\Delta{m}^{2}$ were searched for.
Neutrinos mainly from decays of
pions, produced by 12 GeV protons hitting a beam-dump target at the KEK proton accelerator,
were detected by the S-K detector at the distance of about 250 km from the
source. The average neutrino energy is 1.3 GeV.

In the K2K experiment
there are two near detectors at the distance of about 300 m 
from the beam-dump target: a 1 kt water Cherenkov
detector and a fine-grained detector. The number and the spectrum of 
muon neutrinos detected in S-K are compared with the expected 
quantities, calculated on the basis of the results of the near detectors.
Quasielastic one-ring events
$ \nu_{\mu} + n \to \mu^{-} + p $ are selected
for the measurement of the energy of the neutrinos. 

The total number of muon events observed in the S-K experiment is 56.
The expected number of events is $80.1^{+6.2}_{-5.4}$.
The observed number of one-ring muon events that was used for 
the calculation of the neutrino spectrum is 29.
The expected number of one-ring events is 44.

The regions of the allowed values of the oscillation parameters
obtained from a maximum likelihood two-neutrino oscillation analysis
of the K2K data are presented in Fig.~\ref{k2k-0212007-f3}.
The best-fit
values of the parameters are
\begin{equation}
\sin^{2}2\vartheta_{\mathrm{K2K}}=1
\,,
\qquad
\Delta{m}^{2}_{\mathrm{K2K}} = 2.8 \times 10^{-3} \, \mathrm{eV}^{2}
\,.
\label{K2K01}
\end{equation}
These values are in agreement with the values of the oscillation parameters
obtained from the analysis of the S-K atmospheric neutrino data
(see Eqs.~(\ref{017}) and (\ref{018})). 

Thus, the K2K experiment confirms the evidence for neutrino oscillations
that was found in the atmospheric Super-Kamiokande experiment.
The K2K data reported in Ref.~\cite{Ahn:2002up}
have been obtained with $4.8 \times 10^{19}$ protons on target (POT).
The K2K experiment is planned to continue until about
$10^{20}$ POT will be reached.

\subsection{Solar neutrinos}
\label{Solar neutrinos}

The event rates measured in 
all solar neutrino experiments 
are significantly smaller than the event rates predicted by the Standard Solar
models.
The following values were obtained, respectively,
for the ratio of the rates observed
in the Homestake \cite{Cleveland:1998nv}, 
GALLEX-GNO \cite{Hampel:1998xg,Altmann:2000ft},
SAGE \cite{Abdurashitov:1999zd}
and
S-K \cite{Fukuda:2002pe,Smy:2002rz}  
experiments
and those 
predicted by the BP00 \cite{BP2000} Standard Solar Model (SSM):
$0.34 \pm 0.03$,
$0.58 \pm 0.05$,
$0.60 \pm 0.05$,
$0.465 \pm 0.018$.
It has been known during many years that these data can be explained by
transitions of the initial solar
$\nu_{e}$'s into other neutrinos, which cannot be detected in
the radiochemical Homestake, SAGE,
GALLEX and GNO experiments.
The S-K experiment is sensitive mainly to $\nu_e$'s.

Recently,
strong model independent evidence in favor of the 
transitions of solar $\nu_{e}$'s
into $\nu_{\mu}$'s and $\nu_{\tau}$'s
has been obtained in the SNO experiment \cite{Ahmad:2001an,Ahmad:2002jz,Ahmad:2002ka}.
In this experiment solar neutrinos are detected via the observation
of the following three reactions\footnote{
Here $\nu_{x}$ stands for \emph{any} active neutrino.
}:

\begin{itemize}

\item
The charged-current
(CC) reaction
\begin{equation}
\nu_e + d \to e^{-}+ p +p
\,,
\label{019}
\end{equation}

\item
The neutral-current
(NC) reaction 
\begin{equation}
\nu_{x} + d \to \nu_{x}+ n +p
\,,
\label{020}
\end{equation}

\item
The elastic-scattering
(ES) reaction 
\begin{equation}
\nu_{x} + e \to \nu_{x} + e
\,. 
\label{021}
\end{equation}

\end{itemize}

The kinetic energy threshold for the detection of
electrons in the CC and ES processes in the SNO experiment is 5 MeV.  
The NC process has been detected through the observation of $\gamma$ rays
from the capture of neutrons by deuterium. The NC threshold is 2.2 MeV.
Thus, practically only neutrinos from $^{8}\mathrm{B}$ decay
are detected in the experiment\footnote{
According to the BP00 SSM \cite{BP2000},
the flux of high energy $hep$ neutrinos
is about three orders of magnitude smaller than the flux of $^{8}\mathrm{B}$ neutrinos.
Looking at solar neutrino events beyond the $^{8}\mathrm{B}$ spectrum endpoint,
the Super-Kamiokande Collaboration found that the flux of
$hep$ neutrinos
is smaller than 7.9 times the BP00 SSM $hep$ flux
at 90\% C.L.
\cite{Smy:2002rz}.
}.

The measurement of the total CC event rate allows to determine the
flux of $\nu_{e}$ on the Earth,
\begin{equation}
\Phi_{\nu_{e}}^{\mathrm{CC}}
=
\langle P(\nu_e \to\nu_e) \rangle_{\mathrm{CC}} \, \Phi_{\nu_{e}}^{0}
\,, 
\label{022}
\end{equation}
where $\Phi_{\nu_{e}}^{0}$ is the total initial flux of $\nu_e$'s
and $\langle P(\nu_e \to\nu_e) \rangle_{\mathrm{CC}}$ is the
probability of $\nu_e$ to survive
averaged over the CC cross section
and the known initial spectrum of $^{8}\mathrm{B}$ neutrinos.
In the SNO experiment it was found that
\begin{equation}
(\Phi_{\nu_{e}}^{\mathrm{CC}})_{\mathrm{SNO}}
=
\left( 1.76 {}^{+0.06}_{-0.05} \mbox{(stat.)} {}^{+0.09}_{-0.09} \mbox{(syst.)} \right)
\times 10^{6} \, \mathrm{cm}^{-2} \, \mathrm{s}^{-1}
\,.
\label{023}
\end{equation}

All active neutrinos $ \nu_{e}$, $\nu_{\mu}$ and $\nu_{\tau}$ are recorded by
the detection of the NC process (\ref{020}).
Taking into account the universality of neutral currents
(see the recent analysis in Ref.~\cite{Masso:2002vj}),
the total flux of
all active neutrinos on the Earth measured in the SNO experiment is
\begin{equation}
(\Phi_{\nu}^{\mathrm{NC}})_{\mathrm{SNO}}
=
\sum_{l=e,\mu,\tau} (\Phi_{\nu_l}^{\mathrm{NC}})_{\mathrm{SNO}}
=
\left( 5.09 {}^{+0.44}_{-0.43} \mbox{(stat.)} {}^{+0.46}_{-0.43} \mbox{(syst.)} \right)
\times 10^{6} \, \mathrm{cm}^{-2} \, \mathrm{s}^{-1}
\,,
\label{024}
\end{equation}
which is about three times larger than the CC flux in Eq.~(\ref{023}).

All active neutrinos are detected also via the observation of the ES process
(\ref{021}).
However, the cross section of the neutral-current
$\nu_{\mu,\tau}e\to\nu_{\mu,\tau}e$
scattering is about six times smaller
than the cross section of the charged-current and neutral-current
$\nu_{e}e\to\nu_{e}e$
scattering.

The event rate
$R^{\mathrm{ES}}$
of the ES process (\ref{021})
can be written as
\begin{equation}
R^{\mathrm{ES}}
=
\langle \sigma_{\nu_{e} e} \rangle
\,
\Phi_{\nu}^{\mathrm{ES}}
\,.
\label{0241}
\end{equation}
Here
$\langle \sigma_{\nu_{e} e} \rangle$
is the $\nu_{e} e \to \nu_{e} e$ cross section
averaged over the initial spectrum of $^{8}\mathrm{B}$ neutrinos
and
the ES flux $\Phi_{\nu}^{\mathrm{ES}}$
is given by
\begin{equation}
\Phi_{\nu}^{\mathrm{ES}}
=
\Phi_{\nu_{e}}^{\mathrm{ES}}
+
\frac{\langle \sigma_{\nu_{\mu} e} \rangle}{\langle \sigma_{\nu_{e} e} \rangle}
\,
\Phi_{\nu_{\mu,\tau}}^{\mathrm{ES}}
\,,
\label{025}
\end{equation}
where
$\Phi_{\nu_{e}}$
and
$\Phi_{\nu_{\mu,\tau}}$
are,
respectively,
the fluxes of
$\nu_e$ and $\nu_{\mu,\tau}$
on the Earth
averaged over the ES cross section
and the initial spectrum of $^{8}\mathrm{B}$ neutrinos.
The ratio of the averaged
$\nu_{\mu,\tau} e \to \nu_{\mu,\tau} e$
and
$\nu_e e \to \nu_e e$
cross sections is given by
\begin{equation}
\frac
{\langle \sigma_{\nu_{\mu} e} \rangle}
{\langle \sigma_{\nu_{e} e} \rangle}
\simeq 0.154
\,.
\label{026}
\end{equation}

In the SNO experiment \cite{Ahmad:2002ka} it was found
\begin{equation}
(\Phi_{\nu}^{\mathrm{ES}})_{\mathrm{SNO}}
=
\left( 2.39 {}^{+0.24}_{-0.23} \mbox{(stat.)} \pm 0.12 \mbox{(syst.)} \right)
\times 10^{6} \, \mathrm{cm}^{-2} \, \mathrm{s}^{-1}
\,.
\label{027}
\end{equation}
This value is in good agreement with the value of the ES flux
determined in the S-K experiment \cite{Fukuda:2002pe,Smy:2002rz}.
In the S-K experiment solar neutrinos are detected
via the observation of the ES process (\ref{021}).
During 1496 days of running a large number,
$22400 \pm 800$,
of solar neutrino events
with recoil energy above the 5 MeV threshold
were recorded
(the uncertainty is due to the statistical subtraction of background events). 
From the data of the 
S-K experiment it was obtained 
\begin{equation}
(\Phi_{\nu}^{\mathrm{ES}})_{\mathrm{S-K}}
=
\left( 2.35 \pm 0.02 \mbox{(stat.)} \pm 0.08 \mbox{(syst.)} \right)
\times 10^{6} \, \mathrm{cm}^{-2} \, \mathrm{s}^{-1}
\,.
\label{028}
\end{equation}
In the S-K experiment also the spectrum of the recoil electrons
was measured. No significant distortion of the spectrum with
respect to the expected one (calculated under the assumption that the shape
of the
spectrum of $\nu_{e}$ on the Earth is given by the known initial 
$^{8}\mathrm{B}$ spectrum) was observed. 
Furthermore, no distortion of the spectrum of the electrons
produced in the CC process (\ref{019}) was observed in the
SNO experiment. These data are compatible with the 
assumption that the probability of solar neutrinos to survive
is a constant in the high energy $^{8}\mathrm{B}$ region. 
Thus, we have
\begin{equation}
\Phi_{\nu_{e}}^{\mathrm{NC}}\simeq \Phi_{\nu_{e}}^{\mathrm{CC}} \simeq \Phi_{\nu_{e}}^{\mathrm{ES}}\,.
\label{029}
\end{equation}

\begin{figure}[t]
\begin{center}
\includegraphics*[bb=5 0 521 390, height=7cm]{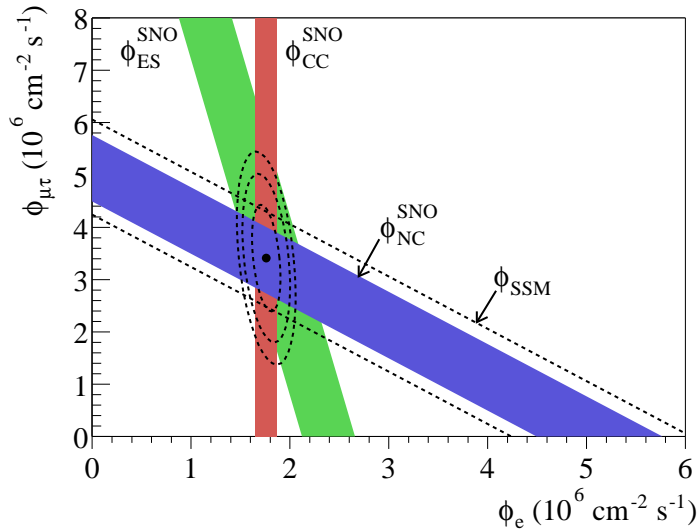}
\end{center}
\caption{ \label{sno-0204008-fig3}
Flux of $\nu_\mu$ and $\nu_\tau$ vs flux of $\nu_e$
in the ${}^{8}$B energy range
deduced from the three neutrino reactions in SNO.
The diagonal bands show the total ${}^{8}$B flux
as predicted by the BP00 SSM \cite{BP2000} (dashed lines)
and that measured with the NC reaction in SNO
(solid band).  The intercepts of these bands with the axes represent the $\pm 1\sigma$ errors.
The
bands intersect at the fit values for
$\phi_{e} \equiv \Phi_{\nu_e}$ and $\phi_{\mu\tau} \equiv \Phi_{\nu_{\mu,\tau}}$,
indicating that the combined
flux results are consistent with neutrino flavor transformation assuming no distortion in the
${}^{8}$B neutrino energy spectrum.
Figure taken from Ref.~\cite{Ahmad:2002jz}.
}
\end{figure}

Obviously, the NC flux can be presented in the form
\begin{equation}
\Phi_{\nu}^{\mathrm{NC}} = \Phi_{\nu_{e}}^{\mathrm{NC}} + \Phi_{\nu_{\mu,\tau}}^{\mathrm{NC}}
\,.
\label{030}
\end{equation}
Combining the CC and NC fluxes and using the relation (\ref{030}),
we can determine now 
the flux $\Phi_{\nu_{\mu,\tau}}^{\mathrm{NC}}$.
In Ref.~\cite{Ahmad:2002jz} the ES flux
(\ref{027}) was also taken into account as an additional constraint
(see Fig.~\ref{sno-0204008-fig3}).
The resulting
flux of $\nu_{\mu}$ and $\nu_{\tau}$ on the Earth is
\begin{equation}
(\Phi_{\nu_{\mu,\tau}})_{\mathrm{SNO}}
=
\left( 3.41 {}^{+0.45}_{-0.45} \mbox{(stat.)} {}^{+0.48}_{-0.45} \mbox{(syst.)} \right)
\times 10^{6} \, \mathrm{cm}^{-2} \, \mathrm{s}^{-1}
\,.
\label{031}
\end{equation}
Thus, the detection of the solar neutrinos  
through the simultaneous observation of
CC, NC and ES processes allowed the SNO collaboration
to obtain a \emph{direct model independent $5.3\,\sigma$ evidence 
of the presence of
$\nu_{\mu}$ and $\nu_{\tau}$ in the flux of the solar neutrinos on the Earth.}

Before the publication of the first results \cite{Ahmad:2001an}
of the SNO experiment,
from the global fit of the data of the
Homestake \cite{Cleveland:1998nv},
SAGE \cite{astro-ph/0204245},
GALLEX \cite{Hampel:1998xg},
GNO \cite{Altmann:2000ft}
and
S-K \cite{Fukuda:1998ua,Fukuda:1998rq}
experiments
several allowed regions in the plane 
of the two-neutrino oscillation parameters
$\Delta{m}^{2}_{\mathrm{sol}}$ and $\tan^{2}\vartheta_{\mathrm{sol}}$
had been found:
the
large mixing angle (LMA),
low mass (LOW)
and
small mixing angle (SMA)
Mikheev-Smirnov-Wolfenstein
(MSW) \cite{Wolfenstein:1978ue,Mikheev:1985gs} regions,
the vacuum oscillation (VAC) region
and others (see, for example, Ref.~\cite{BGG-review-98,Bahcall:2001hv}).
The situation changed after
the publication of the first SNO data \cite{Ahmad:2001an},
which,
together with the recoil electron spectrum measured in the S-K experiment
\cite{Fukuda:1998ua,Fukuda:1998rq,Fukuda:2001nj},
disfavored the SMA-MSW region
(see, for example, Ref.~\cite{Bahcall:2001zu}).
The most recent data from the SNO \cite{Ahmad:2002jz,Ahmad:2002ka}
and
S-K \cite{Fukuda:2002pe,Smy:2002rz}
experiments
strongly disfavor the SMA-MSW region
(see, for example, Ref.~\cite{Bahcall:2002hv}).
All global analyses of the present solar neutrino data favor the LMA-MSW region
\cite{Ahmad:2002ka,%
Barger:2002iv,%
Bahcall:2002hv,%
deHolanda:2002pp,%
Fukuda:2002pe,%
Bandyopadhyay:2002xj,%
Strumia:2002rv,%
Fogli:2002pt,%
Fogli:2002pb}.
%(see Ref.~\cite{Lisi-Nu2002}).
The 
best-fit values of the oscillation parameters 
in the LMA-MSW region found in Ref.~\cite{Ahmad:2002ka} are
\begin{equation}
\Delta{m}^{2}_{\mathrm{sol}} = 5 \times 10^{-5}\,\mathrm{eV}^{2}
\,,
\qquad
\tan^{2}\vartheta_{\mathrm{sol}} = 3.4 \times 10^{-1}
\,,
\qquad
(\chi^{2}_{\mathrm{min}} = 57.0 \,, \, 72 \, \mbox{d.o.f.})
\label{032}
\end{equation}
In the next Subsection we will discuss the recent results
of the long-baseline reactor experiment KamLAND \cite{hep-ex/0212021}.
The data of this experiment allow to exclude the
SMA, LOW and VAC regions,
leaving the LMA region as
the only viable solution of the solar neutrino problem.

\begin{figure}[t]
\begin{minipage}[l]{0.45\textwidth}
\begin{center}
\includegraphics*[bb=0 0 434 360, height=7cm]{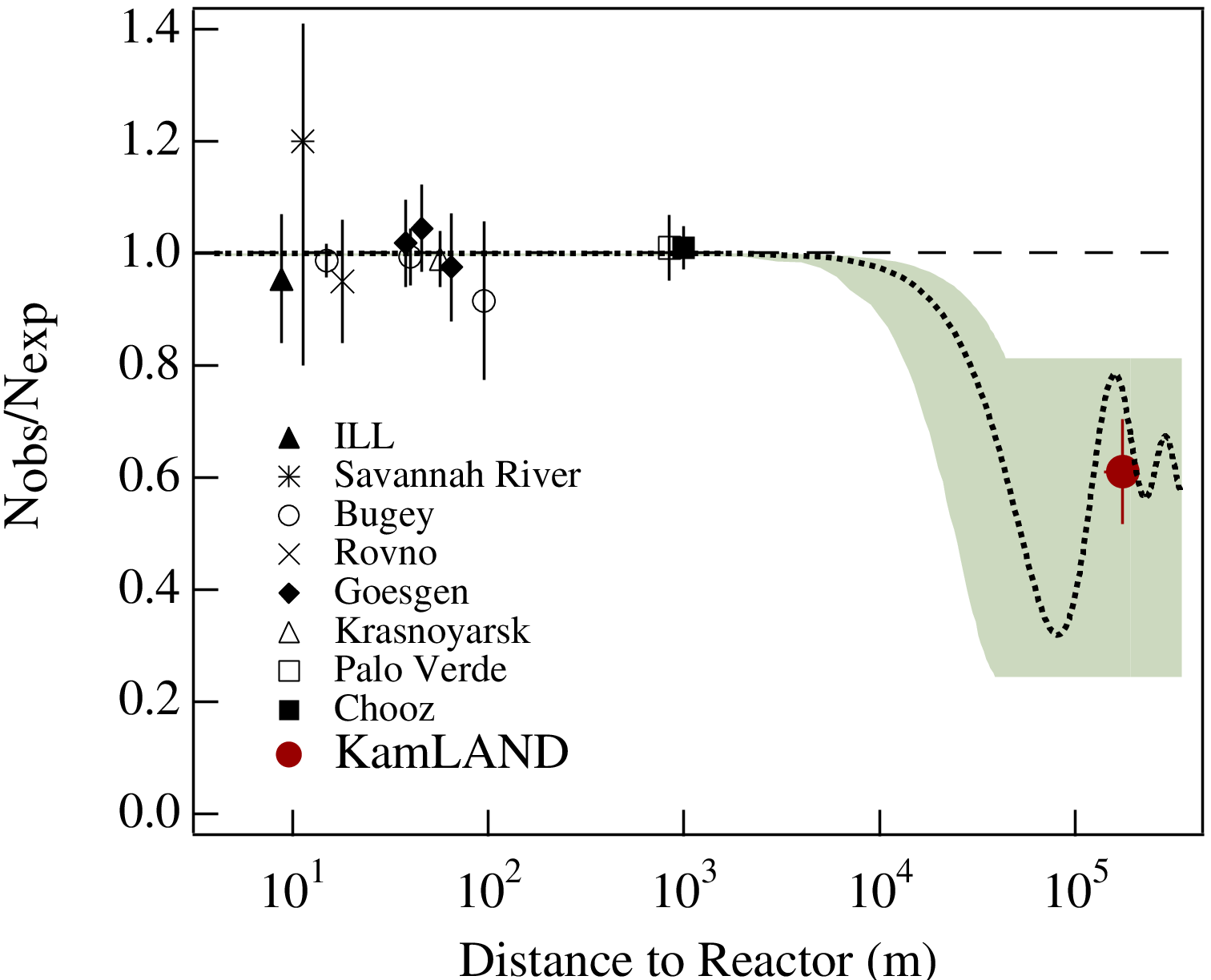}
\end{center}
\end{minipage}
\hfill
\begin{minipage}[l]{0.45\textwidth}
\begin{center}
\includegraphics*[bb=69 50 520 494, height=7cm]{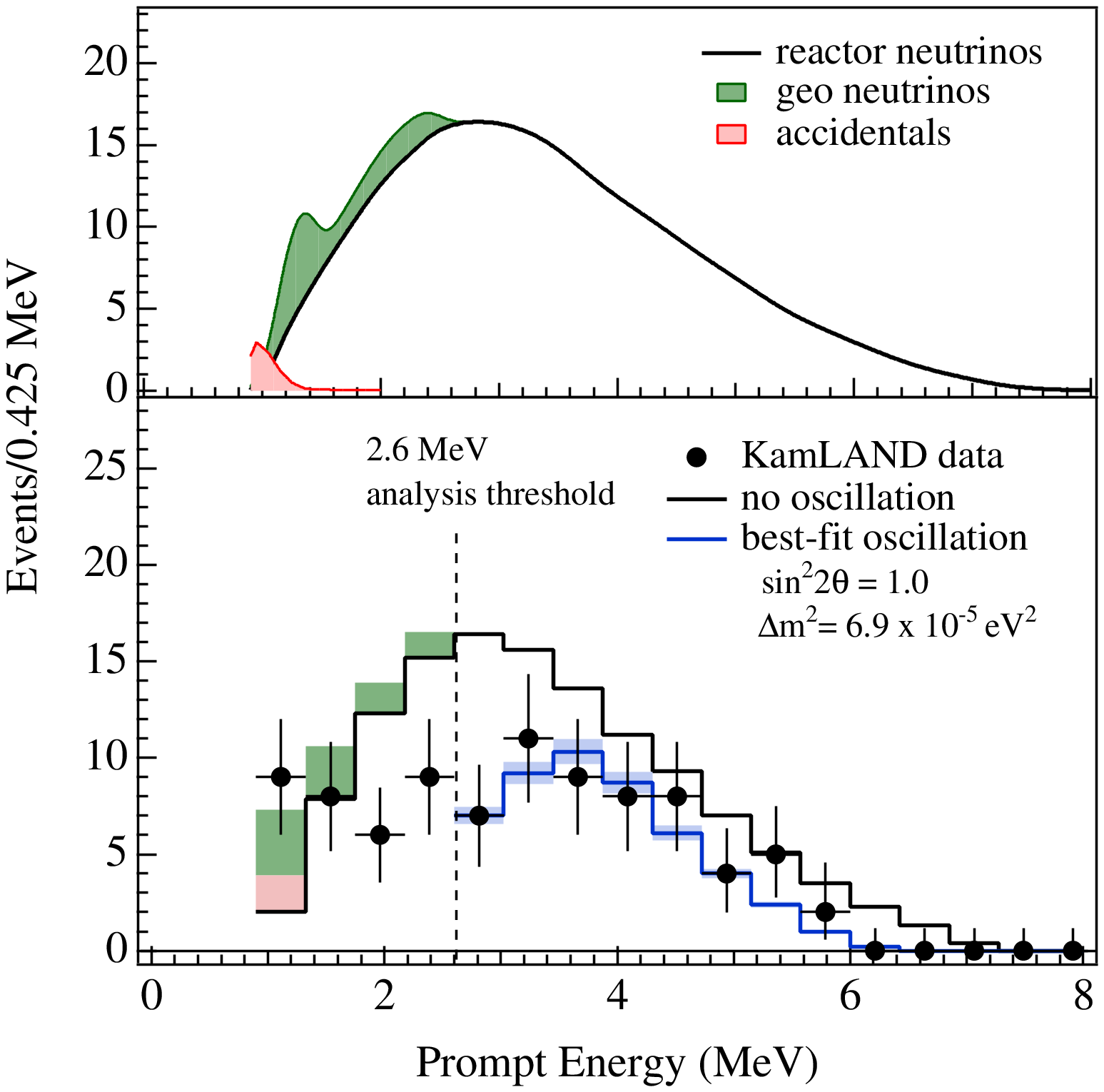}
\end{center}
\end{minipage}
\caption{ \label{kamland-0212021-1}
Left:
The ratio of measured to expected $\bar \nu_e$ flux from reactor 
experiments.
The shaded region indicates the range of flux 
predictions corresponding to the 95\% C.L. LMA region found in a 
global analysis of the solar neutrino data \cite{Fogli:2002pt}.
The dotted 
curve corresponds to the best-fit values
$\Delta{m}^{2}_{\mathrm{sol}} = 5.5 \times 10^{-5} \, \mathrm{eV}^{2}$
and
$\sin^{2}2\vartheta_{\mathrm{sol}} = 0.83$
found in Ref.~\cite{Fogli:2002pt}.
Right:
Upper panel: Expected reactor $\bar\nu_e$ energy spectrum with contributions
of $\bar\nu_{\mathrm{geo}}$
(antineutrinos emitted by $^{238}$U and $^{232}$Th decays in the Earth)
and accidental backround.
Lower panel: Energy spectrum of the observed prompt
events (solid circles with error bars),
along with the expected no oscillation spectrum (upper histogram, with $\bar\nu_{\mathrm{geo}}$
and accidentals shown) and best
fit (lower histogram) including neutrino
oscillations. The shaded band indicates the 
systematic error in the best-fit spectrum.
The vertical dashed line corresponds to the 
analysis threshold at 2.6 MeV.
Figures taken from Ref.~\cite{hep-ex/0212021}.
}
\end{figure}

\begin{figure}[t]
\begin{minipage}[l]{0.45\textwidth}
\begin{center}
\includegraphics*[bb=0 0 567 539, height=7cm]{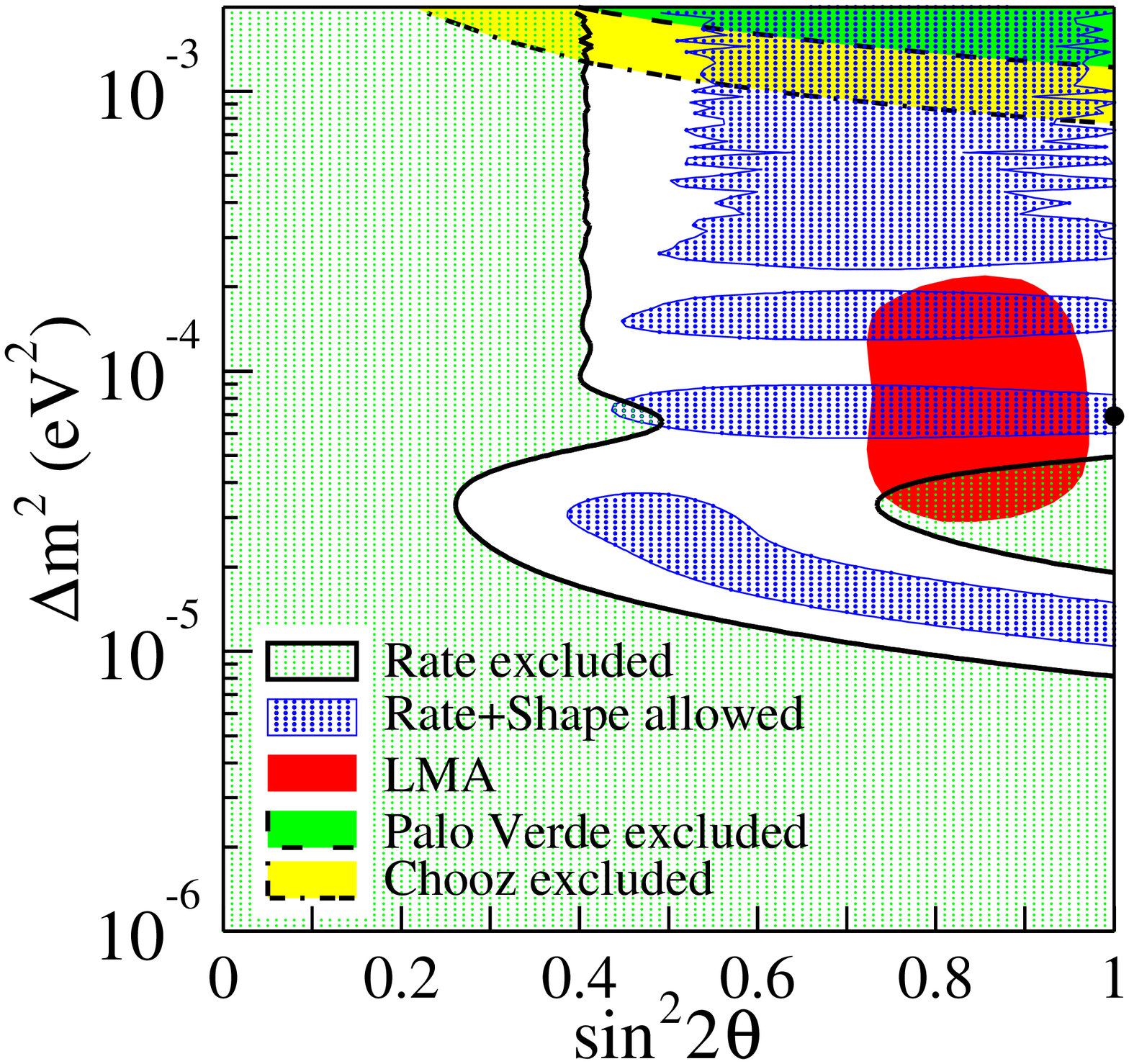}
\end{center}
\end{minipage}
\hfill
\begin{minipage}[l]{0.45\textwidth}
\begin{center}
\includegraphics*[bb=104 224 457 594, height=7cm]{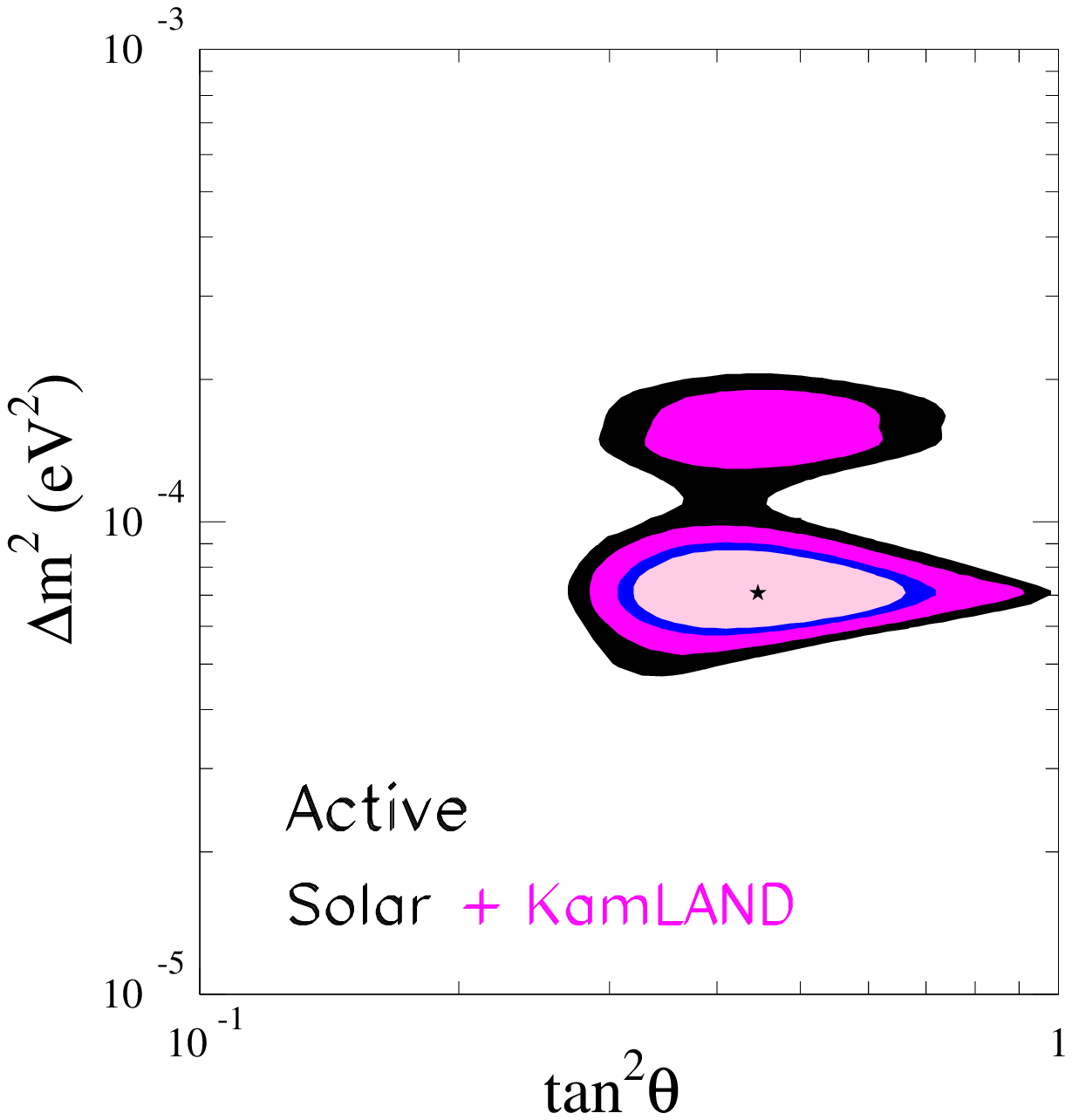}
\end{center}
\end{minipage}
\caption{ \label{kamland-0212021-2}
Left:
KamLAND
excluded regions of neutrino oscillation parameters
$\mathsf{\Delta{m}^{2}} = \Delta{m}^{2}_{\mathrm{KamLAND}}$
and
$\mathsf{\sin^{2}2\theta} = \sin^{2}2 \vartheta_{\mathrm{KamLAND}}$
for the rate analysis and allowed regions for the combined rate and
energy spectrum analysis at 95\% C.L.
At the top are the 95\% C.L. excluded region from CHOOZ \cite{Apollonio:1999ae}
and Palo Verde \cite{Boehm:2001ik} experiments, respectively.
The dark area is the 95\% C.L. LMA allowed
region obtained in Ref.~\cite{Fogli:2002pt}.
The thick dot indicates the best fit of
the KamLAND data in Eq.~(\ref{203}).
Figure taken from Ref.~\cite{hep-ex/0212021}.
Right:
Allowed region at
90\%, 95\%, 99\%, 99.73\% ($3\sigma$) C.L.
for the neutrino oscillation parameters
$\mathsf{\Delta{m}^{2}} = \Delta{m}^{2}_{\mathrm{sol}} = \Delta{m}^{2}_{\mathrm{KamLAND}}$
and
$\mathsf{\tan^{2}\theta} = \tan^{2}\vartheta_{\mathrm{sol}} = \tan^{2}\vartheta_{\mathrm{KamLAND}}$
obtained in Ref.~\cite{hep-ph/0212147}
from the combined analysis of
solar and KamLAND data
(see also Refs.~\cite{hep-ph/0212126,%
hep-ph/0212127,%
hep-ph/0212129,%
hep-ph/0212146,%
hep-ph/0212202,%
hep-ph/0212212,%
hep-ph/0212270,%
hep-ph/0301072}).
}
\end{figure}

\subsection{The first results of the KamLAND experiment}
\label{The first results of the KamLAND experiment}

Recently the first results of the KamLAND experiment,
started in January 2002,
have been published
\cite{hep-ex/0212021}.
In this experiment electron antineutrinos 
from many reactors in Japan and Korea are 
detected via the observation of the process
\begin{equation}
\bar\nu_{e}+p \to e^{+}+n
\,.
\label{201}
\end{equation}
The threshold energy of this process is
$ E_{\nu}^{\mathrm{th}} \simeq m_e + m_n - m_p = 1.8 \, \mathrm{MeV} $.
About 80\% of the $\bar\nu_{e}$ flux is expected  from 26 reactors 
with distances in the range 138-214 km.
The 1 kt liquid scintillator detector of the  KamLAND experiment is 
located in the Kamioka mine, in Japan, at a depth of about 1 km.
Both prompt photons from the annihilation of $e^{+}$ in the scintillator 
and
2.2 MeV delayed photons from the neutron capture $n+p\to d +\gamma$  
are detected (the mean neutron capture time is $188\pm 23\, \mu\mathrm{sec}$).
In order to avoid background, mainly from the decays of
$^{238}\mathrm{U}$ and $^{232}\mathrm{Th}$ in the Earth, the cut 
$E_{\mathrm{prompt}}>2.6\, \mathrm{MeV}$ was applied.

During 145.1 days of running 54 events were observed. The number of events
expected in the case of no neutrino oscillations is $86.8 \pm 5.6$.
The ratio of observed and expected $\bar\nu_{e}$ events is
\begin{equation}
\frac{N_{\mathrm{obs}}-N_{\mathrm{BG}}}{N_{\mathrm{expected}}} = 0.611 \pm 0.085 \pm 0.041
\,,
\label{202}
\end{equation}
where
$N_{\mathrm{BG}} = 0.95 \pm 0.99$
is the estimated number of background events.

The left figure in Fig.~\ref{kamland-0212021-1}
shows the dependence of the ratio of observed and expected 
$\bar\nu_{e}$ events on the average distance between reactors
and detectors for all reactor neutrino experiments.
The dotted curve was obtained with the best-fit solar neutrino LMA values
of the oscillation parameters
$\Delta{m}^{2}_{\mathrm{sol}} = 5.5 \times 10^{-5} \, \mathrm{eV}^{2}$
and
$\sin^{2}2\vartheta_{\mathrm{sol}} = 0.83$
obtained in Ref.~\cite{Fogli:2002pt}.

In the KamLAND experiment the prompt energy spectrum 
was also measured (see the right figure in Fig.~\ref{kamland-0212021-1}).
The prompt energy is connected with
the energy of $\bar\nu_{e}$ by the relation
$ E_{\mathrm{prompt}} = E_{\bar\nu_{e}} - 0.8\,\mathrm{MeV} - \overline{E_{n}} $
($\overline{E_{n}}$ is the average kinetic energy of the neutron
and
$ 0.8\,\mathrm{MeV} = m_n - m_p - m_e $,
with the electron neutrino mass coming from the annihilation
of the final positron in Eq.~(\ref{201}) with an electron in the medium).
From the two-neutrino analysis of the KamLAND spectrum the following best-fit
values of the oscillation parameters were obtained:
\begin{equation}
\Delta{m}^{2}_{\mathrm{KamLAND}} = 6.9 \times 10^{-5}\mathrm{eV}^{2}
\,,
\qquad
\sin^{2}2\vartheta_{\mathrm{KamLAND}} = 1
\,.
\label{203}
\end{equation}
The 95\% C.L. allowed regions in the plane of the oscillation parameters
obtained from the
analysis of the measured rate and energy spectrum are shown in
the left figure in Fig.~\ref{kamland-0212021-2}.
The region outside the solid line is excluded from the rate analysis.
The dark region is the solar neutrino LMA allowed region, obtained in
Ref.~\cite{Fogli:2002pt}.
One can see that two of the KamLAND allowed regions
overlap with the
solar neutrino LMA region.

The KamLAND result provides strong evidence of neutrino oscillations,
obtained for the first time with terrestrial reactor 
antineutrinos
with the initial flux well under control.
The KamLAND result allows to exclude the
SMA, LOW and VAC solutions of the solar neutrino problem.
It proves that the only viable solution of the problem is LMA.
The right figure in Fig.~\ref{kamland-0212021-2}
shows the allowed region for the oscillation parameters
obtained in Ref.~\cite{hep-ph/0212147}
from the combined analysis of
solar and KamLAND data
(see also Refs.~\cite{hep-ph/0212126,%
hep-ph/0212127,%
hep-ph/0212129,%
hep-ph/0212146,%
hep-ph/0212202,%
hep-ph/0212212,%
hep-ph/0212270,%
hep-ph/0301072}).

\subsection{CHOOZ and Palo Verde}
\label{CHOOZ and Palo Verde}

The results of the long-baseline 
reactor experiments CHOOZ \cite{Apollonio:1999ae}
and Palo Verde \cite{Boehm:2001ik},
in which
$\bar\nu_e$ disappearance
due to neutrino oscillations in the atmospheric range of $\Delta{m}^{2}$
was searched for, are very
important for the issue of neutrino mixing.
In these experiments
electron antineutrinos were detected 
via the observation of the process
\begin{equation}
\bar\nu_{e}+ p \to e^{+}+ n
\,.
\end{equation} 
No indication
in favor of the disappearance of reactor $\bar\nu_{e}$'s
was found.
The ratio $R$ of the measured and expected numbers of $\bar\nu_{e}$ events
in the CHOOZ \cite{Apollonio:1999ae} and the Palo Verde \cite{Boehm:2001ik} experiments
are,
respectively,
\begin{equation}
R =1.01 \pm 2.8 \% \pm 2.7 \%
\,,
\qquad
R =1.01 \pm 2.4 \% \pm 5.3 \%
\,.
\label{033}
\end{equation}
From the 95\% C.L.
exclusion plot obtained in Ref.~\cite{Apollonio:1999ae}
from the two-neutrino analysis of CHOOZ data,
for
$\Delta{m}^{2}_{\mathrm{CHOOZ}} = 2.5 \times 10^{-3} \, \mathrm{eV}^{2}$
(the S-K best-fit value for $\Delta{m}^{2}_{\mathrm{atm}}$,
see Eq.~(\ref{018}))
we have
\begin{equation}
\sin^{2}2\vartheta_{\mathrm{CHOOZ}} \lesssim 1.5 \times 10^{-1}
\,.
\label{034}
\end{equation}

\subsection{Phenomenology}
\label{Phenomenology}

In the minimal scheme with mixing of three massive neutrinos,
the $3 \times 3$
Pontecorvo-Maki-Nakagawa-Sakata \cite{Pontecorvo-58,Pontecorvo-68,MNS-62}
mixing matrix $U$ is characterized by
three mixing angles and one CP phase
(in the case of Dirac neutrinos;
in the case of Majorana neutrinos,
in the mixing matrix
there are two additional phases
which are irrelevant for neutrino oscillations).
Let us discuss now neutrino oscillations in the atmospheric and solar ranges  
of neutrino mass-squared differences in the framework of this scheme,
which provides two independent $\Delta{m}^{2}$'s: 
$\Delta{m}^{2}_{21}=m_{2}^{2}- m_{1}^{2}$ and  
$\Delta{m}^{2}_{32}=m_{3}^{2}- m_{2}^{2}$.

Two important features of the neutrino mixing, which were revealed in the
recent solar, atmospheric, long-baseline reactor and accelerator experiments,
determine neutrino oscillations.

\begin{figure}[t]
\begin{minipage}[l]{0.45\textwidth}
\begin{center}
\includegraphics*[bb=180 466 425 773, height=7cm]{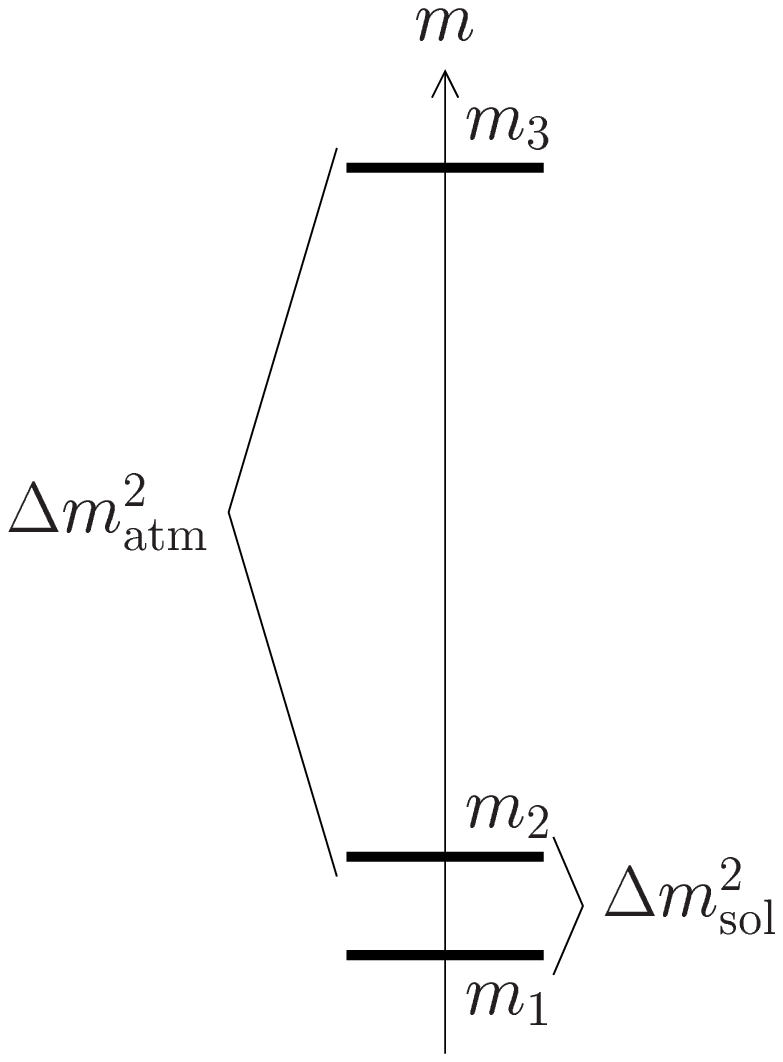}
\\
``normal''
\end{center}
\end{minipage}
\hfill
\begin{minipage}[l]{0.45\textwidth}
\begin{center}
\includegraphics*[bb=183 466 430 773, height=7cm]{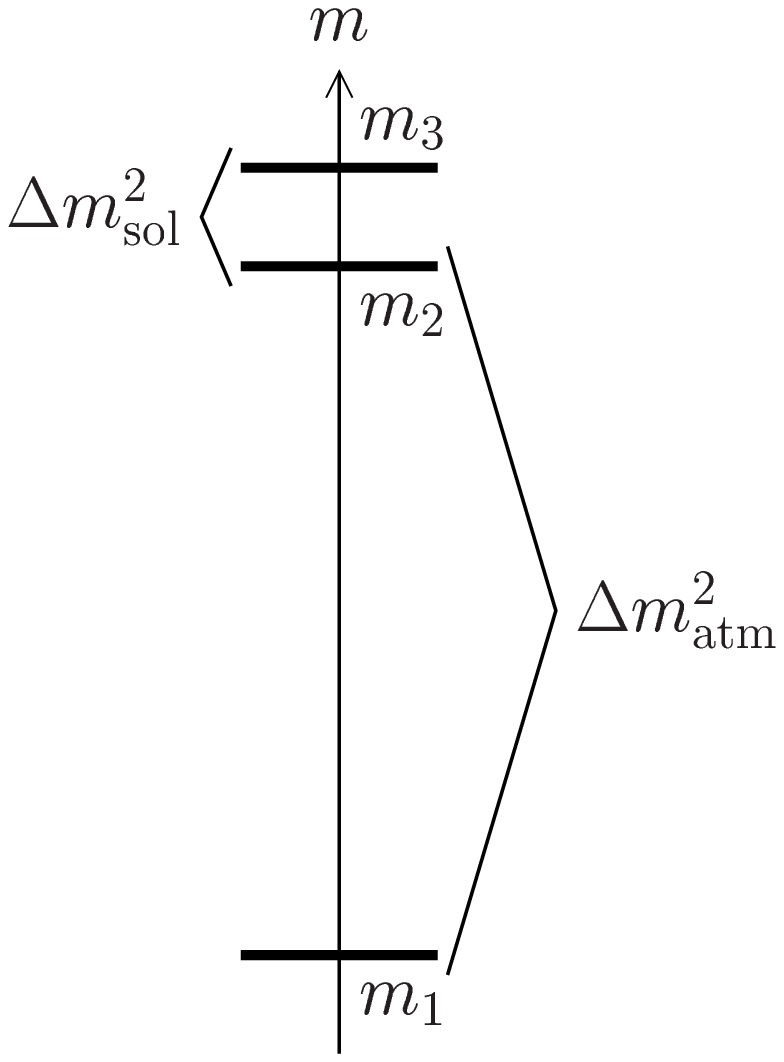}
\\
``inverted''
\end{center}
\end{minipage}
\caption{ \label{3nu}
The two types of three-neutrino mass schemes compatible
with the hierarchy
$\Delta{m}^{2}_{\mathrm{sol}} \ll \Delta{m}^{2}_{\mathrm{atm}}$.
}
\end{figure}

\begin{figure}[t]
\begin{minipage}[l]{0.45\textwidth}
\begin{center}
\includegraphics*[bb=118 426 465 753, height=7cm]{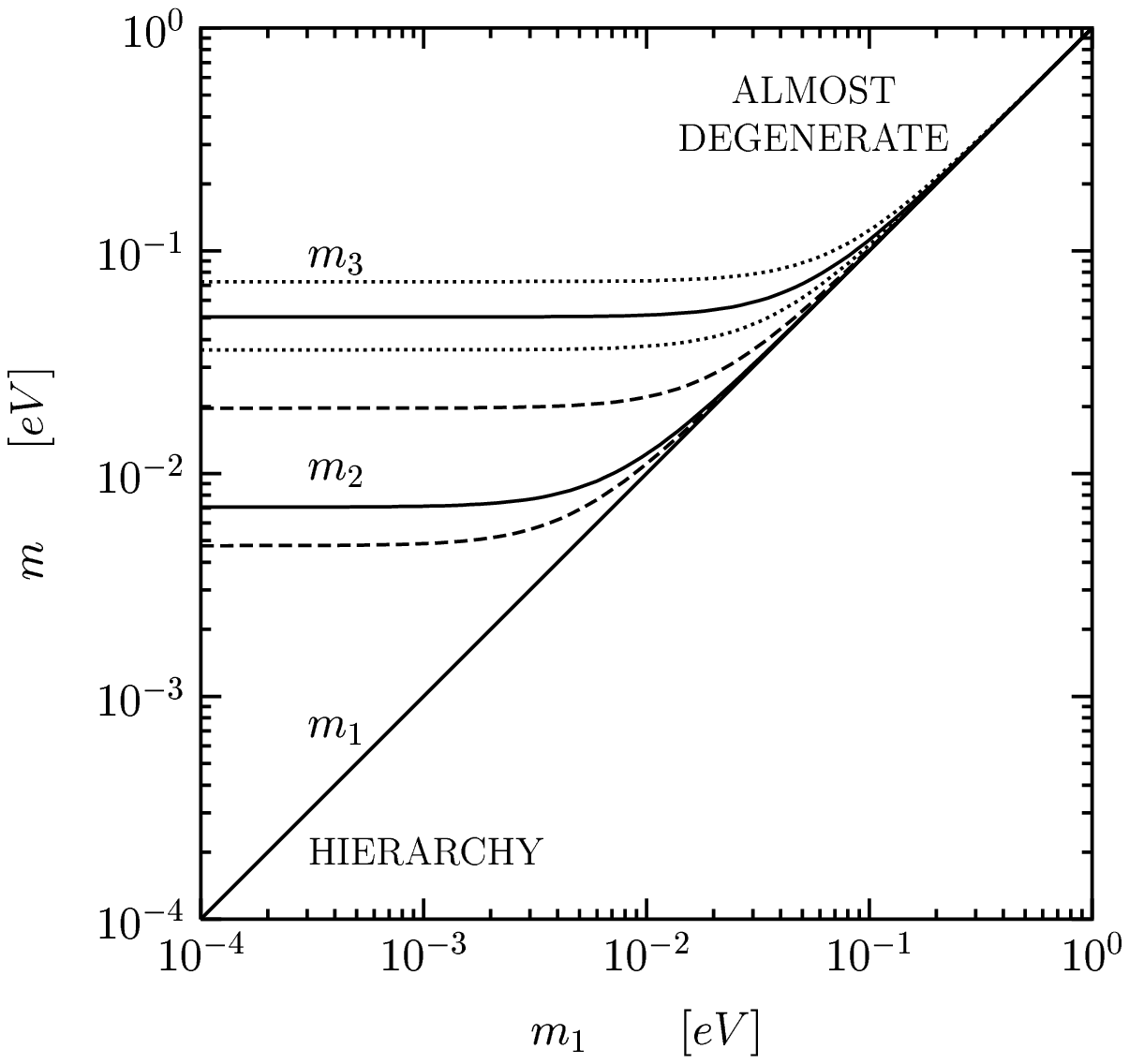}
\\
``normal''
\end{center}
\end{minipage}
\hfill
\begin{minipage}[l]{0.45\textwidth}
\begin{center}
\includegraphics*[bb=118 426 465 753, height=7cm]{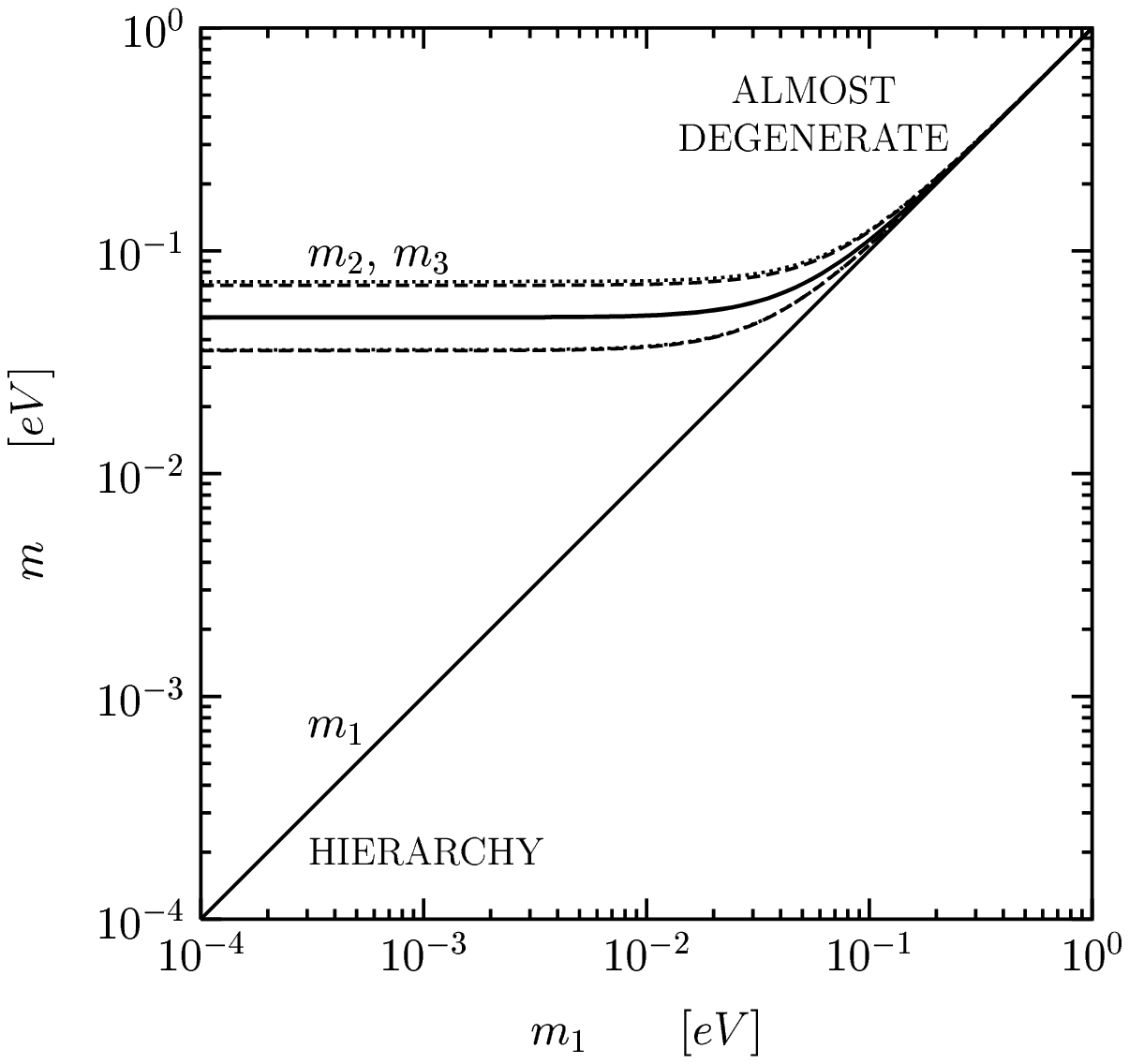}
\\
``inverted''
\end{center}
\end{minipage}
\caption{ \label{3m}
Neutrino masses
as functions of the lightest mass $m_1$
in the
two three-neutrino mass schemes
in Fig.~\ref{3nu}.
The dashed and dotted lines represent, respectively, the limits for $m_2$ and $m_3$
from solar and atmospheric neutrino data.
}
\end{figure}

The first feature is
\emph{the hierarchy of the neutrino mass-squared differences}:
from the analyses of the data of the solar and atmospheric neutrino experiments
it follows that
$\Delta{m}^{2}_{\mathrm{sol}} \ll \Delta{m}^{2}_{\mathrm{atm}}$.
This hierarchy can be realized only with
the two types of three-neutrino mass schemes\footnote{
Independently from the type of neutrino mass spectrum
(``normal'' or ``inverted''),
we label neutrino masses in such a way that $m_1 < m_2 < m_3$.
Another convention is often used in the literature,
such that in both ``normal'' and ``inverted''
mass spectra
$\Delta{m}^{2}_{\mathrm{sol}} \simeq \Delta{m}^{2}_{21} > 0$
and
$\Delta{m}^{2}_{\mathrm{atm}} \simeq |\Delta{m}^{2}_{32}|$.
In this notation,
$m_1 < m_2 < m_3$
in ``normal'' schemes
and
$m_3 < m_1 < m_2$
in ``inverted'' schemes.
}
shown in Fig.~\ref{3nu}.
The absolute scale of the neutrino masses in the two
schemes in Fig.~\ref{3nu} is not
fixed by neutrino oscillation experiments.
Figure~\ref{3m}
shows the neutrino masses as functions of the lightest mass $m_1$
(see Ref.~\cite{Beacom:2002cb}).
One can see that
the ``normal'' scheme in Fig.~\ref{3nu}
is compatible with the natural mass hierarchy (\ref{008})
if $m_1 \ll m_2$,
whereas in the ``inverted'' scheme
$\nu_2$ and $\nu_3$ are always almost degenerate.

Let us
first consider the
``normal'' mass scheme in Fig.~\ref{3nu},
with $\Delta{m}^{2}_{21} \simeq \Delta{m}^{2}_{\mathrm{sol}}$
and $\Delta{m}^{2}_{32} \simeq \Delta{m}^{2}_{\mathrm{atm}}$.
In this case we have
\begin{equation}
\Delta{m}^{2}_{21}\ll\Delta{m}^{2}_{32}
\,.
\label{035}
\end{equation}
For the values $L/E$ relevant for neutrino oscillations in the atmospheric
range of $\Delta{m}^{2}$
(atmospheric and long-baseline accelerator and reactor experiments)
we have
$\Delta{m}^{2}_{21} L / E \ll 1$.
Hence, we can neglect
the contribution of
$\Delta{m}^{2}_{21}$ to 
the transition probability in Eq.~(\ref{012}).
For the probability of the transition 
$\nu_\alpha\to\nu_{\alpha'}$,
we obtain
(see Ref.~\cite{BGG-review-98})
\begin{equation}
\mathrm{P}(\nu_\alpha\to\nu_{\alpha'}) =
\frac{1}{2} \, \mathrm{A}_{{\alpha'};\alpha}
\left( 1 - \cos \Delta{m}^{2}_{32} \frac{L}{2E} \right)
\qquad
(\alpha\neq\alpha')
\,,
\label{036}
\end{equation}
where the oscillation amplitude is given by
\begin{equation}
\mathrm{A}_{{\alpha'};\alpha}
=
4 \, |U_{\alpha' 3}|^{2} \, |U_{\alpha 3}|^{2}
\label{037}
\end{equation}
In the standard parametrization of the mixing matrix
(see Ref.~\cite{PDG})
we have
\begin{equation}
U_{e3} = \sin\vartheta_{13} \, e^{-i\delta}
\,,
\qquad
U_{\mu 3} = \sqrt{ 1 -|U_{e3}|^{2}}\, \sin\vartheta_{23}
\,,
\qquad
U_{\tau 3} = \sqrt{ 1 -|U_{e3}|^{2}}\, \cos\vartheta_{23}
\,,
\label{038}
\end{equation}
where $\vartheta_{13}$ and $\vartheta_{23}$ are mixing angles
and $\delta$ is the CP-violating phase.
Hence, for the amplitude of the transitions $\nu_{\mu}\to\nu_{\tau}$
and $\nu_{\mu}\to\nu_{e}$ we obtain, respectively,
\begin{equation}
\mathrm{A}_{\tau;\mu}
=
\left(1 -|U_{e3}|^{2}\right)^{2} \sin^{2}2\vartheta_{23}
\,,
\qquad
\mathrm{A}_{e;\mu}
=
4 \, |U_{e3}|^{2} \left( 1 -|U_{e3}|^{2} \right) \sin^{2}\vartheta_{23}
\,.
\label{039}
\end{equation}
The probability of $\nu_{\alpha}$ to survive is given by
\begin{equation}
\mathrm{P}(\nu_\alpha\to\nu_\alpha)
=
1 - \sum_{\alpha'\neq\alpha} \mathrm{P}(\nu_\alpha\to\nu_{\alpha'})
=
1 - \frac{1}{2} \, \mathrm{B}_{\alpha;\alpha}
\left( 1 - \cos \Delta{m}^{2}_{32} \frac{L}{2E} \right)
\,.
\label{040}
\end{equation}
where
\begin{equation}
\mathrm{B}_{\alpha;\alpha}
=
\sum_{\alpha'\neq\alpha}\,
\mathrm{A}_{{\alpha'};\alpha}=
4 \, |U_{\alpha 3}|^{2} \left(1 -|U_{\alpha 3}|^{2}\right)
\,.
\label{041}
\end{equation}
Thus, due to the hierarchy in Eq.~(\ref{035}),
the transition probabilities in the
atmospheric range of $\Delta{m}^{2}$ have a two-neutrino form.
Taking into account that the elements  $|U_{\alpha 3}|^{2}$,
which determine the oscillation amplitudes,
satisfy the unitarity condition $\sum_{\alpha}|U_{\alpha 3}|^{2}=1$,
we conclude that the transition probabilities are characterized by three parameters:
$\Delta{m}^{2}_{32}$,
$\sin^2 2\vartheta_{23}$
and
$|U_{e3}|^2$.

Let us consider now neutrino oscillations in the solar range of 
$\Delta{m}^{2}$. 
The $\nu_{e}$ survival probability in vacuum
can be written in the form
\begin{equation}
\mathrm{P}(\nu_e\to\nu_e)
=
\left|
\sum_{i=1, 2}| U_{e i}|^2 \, 
e^{ - i
\, 
\Delta{m}^2_{i1} \frac{L}{2 E} }
+ | U_{e3}|^2  \, 
e^{ - i
\, 
\Delta{m}^2_{31} \frac{L}{2 E} }\,\right|^2
\label{042}
\end{equation}
We are interested in the survival probability averaged over
the region where neutrinos are produced,
over the detector energy resolution,
etc..
Because of the neutrino mass-squared hierarchy in Eq.~(\ref{035}), in 
the expression for the averaged survival probability the interference between
the first and second terms in (\ref{042}) vanishes and we obtain
\begin{equation}
\mathrm{P}(\nu_{e}\to\nu_{e})
=
|U_{e3}|^{4} + \left(1-|U_{e3}|^{2}\right)^{2}
P^{(1,2)}(\nu_{e}\to\nu_{e})
\,,
\label{043}
\end{equation}
where
\begin{equation}
\mathrm{P}^{(1,2)}(\nu_e\to\nu_e)
=
1 - \frac{1}{2} \, A^{(1,2)}
\left( 1 - \cos \Delta{m}^{2}_{21} \frac{L}{2E} \right)
\,,
\label{044}
\end{equation}
and
\begin{equation}
A^{(1,2)}
=
4
\,
\frac
{ |U_{e1}|^{2} \, |U_{e2}|^{2}} 
{ \left( 1-   |U_{e3}|^{2} \right)^{2} }
=
\sin^{2}2\vartheta_{12}
\,.
\label{045}
\end{equation}
We have used the standard parametrization of the neutrino mixing matrix
(see Eq.~(\ref{038}) and Ref.~\cite{PDG}),
with
\begin{equation}
U_{e1} = \sqrt{ 1 -|U_{e3}|^{2}} \, \cos\vartheta_{12}
\,,
\qquad
U_{e2} = \sqrt{ 1 -|U_{e3}|^{2}} \, \sin\vartheta_{12}
\,,
\label{046}
\end{equation}
where $\vartheta_{12}$
is a mixing angle.
The expression (\ref{043}) is also valid in the case
of oscillations in matter \cite{Shi:1992zw}.
In this case
$P^{(1,2)}(\nu_{e}\to\nu_{e})$
is the two-neutrino survival probability in matter
calculated with the charged-current matter potential
$V_{\mathrm{CC}}$
multiplied by
$(1-|U_{e3}|^2)$.

The second important feature of the neutrino mixing is the
\emph{smallness of the parameter $|U_{e3}|^{2}$}.
This follows from the results of the CHOOZ and Palo Verde
experiments and from the results of solar neutrino experiments.
In the CHOOZ and Palo Verde experiments,
the probability of 
$\bar\nu_e$ to survive
is
\begin{equation}
\mathrm{P}(\bar\nu_e\to\bar\nu_e)
=
1 - \frac{1}{2} \,
\mathrm{B}_{e; e}
\left( 1 - \cos \Delta{m}^{2}_{32} \frac{L}{2E} \right)
\,,
\label{047}
\end{equation}
where
\begin{equation}
\mathrm{B}_{e;e}
=
4 \, |U_{e3}|^{2} \left( 1 -|U_{e3}|^{2} \right)
\,.
\label{048}
\end{equation}
From the two-neutrino exclusion plots,
obtained in
the CHOOZ and Palo Verde experiments
\cite{Apollonio:1999ae,Boehm:2001ik},
it follows that
\begin{equation}
\mathrm{B}_{e;e} \leq \mathrm{B}_{e;e}^{0}
\,,
\label{049}
\end{equation}
where the upper bound $\mathrm{B}_{e;e}^{0}$
depends on 
$\Delta{m}^{2}_{32}$.

For the S-K
\cite{Fukuda:1998mi,Fukuda:1998ah,Fukuda:2000np}
allowed values of $\Delta{m}^{2}_{32}$,
from the 95\% C.L. CHOOZ exclusion plot we find
\begin{equation}
1 \times 10^{-1} \lesssim \mathrm{B}_{e;e}^{0} \lesssim 2.4 \times 10^{-1}
\,.
\label{050}
\end{equation}
Thus, the parameter $|U_{e3}|^{2}$ can be small or large (close to one).
This last possibility is excluded by the
solar neutrino data (see \cite{BGG-review-98}).
At the
S-K best-fit point 
$\Delta{m}^{2}_{32}=2.5\times 10^{-3}\mathrm{eV}^{2}$
we have
\begin{equation}
|U_{e3}|^{2} \leq 4 \times 10^{-2}
\qquad
(95\% \, \mathrm{C.L.})
\,.
\label{051}
\end{equation}
A combined fit of all data leads to
\cite{Fogli:2002pb}
\begin{equation}
|U_{e3}|^{2} \leq 5 \times 10^{-2}
\qquad
(99.73\% \, \mathrm{C.L.})
\,.
\label{0511}
\end{equation}

There are three major consequences of the neutrino mass-squared hierarchy
(\ref{035}) and of the smallness of $|U_{e3}|^{2}$:
\begin{enumerate}

\item
The dominant transition in the atmospheric range of
$\Delta{m}^{2}$ is $\nu_{\mu}\to\nu_{\tau}$.
From Eq.~(\ref{039})
it follows that 
\begin{equation}
\Delta{m}^{2}_{32} \simeq \Delta{m}^{2}_{\mathrm{atm}}
\,,
\qquad
\sin^{2}2\vartheta_{23} \simeq \sin^{2}2\vartheta_{\mathrm{atm}}
\,.
\end{equation}

\item
In the solar range of $\Delta{m}^{2}$ the probability of $\nu_{e}$ 
to survive has the
two-neutrino form
\begin{equation}
\mathrm{P}(\nu_e\to\nu_e) \simeq \mathrm{P}^{(1,2)}(\nu_e\to\nu_e)
\,.
\label{052}
\end{equation}
From Eqs.~(\ref{044}) and (\ref{045}) it follows that
\begin{equation}
\Delta{m}^{2}_{21} \simeq \Delta{m}^{2}_{\mathrm{sol}}
\,,
\qquad
\tan^{2}\vartheta_{12}\simeq \tan^{2}\vartheta_{\mathrm{sol}}
\,.
\label{0521}
\end{equation}

\item
Neutrino oscillations in the atmospheric and solar ranges of $\Delta{m}^{2}$ 
in the leading approximation are decoupled \cite{Bilenky-Giunti-CHOOZ-98}.
Oscillations in both ranges are described by two-neutrino
formulas, which are characterized, respectively, by the parameters 
$\Delta{m}^{2}_{32}$,
$\sin^2 2\vartheta_{23}$
and
$\Delta{m}^{2}_{21}$,
$\tan^2\vartheta_{12}$.

\end{enumerate}

We have considered
the hierarchy (\ref{035}) of the neutrino mass-squared differences,
which is realized in the ``normal'' mass scheme in
Fig.~\ref{3nu}.
The data of neutrino oscillation experiments are compatible also with
the ``inverted''
mass scheme in
Fig.~\ref{3nu},
with
$\Delta{m}^{2}_{21} \simeq \Delta{m}^{2}_{\mathrm{atm}}$ and
$\Delta{m}^{2}_{32} \simeq \Delta{m}^{2}_{\mathrm{sol}}$.
In this case
an inverted hierarchy of the mass-squared differences
takes place\footnote{
The type
of the neutrino mass spectrum (``normal'' or ``inverted'')
may be determined via the investigation of
$\nu_{e}\to\nu_{\mu}$ and $\bar\nu_{e}\to\bar\nu_{\mu}$
oscillations in future long-baseline experiments
if $|U_{e3}|^2$ is not too small 
(see Ref.~\cite{Lindner:2002vt}).
The distance between the neutrino source and the detector
in such experiments must be large enough
for the matter effects to be sizable.
}:
\begin{equation}
\Delta{m}^{2}_{32} \ll \Delta{m}^{2}_{21} 
\label{053}
\end{equation}
The expressions for the transition probabilities
in the case of the inverted hierarchy can be obtained from
Eqs.~(\ref{036}), (\ref{037}), (\ref{040}),
(\ref{041}), (\ref{043})-(\ref{045}) with the change
$\Delta{m}^{2}_{32} \leftrightarrows \Delta{m}^{2}_{21}$,
$|U_{\alpha 3}|^{2} \leftrightarrows |U_{\alpha 1}|^{2}$,
$\vartheta_{12} \leftrightarrows \vartheta_{23}$,
$P^{(1,2)} \to P^{(2,3)}$,
$A^{(1,2)} \to A^{(2,3)}$.

We have discussed up to now evidences in favor of neutrino oscillations that
have been obtained in the atmospheric and solar neutrino experiments.
There exist at present also an indication in favor of
short-baseline
$\bar\nu_{\mu}\to\bar\nu_{e}$
transitions, which has been obtained only in the accelerator
experiment LSND \cite{LSND}.
The LSND data can be explained
by neutrino oscillations.
From a two-neutrino analysis of the data,
the best-fit values of the oscillation parameters are
\begin{equation}
\Delta{m}^{2}_{\mathrm{LSND}}\simeq 1.2\,\mathrm{eV}^{2}
\,,
\qquad
\sin^{2}2\vartheta_{\mathrm{LSND}}\simeq 3\times 10^{-3}
\,.
\label{054}
\end{equation}

In order to describe
the results of the solar, atmospheric and LSND experiments,
which require three
different values of neutrino mass-squared differences
$\Delta{m}^{2}_{\mathrm{sol}}$, $\Delta{m}^{2}_{\mathrm{atm}}$ and
$\Delta{m}^{2}_{\mathrm{LSND}}$, 
it is necessary to assume 
that at least one
sterile neutrino exists in addition to
the three active neutrinos $\nu_{e}$, $\nu_{\mu}$, $\nu_{\tau}$.
In the mass basis,
in addition to the three light neutrinos $\nu_{1}$,
$\nu_{2}$, $\nu_{3}$ there must be
at least
one neutrino with mass of the order
$\sqrt{\Delta{m}^{2}_{\mathrm{LSND}}}\simeq 1\,\mathrm{eV} $
(see, for example, Ref.~\cite{BGG-review-98}).
However,
in spite of the additional degrees of freedom,
schemes with four neutrinos
do not fit well the data
(see Refs.~\cite{BGGS-AB-99,Maltoni:2002xd})\footnote{
Since there is no experimental indication in favor of
transitions of active neutrinos into sterile states,
schemes with more than four neutrinos are disfavored as well.
}.

The result of the LSND experiment requires, however, confirmation. 
The MiniBooNE experiment at Fermilab \cite{MiniBooNE-Nu2002},
that started recently,
is aimed to check the LSND result.

From neutrino oscillation experiments we can obtain information
only on the neutrino mass-squared differences,
not on the absolute values of neutrino masses.
The great advantage of neutrino oscillations experiments,
that was stressed in the early papers on neutrino oscillations
\cite{Pontecorvo-58,Pontecorvo-68,Bilenky-Pontecorvo-PR-78},
is that they are sensitive to very small values of $\Delta{m}^{2}$. 
This is connected with the fact that 
neutrino oscillations are an
interference phenomenon. It is also important that there is
the possibility to perform experiments with detectors at very large
distances from neutrino sources (solar, atmospheric and long-baseline experiments)
and for small neutrino energies (solar and reactor experiments).

The understanding of the origin of neutrino 
masses and neutrino mixing requires knowledge of \emph{the absolute values of neutrino masses}
(see Refs.~\cite{King:2002gx,hep-ph/0206077}).
The problem of the absolute values of neutrino masses is one of the 
most challenging problems of the physics of massive and mixed neutrinos.
At present there are only upper bounds for the absolute values 
of neutrino masses. The most stringent bound was obtained
from the experiments on the 
measurement of the high energy part of the 
$\beta$-spectrum of $^{3}\mathrm{H}$.
In the next section we will discuss 
the results of these experiments and future prospects.

\section{Neutrino mass from $\beta$-decay experiments}
\label{Neutrino mass from beta-decay experiments}

\subsection{The $\beta$-spectrum in the case of neutrino mixing}
\label{The beta-spectrum in the case of neutrino mixing}

The method of measurement of the
neutrino mass through
the detailed investigation of the high-energy part
of the $\beta$-spectrum
was proposed in 1934 by Fermi in his classical paper on the 
theory of $\beta$-decay \cite{Fermi-ZP-34}
and by Perrin \cite{Perrin-1933}.
The first experiments on the measurement of the neutrino
mass with this method have been done in 1948-49
\cite{Curran-Angus-Cockcroft-1948,Hanna-Pontecorvo-49}.

Usually,
the neutrino mass is measured through the measurement
of the high energy part of the $\beta$-spectrum of tritium
\begin{equation}
{}^{3}\mathrm{H} \to {}^{3}\mathrm{He} + e^{-} + \bar\nu_{e}
\,.
\label{055}
\end{equation}
The investigation of this decay has several advantages.
Since tritium decay is superallowed, the nuclear matrix element
is a constant and the
electron spectrum is determined by the phase space.
It has a relatively small energy release
$E_{0}\simeq 18.6 \, \mathrm{keV}$
and a convenient lifetime ($T_{1/2} \simeq 12.3$ years).
A small value of $E_{0}$
is convenient because
the relative fraction of events in the high energy part of the spectrum,
which is sensitive to the neutrino mass, is
proportional to $E_{0}^{-3}$.

Let us consider the decay (\ref{055}) 
in the case of nonzero neutrino masses and neutrino mixing.
The effective Hamiltonian of the process is
\begin{equation}
\mathcal{H}_{I}^{\mathrm{CC}}
=
\frac{G_{F}}{\sqrt{2}} \, 2
\, \bar e_{L} \gamma _{\alpha} \nu_{eL} \,
j^{\alpha} + \mathrm{h.c.}
\,,
\label{056}
\end{equation}
where $j^{\alpha}$ is the hadronic charged current and
the field $\nu_{eL}$ is given by (see Eq.~(\ref{002}))
\begin{equation}
\nu_{eL} = \sum_{i} U_{ei} \nu_{iL}
\,.
\label{057}
\end{equation}
The state of the final particles in the decay (\ref{055})  
is
\begin{equation}
|f \rangle
=
\sum_{i}
|P'\rangle
\,
|p\rangle
\,
|p_{i}\rangle
\,
U^{*}_{ei}
\,
\langle p,p_{i},P'|(S-1)|P \rangle
\,.
\label{058}
\end{equation}
Here $p$ is the momentum of the electron,  $P$ and $P'$
are the momenta of the initial and final nuclei, $p_{i}$
is the
antineutrino momentum,
$|P' \rangle$,
$|p \rangle$,
$|p_{i} \rangle$
are the normalized states
of the final nucleus, electron, antineutrino with mass $m_{i}$,
and
\begin{equation}
\langle p,p_{i},P'|(S-1)|P \rangle
=
\langle p,p_{i},P'|T|P \rangle \, (2\pi)^{4} \, \delta^4(p+p_{i}+P'-P)
\label{059}
\end{equation}
is the element of the $S$-matrix.

We are interested in the spectrum of electrons. 
After the integration
over the momenta $\vec{P'}$,
$\vec{p_{i}}$
and over the angle of emission of the electron,
we have
\begin{equation}
\frac{\mathrm{d}\Gamma}{\mathrm{d}E}
=
\sum_{i}|U_{ei}|^{2} \,
\frac{\mathrm{d}\Gamma_{i}}{\mathrm{d}E}
\,,
\label{060}
\end{equation}
where
\begin{eqnarray}
\frac{\mathrm{d}\Gamma_{i}}{\mathrm{d}E}
= C\,p\,(E+m_{e})\,(E_{0}-E)\,
\sqrt{(E_{0}-E)^{2}-m_{i}^{2}}\,F(E)\,\theta(E_{0}-E-m_{i})
\,.
\label{061}
\end{eqnarray}

Here $E$
is the kinetic energy of the electron,
$E_{0}$ is the energy released
in the decay, $ m_{e}$ is the mass of the electron and $F(E)$ 
is the Fermi function,
 which describes the
Coulomb interaction of the final particles. The constant $C$ is given by  
\begin{equation}
C = G_{F}^{2}\frac{m_{e}^{5}}{2\,\pi^{3}}\,\cos^{2}\theta_{C}\,|M|^{2}
\,,
\end{equation}
where $G_{F}$ is the Fermi constant, $\theta_{C}$ is the Cabibbo angle,
$M$ is the nuclear matrix element (a constant).

Let us notice that neutrino masses enter
in the expression (\ref{061}) through the neutrino momentum
$|\vec p_{i}|= \sqrt{(E_{0}-E)^{2}-m_{i}^{2}}$. The step function
$\theta(E_{0}-E-m_{i})$ provides the condition $E \leq E_{0}-m_{i}$.
The recoil of the final nucleus
was neglected in Eq.~(\ref{061}).

Two experiments on the measurement of neutrino masses
with the tritium method are going on at present
(Mainz \cite{Weinheimer:1999tn,Bonn:2002jw,hep-ex/0210050}
and
Troitsk \cite{Lobashev:1999tp,Lobashev:2001uu}).
The sensitivity of these experiments to the
neutrino mass is about 2-3 eV.
The sensitivity to the neutrino mass of the future experiment KATRIN
\cite{Osipowicz:2001sq}
is expected to be about one order of magnitude better (0.35 eV).
We will discuss the results of the Mainz and Troitsk experiments later.
Now we consider the possibility to determine the neutrino mass
from the results of the $\beta$-decay experiments
for different neutrino mass spectra,
having in mind these sensitivities.

As it is seen from Eq.~(\ref{061}),
the largest distortion of the $\beta$-spectrum
due to neutrino masses can be observed in the region 
\begin{equation}
E_{0}-E \simeq m_{i}
\,.
\label{062}
\end{equation}
However, for $m_{i} \simeq 1 \, \mathrm{eV}$ 
only a very small part (about $ 10^{-13}$) of the decays of tritium
give contribution to the region (\ref{062}).
This is the reason why in the analysis of the results of the measurement of
the tritium $\beta$-spectrum
a relatively large part of the spectrum is used
(for example, in the Mainz experiment the last
70 eV of the spectrum). Taking this into account,
the tritium $\beta$-spectrum that is used for the fit of the data
can be presented in the  
form
\cite{McKellar:1980cn,Holzschuh:1992xy,Weinheimer:1999tn,Vissani:2000ci,Feruglio:2002af}
(see also the discussion in Ref.~\cite{Farzan:2002zq})
\begin{equation}
\frac{\mathrm{d}\Gamma}{\mathrm{d}E}
=
C\,p\,(E+m_{e})\,(E_{0}-E)\,
\sqrt{(E_{0}-E)^{2}-m_{\beta}^{2}}
\,
F(E)
\,,
\label{063}
\end{equation}
where the effective mass $m_{\beta}$
is given by
\begin{equation}
m^{2}_{\beta}= \sum_{i}|U_{ei}|^{2}\,m^{2}_{i}
\label{064}
\end{equation}

Let us consider first the minimal scheme with three massive 
neutrinos $\nu_1$, $\nu_2$ and  $\nu_3$.
The minimal neutrino mass $m_{1}$ and
the character
(``normal'' or ``inverted'', hierarchical or almost degenerate)
of the neutrino mass spectrum 
are unknown at present.
Neutrino oscillation experiments
allow to determine the
neutrino mass-squared differences $\Delta{m}^{2}_{21}$ and
$\Delta{m}^{2}_{32}$.
Hence,
it is possible to express the values of the neutrino masses $m_{2}$
and $m_{3}$ in terms of the unknown mass $m_1$ as
\begin{equation}
m_{2} = \sqrt{ m^{2}_{1} + \Delta{m}^{2}_{21} }
\,,
\qquad
m_{3} = \sqrt{ m^{2}_{1} + \Delta{m}^{2}_{21} + \Delta{m}^{2}_{32} }
\,.
\label{065}
\end{equation}

In the ``normal'' three-neutrino scheme in Fig.~\ref{3nu},
$\Delta{m}^{2}_{21} = \Delta{m}^{2}_{\mathrm{sol}}$
and
$\Delta{m}^{2}_{32} = \Delta{m}^{2}_{\mathrm{atm}}$.
Using Eqs.~(\ref{046}) and (\ref{0521}),
we obtain
\begin{equation}
m_{\beta}^2
=
m_1^2
+
\left(
\sin^{2}\vartheta_{\mathrm{sol}}
+
\cos^{2}\vartheta_{\mathrm{sol}} \, |U_{e3}|^{2}
\right)
\Delta{m}^{2}_{\mathrm{sol}}
+
|U_{e3}|^{2} \, \Delta{m}^{2}_{\mathrm{atm}}
\,.
\label{0651}
\end{equation}

In the case of the natural neutrino mass hierarchy (\ref{008}), we have
\begin{equation}
m_{2}
\simeq
\sqrt{\Delta{m}^{2}_{\mathrm{sol}}}
\simeq 7 \times 10^{-3} \, \mathrm{eV}
\,,
\qquad
m_{3}
\simeq
\sqrt{\Delta{m}^{2}_{\mathrm{atm}}}
\simeq
5 \times 10^{-2} \, \mathrm{eV}
\,,
\label{066}
\end{equation}
where the best-fit values
(\ref{018}) and (\ref{032})
of the oscillation parameters were used.

The contribution of the heaviest neutrino mass $m_{3}$
to the effective neutrino mass
(\ref{064})
enters with the weight $|U_{e3}|^{2}$, for which
we have the upper bound 
(\ref{0511})
from the results of the CHOOZ experiment \cite{Apollonio:1999ae}.
Taking into account this bound and using the best-fit
value in Eq.~(\ref{032}) for $\vartheta_{\mathrm{sol}}$,
for the effective neutrino mass $m_{\beta}$ 
we obtain
\begin{equation}
m_{\beta}
\simeq
\left(
\sin^{2}\vartheta_{\mathrm{sol}} \, \Delta{m}^{2}_{\mathrm{sol}}
+
|U_{e3}|^{2} \, \Delta{m}^{2}_{\mathrm{atm}} \right)^{1/2}
\lesssim
1.2 \times 10^{-2}\mathrm{eV}
\,,
\label{067}
\end{equation}
which is about one order of magnitude smaller than the sensitivity of
the future tritium experiment KATRIN \cite{Osipowicz:2001sq}. 

In the ``inverted'' three-neutrino scheme in Fig.~\ref{3nu},
with
$\Delta{m}^{2}_{21} = \Delta{m}^{2}_{\mathrm{atm}}$
and
$\Delta{m}^{2}_{32} = \Delta{m}^{2}_{\mathrm{sol}}$,
using Eq.~(\ref{046}),
we obtain
\begin{equation}
m_{\beta}^2
=
m_1^2
+
\left( 1 - |U_{e1}|^{2} \right)
\left(
\Delta{m}^{2}_{\mathrm{atm}}
+
\sin^{2}\vartheta_{\mathrm{sol}}
\,
\Delta{m}^{2}_{\mathrm{sol}}
\right)
\,.
\label{0652}
\end{equation}
The value of $|U_{e1}|^{2}$ in the ``inverted'' neutrino scheme
is bounded by the results of the CHOOZ experiment as
the value of $|U_{e3}|^{2}$ in the ``normal'' neutrino scheme
(see Eq.~(\ref{0511}) and the remark after Eq.~(\ref{053})):
\begin{equation}
|U_{e1}|^{2} \leq 5 \times 10^{-2}
\qquad
(99.73\% \, \mathrm{C.L.})
\,.
\label{0512}
\end{equation}

In the case of the ``inverted'' neutrino mass hierarchy
$m_{1}\ll m_{2}< m_{3}$
we have 
\begin{equation}
m_{2}
\simeq
m_{3}
\simeq
\sqrt{\Delta{m}^{2}_{\mathrm{atm}}}
\simeq
5 \times 10^{-2} \, \mathrm{eV}
\,,
\qquad
m_{1}
\ll
\sqrt{\Delta{m}^{2}_{\mathrm{atm}}}
\,.
\label{068}
\end{equation}
For the effective neutrino mass $m_{\beta}$
we obtain
\begin{equation}
m_{\beta}
\simeq
\sqrt{\Delta{m}^{2}_{\mathrm{atm}}}
\simeq
5 \times 10^{-2} \, \mathrm{eV}
\,,
\label{069}
\end{equation}
which is also much smaller than the sensitivity of the 
KATRIN experiment.

\begin{figure}[t]
\begin{minipage}[l]{0.45\textwidth}
\begin{center}
\includegraphics*[bb=118 426 465 753, height=7cm]{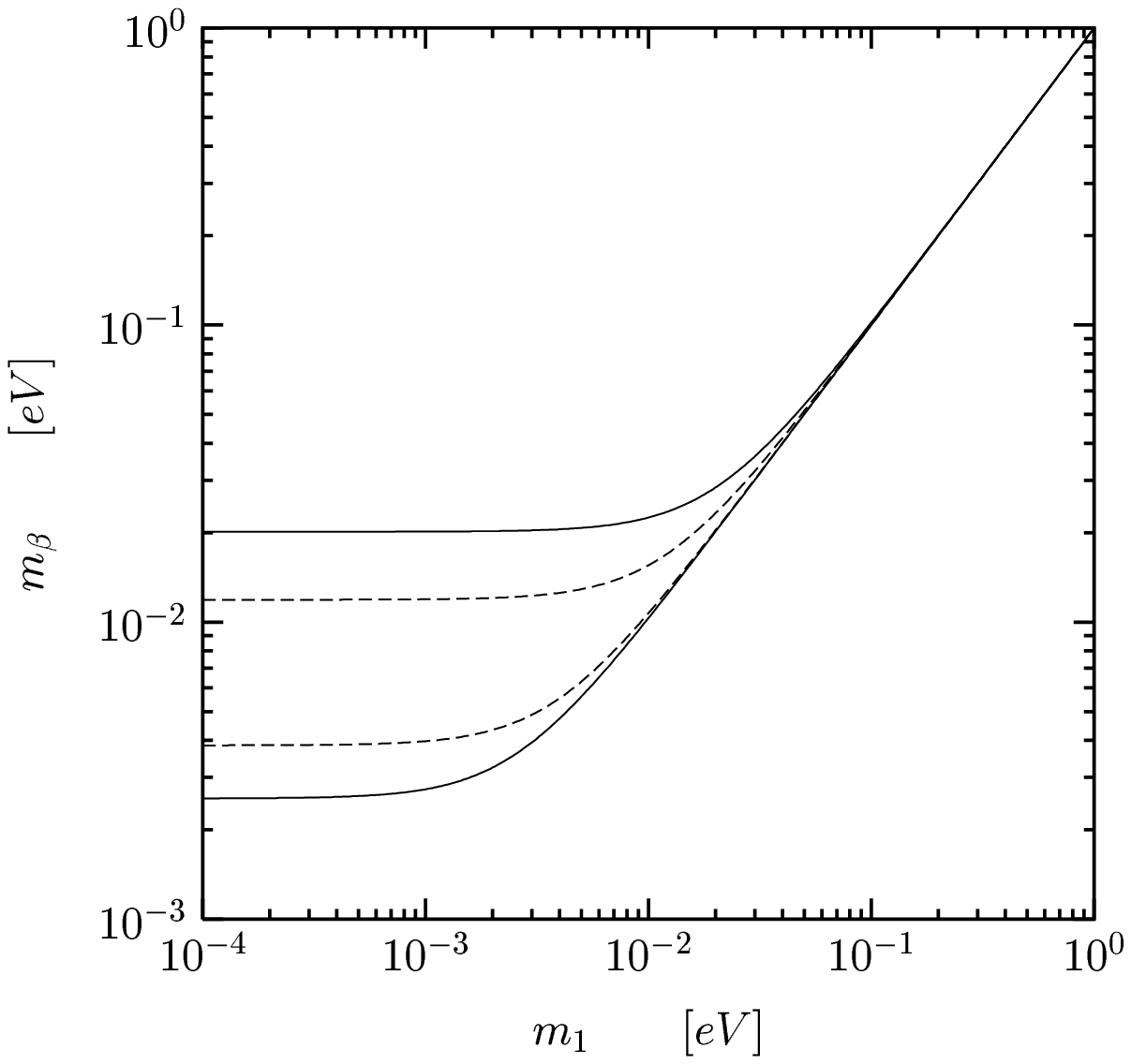}
\\
``normal''
\end{center}
\end{minipage}
\hfill
\begin{minipage}[l]{0.45\textwidth}
\begin{center}
\includegraphics*[bb=118 426 465 753, height=7cm]{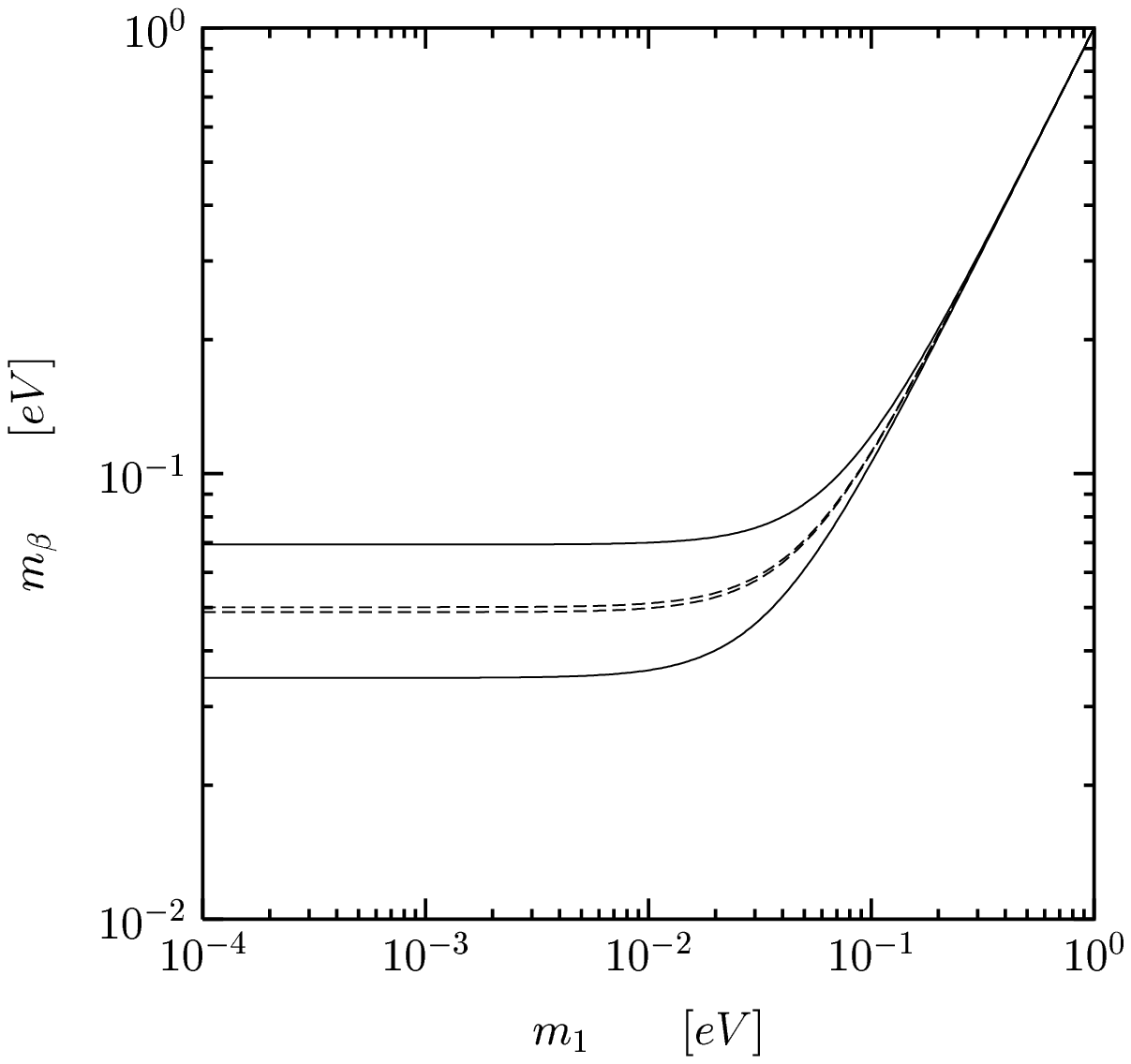}
\\
``inverted''
\end{center}
\end{minipage}
\caption{ \label{mb}
Value of $m_{\beta}$ as a function of $m_1$
(see Eqs.~(\ref{0651}) and (\ref{0652})).
The two dashed curves
have been calculated assuming the best-fit values
of
$\Delta{m}^{2}_{\mathrm{sol}}$
and
$\vartheta_{\mathrm{sol}}$
in Eq.~(\ref{032})
and the best-fit value of
$\Delta{m}^{2}_{\mathrm{atm}}$
in Eq.~(\ref{018}).
The lower dashed curves correspond to
$|U_{e3}|^{2} = 0$ in the ``normal'' scheme
and
$|U_{e1}|^{2} = 0$ in the ``inverted'' scheme.
The upper dashed curves correspond to the upper limits
$|U_{e3}|^{2} = 5 \times 10^{-2}$
(Eq.~(\ref{0511}))
in the ``normal'' scheme
and
$|U_{e1}|^{2} = 5 \times 10^{-2}$
(Eq.~(\ref{0512}))
in the ``normal'' scheme.
The two solid curves represent the lower and upper limits for
$m_{\beta}$
obtained from
the 99.73\% C.L. solar LMA-MSW region in
Ref.~\protect\cite{Bahcall:2002hv}
and the 99\% C.L. atmospheric region in
Ref.~\protect\cite{SK-atm-Nu2002},
and
$|U_{e3}|^{2}$ and $|U_{e1}|^{2}$ bounded by Eqs.~(\ref{0511}) and (\ref{0512}),
respectively,
in the ``normal'' and ``inverted'' schemes.
}
\end{figure}

Figure~\ref{mb} shows the allowed values of $m_{\beta}$
as a function of $m_1$.
One can see that in both the ``normal'' and ``inverted''
three-neutrino schemes in Fig.~\ref{3nu},
the KATRIN experiment
may obtain a positive result only if
the three neutrino masses are almost degenerate
and $m_1$ is of the same order or larger than the sensitivity
of the experiment
($ m_1 \gtrsim 0.3 \, \mathrm{eV} $).
In this case
$m_1 \simeq m_2 \simeq m_3$,
and
from the unitarity of the mixing matrix
we obtain
\begin{equation}
m_{\beta} \simeq m_{1}
\,,
\label{071}
\end{equation}
as shown in Fig.~\ref{mb}.

% from the hierarchy of neutrino mass-squared differences we have
% \begin{equation}
% m_{2} - m_{1}
% \simeq
% \frac{\Delta{m}^{2}_{\mathrm{21}}}{2\,m_{1}}
% \ll
% m_{1}
% \,,
% \qquad
% m_{3} - m_{1}
% \simeq
% \frac{\Delta{m}^{2}_{\mathrm{31}}}{2\,m_{1}}
% \ll
% m_{1}
% \,.
% \label{070}
% \end{equation}

If the  LSND result \cite{LSND} is confirmed by the 
MiniBooNE  \cite{MiniBooNE-Nu2002} experiment,
it will mean that (at least) four massive and mixed neutrinos
exist in nature\footnote{
Although
the relatively bad fit of the data
in the framework of four-neutrino schemes
(see Refs.~\cite{BGGS-AB-99,Maltoni:2002xd})
may suggest the possibility of more exotic
explanations.
}.

Let us discuss now the possibilities to measure the neutrino mass with the 
tritium method in the case of four massive neutrinos.
In this case,
we have three different neutrino mass-squared differences 
$\Delta{m}^{2}_{\mathrm{sol}}$,
$\Delta{m}^{2}_{\mathrm{atm}}$
and 
$\Delta{m}^{2}_{\mathrm{LSND}}$,
given by (\ref{018}), (\ref{032}) and (\ref{054}). 
Let us assume that 
$ m_{1}\ll \sqrt{\Delta{m}^{2}_{\mathrm{LSND}}}$.

\begin{figure}[t]
\begin{center}
\includegraphics*[bb=88 637 510 754, width=\textwidth]{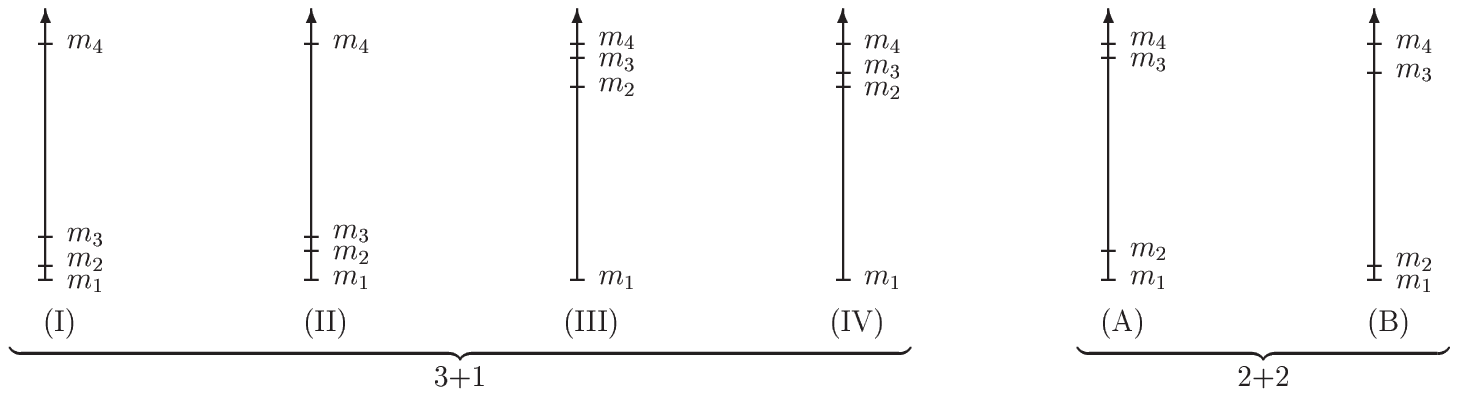}
\end{center}
\caption{ \label{4nu}
The six allowed types of four-neutrino mass schemes.
}
\end{figure}

Figure~\ref{4nu} shows
the six four-neutrino mass spectra
compatible with the mass-squared hierarchy
$
\Delta{m}^{2}_{\mathrm{sol}}
\ll
\Delta{m}^{2}_{\mathrm{atm}}
\ll
\Delta{m}^{2}_{\mathrm{LSND}}
$.
In all spectra there are two groups of close masses, separated by 
the LSND gap of the order of 1 eV.
There are two possibilities for the groups: 
2+2 and 3+1.

In order to calculate the contribution of neutrino masses  to the 
$\beta$-spectrum it is necessary 
to take into account
the constraints on the elements of the neutrino mixing matrix that
can be obtained from the data of the short-baseline 
reactor experiment
Bugey \cite{Declais:1995su},
in which no indication in favor of neutrino oscillations
was found.

In the framework of four-neutrino mixing,
the probability of the reactor $\bar\nu_{e}$'s to survive 
is given by
(see \cite{BGG-review-98})
\begin{equation}
\mathrm{P}(\bar\nu_e \to\bar\nu_e)
=
1 - \frac{1}{2}\mathrm{B}_{e;e}
\left( 1 - \cos \Delta{m}^{2}_{41} \frac{L}{2E} \right)
\,.
\label{072}
\end{equation}
Here
\begin{equation}
\mathrm{B}_{e;e}
=
4 \, \sum_{i} |U_{ei}|^{2}
\left(1-\sum_{i}|U_{ei}|^{2}\right)
\,,
\label{073}
\end{equation}
where
$i$ runs over the mass indices of the first (or second) group.

From the exclusion curve obtained in the Bugey experiment
\cite{Declais:1995su}, 
we have
\begin{equation}
\mathrm{B}_{e;e}\leq \mathrm{B}^{0}_{e;e},
\label{074}
\end{equation}
where the upper bound $\mathrm{B}^{0}_{e; e}$ depends on 
$\Delta{m}^{2}_{41}$.

Let us consider first
the neutrino mass spectra
I, II and B
in Fig.~\ref{4nu},
in which
the solar neutrino mass-squared difference
$\Delta{m}^{2}_{\mathrm{sol}}$
belongs to the light group.
Taking into account the solar neutrino data, from the Bugey exclusion plot
\cite{Declais:1995su}
we obtain
\begin{equation}
\sum_{i}|U_{ei}|^{2} \lesssim 2 \times 10^{-2}\,,
\label{075}
\end{equation}
where $i$ runs over indices of neutrinos belonging to the heavy group
($i = 3,4$ for 2+2 scheme
and $i = 4$ for 3+1 schemes).
Thus, in this case the contribution of heavy neutrinos to the
$\beta$ spectrum
is suppressed.
For the effective neutrino mass we have the upper bound
\begin{equation}
m_{\beta}
\lesssim
\left(\sum_{i}|U_{ei}|^{2}\,\Delta{m}^{2}_{\mathrm{LSND}}
+\Delta{m}^{2}_{\mathrm{atm}}\right)^{1/2}
\,.
\label{076}
\end{equation}
Taking into account (\ref{075}) and using the best fit values of 
the parameters $\Delta{m}^{2}_{\mathrm{LSND}}$ and $\Delta{m}^{2}_{\mathrm{atm}}$
(see (\ref{018}) and (\ref{054})),
for the effective neutrino mass $m_{\beta}$ we obtain the bound
\begin{equation}
m_{\beta}
\lesssim
1.6 \times 10^{-1} \mathrm{eV}
\,,
\label{077}
\end{equation}
which is smaller than the sensitivity of the future tritium experiment 
KATRIN.

If $\Delta{m}^{2}_{\mathrm{sol}}$ is the mass-squared difference of neutrinos
belonging to the heavy group,
in Eq.~(\ref{073}) the index $i$
takes the values $i=1,2$ for 2+2 scheme and $i=1$ for 3+1 schemes. 

In this case,
taking into account the unitarity of the mixing matrix, we have 
\begin{equation}
m_{\beta} 
\simeq
\sqrt{\Delta{m}^{2}_{\mathrm{LSND}}}
\,.
\label{078}
\end{equation}
The allowed range
for the parameter $\Delta{m}^{2}_{\mathrm{LSND}}$
found in Ref.~\cite{LSND}
is
(see also Ref.~\cite{Church:2002tc})
\begin{equation}
0.2 \, \mathrm{eV}^{2}
\lesssim
\Delta{m}^{2}_{\mathrm{LSND}}
\lesssim
2 \, \mathrm{eV}^{2}
\,.
\label{079}
\end{equation}
From Eqs.~(\ref{078}) and (\ref{079}),
for the effective neutrino mass we have the limits
\begin{equation}
0.45 \, \mathrm{eV}
\lesssim
m_{\beta}
\lesssim
1.4 \, \mathrm{eV}
\,.
\label{080}
\end{equation}
In this case,
if the data of the LSND experiment are confirmed by the MiniBooNE
experiment,
there is a chance to observe the effect of neutrino mass in the future KATRIN
experiment \cite{Osipowicz:2001sq} with the expected sensitivity
of about 0.35 eV.

\subsection{Mainz experiment}
\label{Mainz experiment}

The source in the Mainz experiment
\cite{Weinheimer:1999tn,Bonn:2002jw,hep-ex/0210050}
is frozen molecular tritium
condensed on a graphite substrate.
The spectrum of the electron 
is measured
by an integral MAC-E-Filter spectrometer (Magnetic Adiabatic Collimator
with a retarding Electrostatic filter).
This spectrometer  
combines high luminosity
with high resolution. 
The resolution of the spectrometer
is $4.8\,\mathrm{eV}$.
In the analysis of the experimental data
four variable parameters are used: the normalization $C$,
the background $B$,
the released energy $E_{0}$ and the effective neutrino mass-squared 
$m^{2}_{\beta}$.
From the fit of the data it was found that
$E_{0}= 18.575\,\mathrm{keV}$.

\begin{figure}[t]
\begin{minipage}[l]{0.45\textwidth}
\begin{center}
\includegraphics*[bb=0 0 567 567, height=7cm]{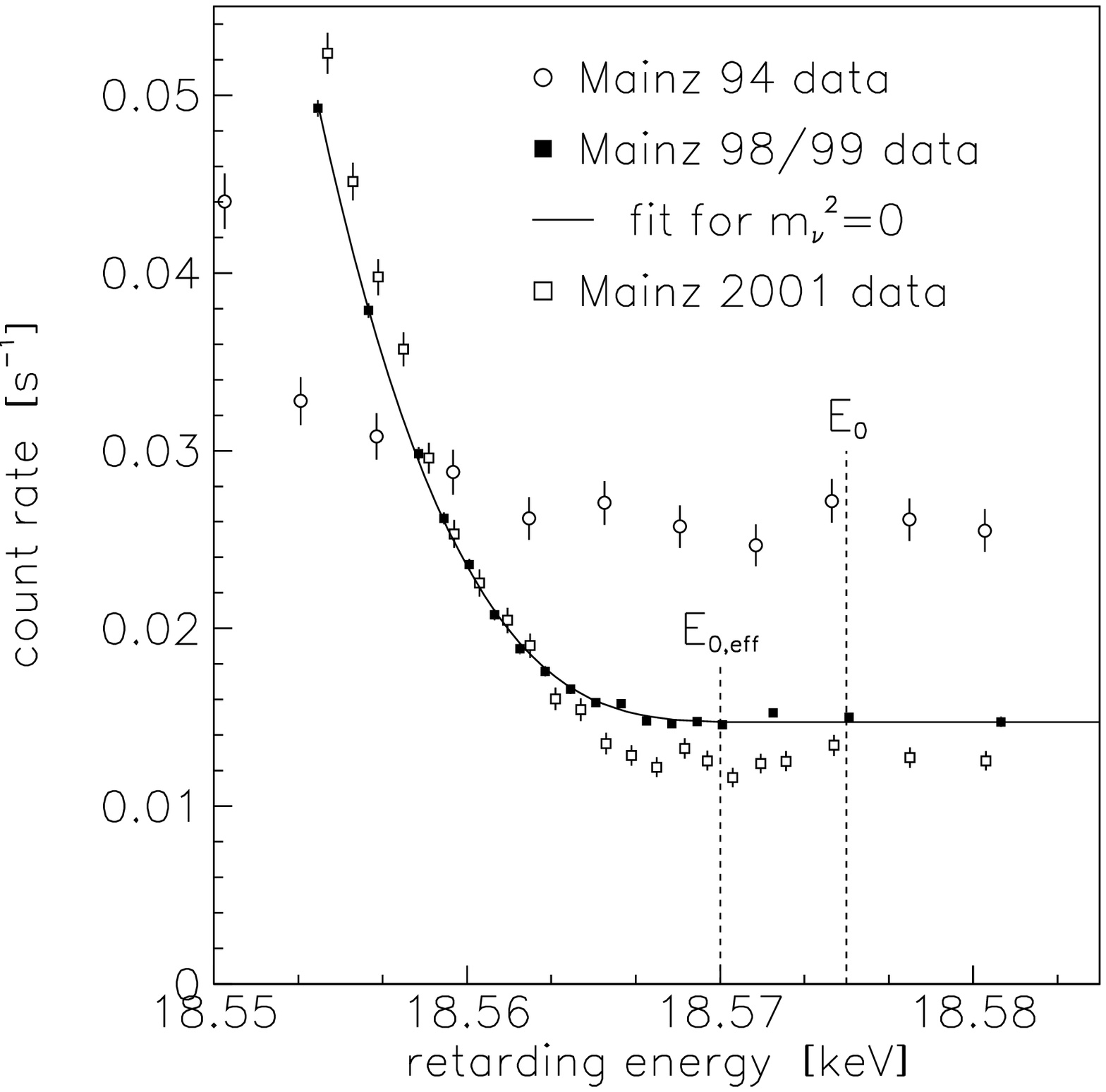}
\end{center}
\end{minipage}
\hfill
\begin{minipage}[l]{0.45\textwidth}
\begin{center}
\includegraphics*[bb=0 0 567 567, height=7cm]{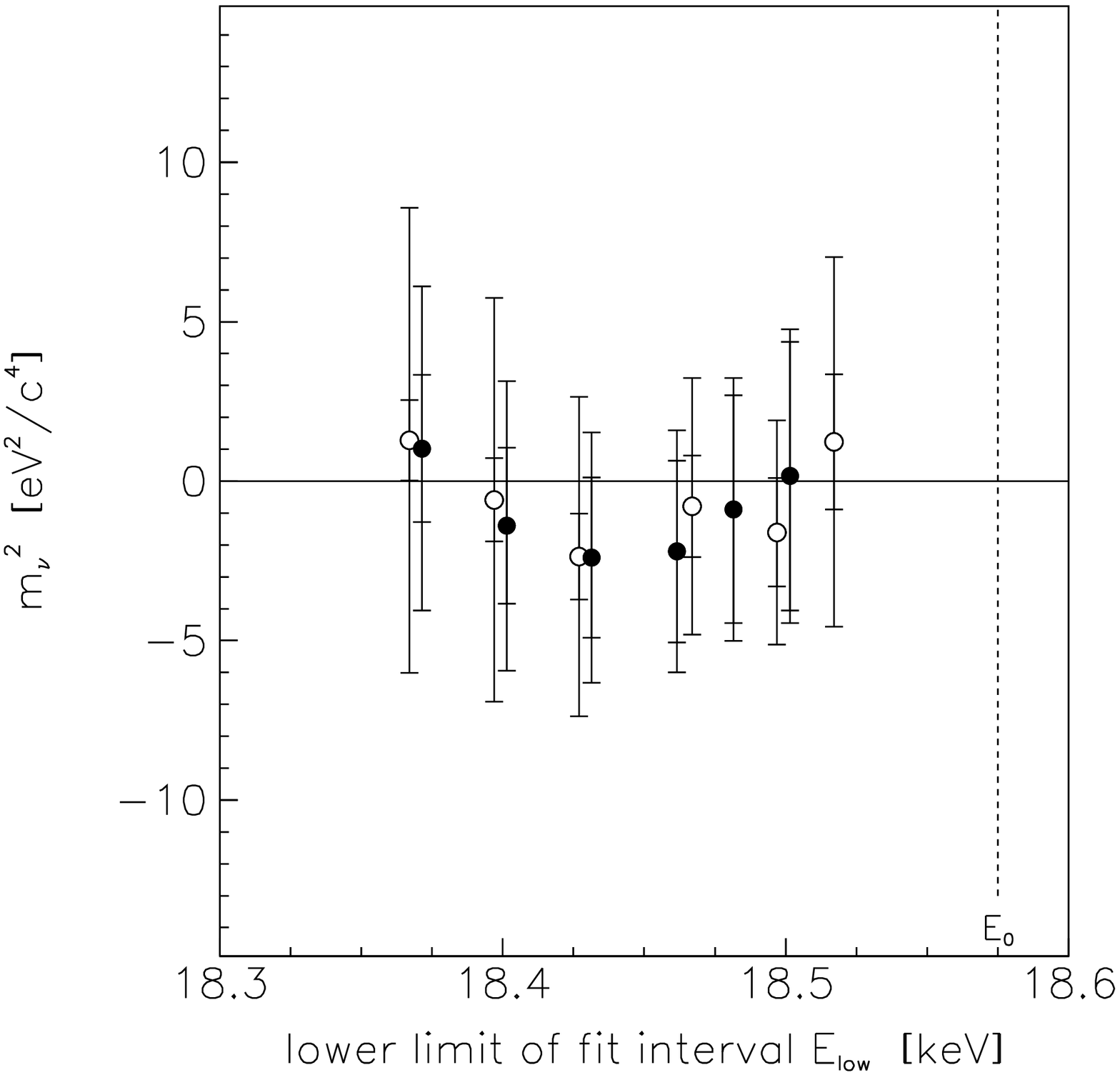}
\end{center}
\end{minipage}
\caption{ \label{Mainz}
Left:
Integral spectrum measured in
1994 (open circles),
1998/1999 (filled squares) with fit (line)
and
2001 (open squares)
in the Mainz experiment.
Right:
Mainz fit results on
$ m^{2}_{\nu} \equiv m^{2}_{\beta} $
as a function of
the lower boundary $E_{\mathrm{low}}$ of the fit interval
for data from 1998 and 1999
(open circles) and from 2001
(filled circles).
Figures taken from Ref.~\cite{hep-ex/0210050}.
}
\end{figure}

The left figure in Fig.~\ref{Mainz} shows the integral spectrum, measured
in 1994, in 1998-1999, and in 2001,
as a function of the retarding
energy near the endpoint $E_0$, and the effective endpoint $E_{0,\mathrm{eff}}$.
The position of $E_{0,\mathrm{eff}}$
takes into account the width of the response function of the
setup and
the mean rotation-vibration excitation energy of the electronic
ground state of the $\mathrm{^3HeT^+}$ daughter molecule.
The solid curve was obtained from the fit
of the data under the assumption $m_{\beta}= 0$.
Different parts of the spectrum were used in the analysis of the data.
The right figure in Fig.~\ref{Mainz}
shows the dependence of  $m^{2}_{\beta}$ on the lower limit $E_{\mathrm{low}}$ of
the corresponding fit interval
(the upper limit is fixed
at 18.66~keV, well above $E_0$) for data from 1998 and 1999
(open circles) and from the last runs of 2001
(filled circles).
The error bars show the statistical uncertainties
(inner bar) and the total uncertainty (outer bar).
The correlation of data points for large fit intervals
is due to the uncertainties of the systematic corrections, which are
dominant for fit intervals with a lower boundary
$E_{\mathrm{low}} <  18.5 \, \mathrm{keV}$.

In the last $70\,\mathrm{eV}$
of the spectrum the combined statistical and systematical error 
is minimal. From the fit of the 1998-1999 experimental data in this interval,
it was found that
\begin{equation}
m^{2}_{\beta} = (-1.6 \pm 2.5 \pm 2.1) \, \mathrm{eV}^{2}
\,,
\label{081}
\end{equation}
which corresponds to the upper bound
\begin{equation}
m_{\beta}< 2.2 \, \mathrm{eV}
\qquad
(95\%\, \mathrm{C.L.})
\,.
\label{082}
\end{equation}
In 2001 additional measurements of the $\beta$-decay spectrum were
carried on by the Mainz group.
From the analysis of these data it was obtained \cite{hep-ex/0210050}
\begin{equation}
m^{2}_{\beta}= (0.1 \pm 4.2 \pm 2.0)\,\mathrm{eV}^{2}\,.
\label{083}
\end{equation}
From the combined analysis of 1998, 1999 and 2001 data 
it was found \cite{hep-ex/0210050}
\begin{equation}
m^{2}_{\beta}
=
(-1.2 \pm 2.2 \pm 2.1) \, \mathrm{eV}^{2}
\,.
\label{084}
\end{equation}
This value corresponds again to the upper bound \cite{hep-ex/0210050}
\begin{equation}
m_{\beta} < 2.2\,\mathrm{eV}
\qquad
(95\% \, \mathrm{C.L.})
\,,
\label{085}
\end{equation}
showing that the Mainz experiment
has reached his sensitivity limit.

\subsection{Troitsk experiment}
\label{Troitsk experiment}

In the Troitsk neutrino experiment \cite{Lobashev:1999tp,Lobashev:2001uu},
as in the Mainz experiment, an integral electrostatic spectrometer
with a strong inhomogeneous magnetic field, focusing the electrons, is used.
The resolution of the spectrometer is $3.5-4$ eV.
An important difference 
between the Troitsk and Mainz
experiments is that in the Troitsk experiment the tritium source 
is a gaseous molecular source.
Such a source has important advantages in comparison 
with the frozen source: there is no backscattering, there are no
effects of the self-charging, the interaction between tritium molecules 
can be neglected, etc..
In the analysis of the data the same four variable parameters 
$C$, $B$, $E_{0}$ and $m^{2}_{\beta}$ were used.
From the fit of the data, for the parameter
$m^{2}_{\beta}$ large negative values in the range
$-20 \div -10 \,\mathrm{eV}^{2}$ have been obtained.
The investigation of the character of the measured spectrum suggests
that the negative $m^{2}_{\beta}$ is due to a step function
superimposed on the integral continuous spectrum.
The step function in the integral spectrum corresponds to a narrow peak in
the differential spectrum.

In order to describe the data, the authors of the Troitsk experiment
added to the theoretical integral spectrum 
a step function with two additional variable
parameters (position of the step $E_{\mathrm{step}}$
and the height of the step).
From a six-parameter fit of the data,
the Troitsk Collaboration found
\begin{equation}
m^{2}_{\beta}= (-2.3 \pm 2.5 \pm 2.0) \, \mathrm{eV}^{2}
\,.
\label{086}
\end{equation}
This value corresponds to the upper bound
\begin{equation}
m_{\beta} < 2.2\,\mathrm{eV}
\qquad
(95\% \, \mathrm{C.L.})
\,,
\label{087}
\end{equation}
as in the Mainz experiment.

The Troitsk Collaboration found
that the position of the step $E_{0}-E_{\mathrm{step}}$
changes periodically in the interval $5-15$ eV and that
the average value of the height of the step is about
$6 \times 10^{-11}$
of the total number of events.
This effect has been called ``Troitsk anomaly''.
Since the Mainz data do not show any indication of a Troitsk-like anomaly,
it is believed
\cite{hep-ex/0210050}
that the Troitsk anomaly is caused by
some experimental artifact.

\subsection{Other experiments}
\label{Other experiments}

We have discussed up to now tritium experiments for the measurement
of the neutrino mass. The groups
in Genova \cite{Gatti:2001ty} and Milano \cite{Sisti:2002rr}
are developing low temperature
cryogenic detectors for the measurement of the $\beta$-decay spectrum of
$^{187}\mathrm{Re}$.
This element has the lowest known energy release
($E_{0} = 2.5\, \mathrm{keV}$).
The relative fraction of events in the high energy part of the spectrum is
proportional to $E_{0}^{-3}$. Thus, decays with low $E_{0}$ values
are very suitable for calorimetric experiments
in which the full spectrum is measured.
The limit for the neutrino mass that was obtained by the Genova group in 
Ref.~\cite{Gatti:2001ty} is
\begin{equation}
m_{\beta}
<
26 \, \mathrm{eV}
\qquad
(95\% \, \mathrm{C.L.})
\,.
\label{088}
\end{equation}
In the future a sensitivity of about
$10\,\mathrm{eV}$
is expected to be reached.

\subsection{The future KATRIN experiment}
\label{The future KATRIN experiment}

In the future tritium experiment KATRIN \cite{Osipowicz:2001sq},
two tritium sources will be used:
a gaseous molecular source ($\mathrm{T}_{2}$), as in the
Troitsk experiment, and a frozen tritium source, as in the Mainz experiment.
The windowless gaseous tritium source will allow to reach a column
density $ 5 \times 10^{17} \, \mathrm{T}_{2}/\mathrm{cm}^{2} $.

The integral MAC-E-Filter spectrometer will have 
two parts: the pre-spectrometer, which will select electrons in the
last part (about 100\, eV)
of the spectrum, and the main spectrometer. This spectrometer will have
a resolution of 1 eV. 

It is planned that the KATRIN experiment will start to collect data in
2007. After three years of running the  accuracy in the measurement of the 
parameter $m^{2}_{\beta}$
will be $0.08 \, \mathrm{eV}^{2}$.
This will allow to reach a sensitivity of 0.35 eV in the determination
of the effective neutrino mass $m_{\beta}$.

As in the case of the Mainz experiment, in the analysis of the data  
of the future KATRIN experiment four variable parameters are planned to
be used. The value of the parameter $E_{0}$
can be taken, however, from the independent measurement
of the $^{3}\mathrm{H}$ and $^{3}\mathrm{He}$ mass difference.
If the accuracy of such measurements reaches 1 p.p.m.,
the sensitivity of the KATRIN experiment to
the neutrino mass will be significantly improved.

In the KATRIN experiment, not only
the integral spectrum, but also the differential
spectrum is planned to be measured.
These measurements will allow
to clarify the problem of the Troitsk anomaly
in a direct way.

\section{Muon and tau neutrino mass measurements}
\label{Muon and tau neutrino mass measurements}

Information on the ``mass'' of the muon neutrino can be obtained from 
the measurement of the muon momentum in the decay 
\begin{equation}
\pi^{+} \to \mu^{+} + \nu_{\mu}
\,.
\label{MT001}
\end{equation}
We discuss here such measurements from the point of view of neutrino mixing.
From four-momentum conservation in the decay (\ref{MT001})
it follows that
\begin{equation}
m_{i}^{2}
=
m_{\pi}^{2} + m_{\mu}^{2} - 2 \, m_{\pi} \sqrt{ m_{\mu}^{2} + p_{\mu}^{2} }
\,.
\label{MT002}
\end{equation}
Here $m_{i}$ is the mass of $\nu_{i}$, $ m_{\pi}$ and $ m_{\mu}$ are the masses of
the pion and the muon, and $p_{\mu}$ is the muon momentum in the pion rest-frame.

The value of the muon momentum
measured in the most precise PSI experiment \cite{Assamagan:1996wb}
is
\begin{equation}
p_{\mu} = 29.79200 \pm 0.00011 \, \mathrm{MeV}
\,.
\label{MT003}
\end{equation}

Taking into account the resolution in the
measurement of the momentum of the muon
and the values of the neutrino mass-squared differences
measured in neutrino oscillation experiments (see Section~\ref{Status of neutrino oscillations}), 
we come to the conclusion that the effect of neutrino masses in the decay
(\ref{MT001})
can be observed only in the case of an almost degenerate neutrino mass spectrum with
$m_{1}\gg \sqrt{\Delta m_{\mathrm{atm}}^{2}}$
(or $m_{1}\gg \sqrt{\Delta m_{\mathrm{LSND}}^{2}}$ in the case of four neutrinos).

In this case,
from the unitarity condition $\sum_{i}|U_{\mu i}|^{2}=1$
it follows that the experiments on the measurement of the momentum 
of the muon produced in the decay (\ref{MT001})
allow to obtain information about the mass $m_{1}$.
The value of
$m_{1}$
found in Ref.~\cite{Assamagan:1996wb} is
\begin{equation}
m_{1}^{2} = -0.016 \pm 0.023 \, \mathrm{MeV}^{2}
\qquad
(\pi^{+} \to \mu^{+} \nu_{\mu})
\,.
\label{MT004}
\end{equation}
For the masses of the muon and pion the following values were used 
\cite{Assamagan:1996wb}:
\begin{align}
\null & \null
m_{\mu} = 105.658389 \pm 0.000034 \, \mathrm{MeV}
\,,
\label{MT0051}
\\
\null & \null
m_{\pi} = 139.56995\pm 0.00037 \, \mathrm{MeV}
\,.
\label{MT0052}
\end{align}
The upper bound of the neutrino mass given by the Particle Data Group \cite{PDG}
is
\begin{equation} 
m_{1} < 190 \, \mathrm{keV}
\quad
(90\% \, \mathrm{C.L.})
\quad
( \pi^{+} \to \mu^{+} \nu_{\mu})
\,.
\label{MT006}
\end{equation}

The most stringent upper bound on the ``mass'' of the
tau neutrino was obtained in the ALEPH experiment \cite{Barate:1998zg}.
In this experiment the decays
\begin{equation} 
\tau^{-} \to 2\pi^{-}\,\pi^{+}\,\nu_{\tau}
\,,
\qquad
\tau^{-} \to 3\pi^{-}\,2\pi^{+}(\pi^{0})\,\nu_{\tau}
\label{MT007}
\end{equation}
were studied.
From the conservation of the four-momentum in the decay
\begin{equation}
\tau^{-}\to n\pi +\nu_{\tau}
\,,
\label{MT0071}
\end{equation}
we have 
\begin{equation} 
E_{h}^{*} = \frac{m_{\tau}^{2}+m_{h}^{2}-m_{i}^{2}}{2\,m_{\tau}}
\,.
\label{MT008}
\end{equation}
Here $m_{\tau}$ is the tau mass, $ m_{h}$ is the invariant mass of the
$n$ pions and $E_{h}^{*}$ is the total energy of the $n$ pions in the
rest frame of the tau.
Information on the neutrino mass was obtained in
Ref.~\cite{Barate:1998zg} from the fit of the distribution
$\mathrm{d}^{2}\Gamma/\mathrm{d}E_{h}\mathrm{d}m_{h}$
($E_{h}$ is the total energy of the pions in the laboratory system).
In the case of neutrino mixing, information 
about the minimal (common) mass $m_{1}$ of an almost 
degenerate neutrino mass spectrum can be obtained
from such experiments.
The bound obtained in the ALEPH experiment \cite{Barate:1998zg} is
\begin{equation} 
m_{1} < 18.2 \, \mathrm{MeV}
\quad
(95\% \, \mathrm{C.L.})
\quad
(\tau^{-}\to n\pi +\nu_{\tau})
\,.
\label{MT009}
\end{equation}

Thus, the experiments on the measurement of the muon momentum in
the decays of pions and
the $\mathrm{d}^{2}\Gamma/\mathrm{d}E_{h}\mathrm{d}m_{h}$
distribution in the decays of taus are much less sensitive
to the absolute neutrino mass $m_{1}$ than tritium experiments.
These experiments could, however, reveal the existence of particles with 
masses much larger than the light neutrino masses.

\section{Neutrinoless double-$\beta$ decay}
\label{Neutrinoless double-beta decay}

The search for neutrinoless double-$\beta$ decay
\begin{equation}
(A,Z) \to (A,Z+2)+ e^{-}+ e^{-}
\label{089}
\end{equation}
of some even-even nuclei is the most sensitive
and direct way of investigation
of the nature of the neutrinos with definite masses
(Majorana or Dirac?).
Neutrinoless double-$\beta$ decay
is allowed only if
massive neutrinos $\nu_{i}$ are Majorana particles.

We will assume that
the Hamiltonian of the process has the standard form
in Eq.~(\ref{056}) and
the flavor field $\nu_{eL}$
is given by the relation (see Eq.~(\ref{002}))
\begin{equation}
\nu_{eL} = \sum_{i} U_{ei} \, \nu_{iL}
\,,
\label{090}
\end{equation}
where $\nu_{i}$ are Majorana fields which
satisfy the
condition
\begin{equation}
\nu_{i}
=
\nu^{c}_{i}
=
\mathcal{C}\,\overline{\nu_{i}}^{T}
\,. 
\label{091}
\end{equation}
Here $\mathcal{C}$ is the charge conjugation matrix
($\mathcal{C} \gamma_{\alpha}^{T} \mathcal{C}^{-1} = -\gamma_{\alpha}$,
$\mathcal{C}^{T}=-\mathcal{C}$).

The neutrinoless
double-$\beta$ decay ($(\beta\beta)_{0\nu}$-decay)
is a process of second order in the Fermi constant
$G_{F}$, with virtual neutrinos.
In the case of the Majorana neutrino mixing in Eq.~(\ref{090}),
the neutrino propagator is given by the expression
\begin{equation}
\langle 0|T(\nu_{eL}(x_{1})\nu^{T}_{eL}(x_{2})|0 \rangle
\simeq
\langle m \rangle
\,
\frac{i}{(2\,\pi)^{4}}
\int
\frac{d^{4}p}{p^{2}}
\,
e^{-ip(x_{1}-x_{2})}
\,
\frac{1-\gamma_{5}}{2}
\,
\mathcal{C}
\,.
\label{092}
\end{equation}
Here 
\begin{equation}
\langle m \rangle = \sum_{i} U^{2}_{ei} \, m_{i}
\,.
\label{093}
\end{equation}
The matrix element of neutrinoless double-$\beta$ decay is proportional
to the nuclear matrix element and to
the effective Majorana mass $\langle m \rangle$,
which depends on neutrino masses $ m_{i}$ and on $ U^{2}_{ei}$.

The elements of the neutrino mixing matrix $U_{ei}$ are complex quantities.
In the case of CP invariance in the lepton sector,
the elements $U_{ei}$
satisfy the condition \cite{Bilenky:1984fg,Kayser:1984ge}
\begin{equation}
U^{*}_{ei} = \eta^{*}_{i} \, U_{ei}
\,,
\label{094}
\end{equation}
where $\eta_{i}= i\,\rho_{i}$ is the CP-parity of the neutrino
$\nu_{i}$ ($\rho_{i}=\pm 1$).
Let us write down
\begin{equation}
U_{ei} = |U_{ei}| \, e^{i\,\alpha_{i}}
\,.
\end{equation}
From Eq.~(\ref{094}) we obtain
\begin{equation}
2 \, \alpha_{i} = \frac{\pi}{2} \, \rho_{i}
\,.
\end{equation}
Thus, in the case of CP invariance in the lepton sector,
the effective Majorana mass is given by
\begin{equation}
\langle m \rangle
=
\sum_{i} |U_{ei}|^{2} \, e^{i\frac{\pi}{2}\rho_{i}} \, m_{i}
\,.
\label{095}
\end{equation}
The results of many experiments on the search for $(\beta\beta)_{0\nu}$-decay
are available at present (see Refs.\cite{Cremonesi:2002is,Tretyak-Zdesenko-02}).
No indication in favor of $(\beta\beta)_{0\nu}$-decay
have been obtained up to
now\footnote{
The recent
claim \cite{Klapdor-Kleingrothaus:2002ke} of an evidence of the $(\beta\beta)_{0\nu}$-decay,
obtained from the reanalysis of the data of the Heidelberg-Moscow experiment,
was strongly criticized in Refs.~\cite{Feruglio:2002af,Aalseth-0202018}.
}.
The most stringent lower bounds for the lifetime
of $(\beta\beta)_{0\nu}$-decay 
were obtained in the Heidelberg-Moscow
\cite{Klapdor-Kleingrothaus:2001yx}
and IGEX
\cite{Aalseth:2002rf}
$^{76} \mathrm{Ge}$ experiments:
\begin{alignat}{3}
\null & \null
T^{0\nu}_{1/2}
\geq
1.9 \times 10^{25} \, \mathrm{yr}
\qquad
\null && \null
(90\% \, \mathrm{C.L.})
\qquad
\null && \null
\mbox{(Heidelberg-Moscow)}
\,,
\label{0951}
\\
\null & \null
T^{0\nu}_{1/2}
\geq
1.57 \times 10^{25} \, \mathrm{yr}
\qquad
\null && \null
(90\% \, \mathrm{C.L.})
\qquad
\null && \null
\mbox{(IGEX)}
\,.
\label{0952}
\end{alignat}
Taking into account different calculations of the nuclear matrix element,
from these results
the following upper bounds for the effective Majorana mass were obtained:
\begin{alignat}{2}
\null & \null
|\langle m \rangle|
\leq
(0.35-1.24) \, \mathrm{eV}
\qquad
\null && \null
\mbox{(Heidelberg-Moscow)}
\,,
\label{0961}
\\
\null & \null
|\langle m \rangle|
\leq
(0.33-1.35) \, \mathrm{eV}
\qquad
\null && \null
\mbox{(IGEX)}
\,.
\label{0962}
\end{alignat}
Many new experiments for the search of neutrinoless
double-$\beta$ decay are in preparation at present
(see Table~\ref{bb} and Ref.~\cite{Cremonesi:2002is}).
In these experiments the sensitivities 
\begin{equation}
|\langle m \rangle|
\simeq
1.5  \times 10^{-2} - 5.6 \times 10^{-1} \, \mathrm{eV}
\label{097}
\end{equation}
are expected to be achieved\footnote{
These sensitivities have been estimated using
the nuclear matrix elements calculated 
in Ref.~\cite{Staudt:1990qi}.
}.

\begin{table}[t]
\begin{center}
\begin{tabular}{|c|c|c|c|}
\hline
Experiment
&
Nucleus
&
\begin{tabular}{c}
Sensitivity
\\
$T^{0\nu}_{1/2} \, [\mathrm{yr}]$
\end{tabular}
&
\begin{tabular}{c}
Sensitivity
\\
$|\langle m \rangle| \, [\mathrm{eV}]$
\end{tabular}
\\
\hline
NEMO 3
\protect\cite{Sarazin:2000xv}
&
$^{100}\mathrm{Mo}$
&
$4 \times 10^{24}$
&
$5.6 \times 10^{-1}$
\\
COBRA
\protect\cite{Zuber:2001vm}
&
$^{130}\mathrm{Te}$
&
$1 \times 10^{24}$
&
$2.4 \times 10^{-1}$
\\
CUORICINO
\protect\cite{Alessandrello:2002sj}
&
$^{130}\mathrm{Te}$
&
$1.5 \times 10^{25}$
&
$1.9 \times 10^{-1}$
\\
XMASS
\protect\cite{XMASS-LowNu2002}
&
$^{136}\mathrm{Xe}$
&
$3.3 \times 10^{26}$
&
$9 \times 10^{-2}$
\\
CAMEO
\protect\cite{Bellini:2000hp}
&
$^{116}\mathrm{Cd}$
&
$1 \times 10^{26}$
&
$6.9 \times 10^{-2}$
\\
EXO
\protect\cite{Danilov:2000pp}
&
$^{136}\mathrm{Xe}$
&
$8 \times 10^{26}$
&
$5.2 \times 10^{-2}$
\\
MOON
\protect\cite{Ejiri:1999rk,Ejiri:2002rt}
&
$^{100} \mathrm{Mo}$
&
$1 \times 10^{27}$
&
$3.6 \times 10^{-2}$
\\
CUORE
\protect\cite{Alessandrello:2002sj}
&
$^{130}\mathrm{Te}$
&
$7 \times 10^{26}$
&
$2.7 \times 10^{-2}$
\\
Majorana
\protect\cite{Aalseth:2002sy}
&
$^{76}\mathrm{Ge}$
&
$4 \times 10^{27}$
&
$2.5 \times 10^{-2}$
\\
GEM
\protect\cite{Zdesenko:2001ee}
&
$^{76}\mathrm{Ge}$
&
$7 \times 10^{27}$
&
$1.8 \times 10^{-2}$
\\
GENIUS
\protect\cite{Klapdor-Kleingrothaus:1998td}
&
$^{76}\mathrm{Ge}$
&
$1 \times 10^{28}$
&
$1.5 \times 10^{-2}$
\\
\hline
\end{tabular}
\end{center}
\caption{ \label{bb}
Future
neutrinoless double-$\beta$ decay
projects.
}
\end{table}

An evidence for neutrinoless double-$\beta$ decay would be a proof
that neutrinos with definite masses $\nu_{i}$
are Majorana particles
and that neutrino masses have an origin beyond the Standard Model.
The
\emph{value of the effective Majorana mass $|\langle m \rangle|$} 
combined with the results of neutrino oscillation experiments
could allow to obtain important information about the character of
the neutrino mass spectrum,
about the minimal neutrino mass $m_1$ and about the
Majorana CP phase (see
Refs.~\cite{Bilenky:1999wz,%
Vissani:1999tu,%
Czakon:2000vz,%
Klapdor-Kleingrothaus:2000gr,%
Bilenky:2001rz,%
hep-ph/0205022}
and references therein).

Let us consider
three typical neutrino mass spectra
in the case of three massive and mixed neutrinos\footnote{
Neutrinoless double-$\beta$ decay in the case of
four-neutrino mixing was considered in detail in
Ref.~\cite{Bilenky:2001xq}.
}:
\begin{enumerate}

\item
$m_{1} \ll m_{2} \ll m_{3}$
(hierarchy of neutrino masses).

For the effective Majorana mass
$|\langle m \rangle|$ we have the upper bound 
\begin{equation}
|\langle m \rangle|
\leq
\sin^{2} \vartheta_{\mathrm{sol}}
\,
\sqrt{\Delta{m}^{2}_{\mathrm{sol}}}
+ 
|U_{e3}|^{2}
\,
\sqrt{\Delta{m}^{2}_{\mathrm{atm}}}
\,.
\label{098}
\end{equation}
Using the best-fit values of the oscillation parameters
$\Delta{m}^{2}_{\mathrm{atm}}$,
$\Delta{m}^{2}_{\mathrm{sol}}$,
$\tan^{2} \vartheta_{\mathrm{sol}}$,
and the CHOOZ bound on $|U_{e3}|^{2}$
(Eqs.~(\ref{018}), (\ref{032}), and (\ref{051})),
we have
\begin{equation}
|\langle m \rangle| \lesssim 3.8 \times 10^{-3} \, \mathrm{eV}
\,.
\label{099}
\end{equation}
Taking into account the upper bounds
for the oscillation parameters,
one obtains
\cite{Pascoli:2002qm}
\begin{equation}
|\langle m \rangle| \lesssim 8.2 \times 10^{-3} \, \mathrm{eV}
\,.
\label{0991}
\end{equation}
The bounds (\ref{099}) and (\ref{0991})
are significantly smaller than the expected sensitivities 
(\ref{097}) of the future $(\beta\beta)_{0\nu}$-decay experiments.
Thus, the observation of
$(\beta\beta)_{0\nu}$-decay in the experiments of the next generation
could pose a problem for 
the natural hierarchy of neutrino masses.

\item
$m_{1} \ll m_{2} < m_{3}$
(inverted hierarchy of neutrino masses).

The effective Majorana mass is given by
\begin{equation}
|\langle m \rangle|
\simeq
\left(
1 - \sin^{2}2\vartheta_{\mathrm{sol}} \, \sin^{2}\alpha
\right)^{\frac{1}{2}}
\,
\sqrt{\Delta{m}^{2}_{\mathrm{atm}}}
\,,
\label{100}
\end{equation}
where $\alpha=\alpha_{3}-\alpha_{2}$ is the the difference of CP phases.
From this expression it follows that
\begin{equation}
\sqrt{\Delta{m}^{2}_{\mathrm{atm}}}
\,
|\cos2\vartheta_{\mathrm{sol}}|
\lesssim
|\langle m \rangle|
\lesssim
\sqrt{\Delta{m}^{2}_{\mathrm{atm}}}
\,,
\label{101}
\end{equation}
where the upper and lower bounds correspond to 
equal and opposite CP parities in the case of CP conservation.

Using the best-fit value of the parameter $\tan^{2}\vartheta_{\mathrm{sol}}$
in Eq.~(\ref{032}), we have
\begin{equation}
\frac{1}{2}\,\sqrt{\Delta{m}^{2}_{\mathrm{atm}}}
\lesssim
|\langle m \rangle|
\lesssim
\sqrt{\Delta{m}^{2}_{\mathrm{atm}}}
\,.
\label{102}
\end{equation}
Thus, in the case of an inverted mass hierarchy,
the scale of $|\langle m \rangle|$ is determined by
$\sqrt{\Delta{m}^{2}_{\mathrm{atm}}} \simeq 5 \times 10^{-2} \, \mathrm{eV}$.
If the value of $|\langle m \rangle|$ is in the range (\ref{102}),
which can be reached in the future experiments 
searching for  
$(\beta\beta)_{0\nu}$-decay,
it would be an argument in favor of an inverted neutrino mass hierarchy.

The measurement of the effective Majorana mass  $|\langle m \rangle|$
could allow to obtain information about the CP phase
$\alpha$ \cite{Bilenky:1996cb,Pascoli:2002qm}.
Indeed, from Eq.~(\ref{100})
we have
\begin{equation}
\sin^{2}\alpha
\simeq
\left(
1 - \frac{ |\langle m \rangle|^{2}}{\Delta{m}^{2}_{\mathrm{atm}}}
\right)
\frac{1}{\sin^{2}2\vartheta_{\mathrm{sol}}}
\,.
\label{103}
\end{equation}

\item
$m_{1} \simeq m_{2} \simeq m_{3}$
(almost degenerate neutrino masses).

Let us assume that $m_1 \gg \sqrt{\Delta{m}^{2}_{\mathrm{atm}}}$.
In this case $m_2 \simeq m_3\simeq m_1$ in both the ``normal'' and ``inverted''
spectra in Fig.~\ref{3nu}.
For the effective Majorana mass,
independently on the character of the mass spectrum, we have
\begin{equation}
|\langle m \rangle|
\simeq
m_{1}
\left|\sum_{i=1}^{3}U_{ei}^{2}\right|
\,.
\label{104}
\end{equation}
Neglecting small contribution of
$|U_{e3}|^{2}$ ($|U_{e1}|^{2}$ in the case of the inverted hierarchy),
for $|\langle m \rangle|$ we obtain the relations
(\ref{100})--(\ref{102}), in which 
$\sqrt{\Delta{m}^{2}_{\mathrm{atm}}}$ must be changed by $ m_1$. 
Thus, if it happens that 
$
|\langle m \rangle|
\gg
\sqrt{\Delta{m}^{2}_{\mathrm{atm}}}
\simeq
5 \times 10^{-2}\mathrm{eV}
$,
it would be a signature of an almost
degenerate neutrino mass spectrum.
In this case,
the neutrino mass $m_{1}$ is limited by
\begin{equation}
|\langle m \rangle|
\leq
m_{1}
\leq
\frac{|\langle m \rangle|}{|\cos 2\vartheta_{\mathrm{sol}}|}
\lesssim
2\,|\langle m \rangle|
\,.
\label{105}
\end{equation}
For the parameter $\sin^{2}\alpha$,
which characterizes the violation of CP invariance in the lepton sector,
we have \cite{Bilenky:1996cb,Pascoli:2002qm}
\begin{equation}
\sin^{2}\alpha
\simeq
\left(
1 - \frac{|\langle m \rangle|^{2}}{m_{\beta}^{2}}
\right)
\frac{1}{\sin^{2}2\,\vartheta_{\mathrm{sol}}}
\,.
\label{106}
\end{equation}
If the mass $m_1$ is measured in the KATRIN experiment \cite{Osipowicz:2001sq}
and the precise value of the parameter $\sin^{2}2\,\vartheta_{\mathrm{sol}}$ 
is determined in the solar,
KamLAND \cite{hep-ex/0212021}, BOREXINO \cite{Bellini-Nu2002}
and other neutrino experiments,
information on the Majorana
CP phase can be inferred
from the results of the future $(\beta\beta)_{0\nu}$-decay experiments.

\end{enumerate}

\begin{figure}[t]
\begin{center}
\includegraphics*[bb=66 322 476 706, height=10cm]{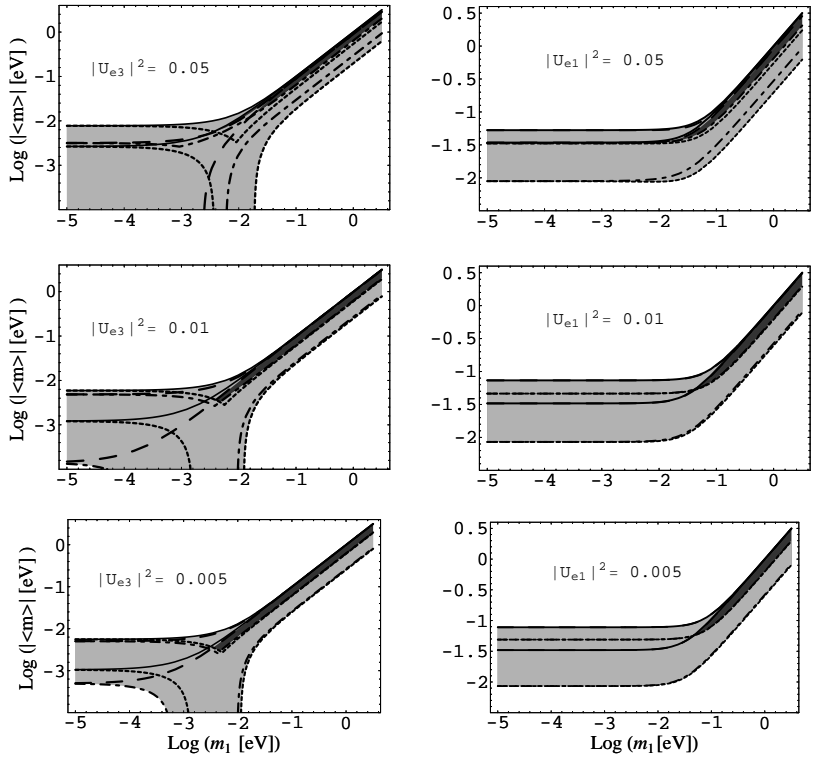}
\end{center}
\caption{ \label{pascoli-0205022}
The dependence of $|\langle m \rangle|$ on $m_1$ 
in the case of the LMA-MSW solution
of the solar neutrino problem \cite{Ahmad:2002ka}
($99.73 \%$~C.L.),
for the ``normal'' scheme in Fig.~\ref{3nu} (left panels),
with
$\Delta{m}^2_{\mathrm{sol}} = \Delta m_{21}^2$,
and
for the ``inverted'' scheme in Fig.~\ref{3nu} (right panels),
with
$\Delta{m}^2_{\mathrm{sol}} = \Delta m_{32}^2$.
Figure taken from Ref.~\cite{Pascoli:2002qm}.
}
\end{figure}

Figure~\ref{pascoli-0205022} \cite{Pascoli:2002qm} shows
the dependence
of $|\langle m \rangle|$ on $m_1$
in the case of the LMA-MSW  solution
of the solar neutrino problem
(99.73\%~C.L. region in Ref.~\cite{Ahmad:2002ka}),
for the ``normal'' scheme in Fig.~\ref{3nu} (left panels)
and
for the ``inverted'' scheme in Fig.~\ref{3nu} (right panels).
For the ``normal'' scheme
with
$\Delta{m}^2_{\mathrm{sol}} = \Delta m_{21}^2$,
in the case of CP-conservation
the allowed values of $|\langle m \rangle|$
are constrained to lie
in the medium-gray regions
a) 
between the two  thick solid lines if
$\eta_{21} = \eta_{31} = 1$,
b) 
between the two long-dashed lines and the axes if
$\eta_{21} = - \eta_{31} = 1$,
c) 
between the dash-dotted lines and the axes
if $\eta_{21} = - \eta_{31} = - 1$,
d) 
between  the short-dashed lines 
if $\eta_{21} = \eta_{31} = - 1$.
For the ``inverted'' scheme
with
$\Delta{m}^2_{\mathrm{sol}} = \Delta m_{32}^2$,
in the case of CP-conservation
the allowed regions
for $|\langle m \rangle|$ correspond:
for $|U_{e1}|^2 = 0.005$ and $|U_{e1}|^2 = 0.01$ to
the medium-gray regions
a)  between the 
solid lines  
if $\eta_{21} = \eta_{31} = \pm 1$,
b)  between the dashed lines
if $\eta_{21} = - \eta_{31} = \pm 1$;
for $|U_{e1}|^2 = 0.05$ to
the medium-gray regions
c) between the 
solid lines  
if $\eta_{21} = \eta_{31} =  1$,
d) between the 
long-dashed lines  
if $\eta_{21} =  \eta_{31} = -  1$,
e)  between the 
dashed-dotted lines  
if $\eta_{21} = - \eta_{31} =   1$,
f)  between the 
short-dashed lines  
if $\eta_{21} = - \eta_{31} = -  1$.
Here
$\eta_{ij}$
is the relative CP-parity
the neutrinos
$\nu_{i}$ and $\nu_{j}$,
given by
\begin{equation}
\eta_{ij}
=
e^{i\frac{\pi}{2}\left(\rho_i-\rho_j\right)}
\,.
\label{0941}
\end{equation}

In the case of CP-violation, the allowed area
for $|\langle m \rangle|$ covers all the gray regions
in Fig.~\ref{pascoli-0205022}. 
Values of $|\langle m \rangle|$ in the dark gray regions
signal CP-violation.

All previous
conclusions are based on the assumption that the value of the effective
Majorana mass $|\langle m \rangle|$ can be obtained
from the measurement of the life-time
of $(\beta\beta)_{0\nu}$-decay.
There is, however, a serious theoretical problem in the determination of
$|\langle m \rangle|$ from experimental data
caused by the necessity to calculate
the nuclear matrix elements.

In the framework of 
Majorana neutrino mixing, 
the total probability of
$(\beta\beta)_{0\nu}$-decay has the general form
(see Ref.~\cite{Doi:1985dx})
\begin{equation}
\Gamma^{0\nu}(A,Z)
=
|\langle m \rangle|^{2}\,|M(A,Z)|^{2}\,G^{0\nu}(E_{0},Z)
\,,
\label{107}
\end{equation}
where $M(A,Z)$ is the nuclear matrix element and
$G^{0\nu}(E_{0},Z)$
is a known phase-space factor ($E_{0}$ is the energy release).
Thus, in order to determine  $|\langle m \rangle|$ from the experimental data we need
to know the nuclear matrix element $M(A,Z)$. This last quantity 
must be calculated.

There are at present large uncertainties in the calculations of the
nuclear matrix elements of $(\beta\beta)_{0\nu}$-decay 
(see Refs.~\cite{Suhonen:1998ck,Faessler:1998zg,Elliott:2002xe}).
Two basic approaches to the calculation are  
used:
quasi-particle random phase approximation and the nuclear shell model.
Different calculations of the
lifetime of the $(\beta\beta)_{0\nu}$-decay differ
by about one order of magnitude.
For example,
for the lifetime of the $(\beta\beta)_{0\nu}$-decay of $^{76}\mathrm{Ge}$,
assuming that
$|\langle m \rangle| = 5 \times 10^{-2} \, \mathrm{eV}$,
the range
\begin{equation}
6.8 \times 10^{26} \, \mathrm{yr}
\leq
T^{0\nu}_{1/2}(^{76}\mathrm{Ge})
\leq
70.8 \times 10^{26} \, \mathrm{yr}
\label{108}
\end{equation}
has been obtained
(see Ref.~\cite{Elliott:2002xe}).

The problem of the calculation of the nuclear matrix elements
of neutrinoless double-$\beta$ decay is a real theoretical challenge.
It is obvious that without a solution of this problem
the effective Majorana neutrino mass $|\langle m \rangle|$
cannot
be determined from the experimental data with reliable accuracy
(see the discussion in Ref.~\cite{Barger:2002vy,Pascoli:2002qm}).

The authors of Ref.~\cite{Bilenky:2002fi} proposed 
a method which allows to check the results
of the calculations
of the nuclear matrix elements of the $(\beta\beta)_{0\nu}$-decay
of different nuclei by confronting them with experimental data.
Let us take into account that
\begin{enumerate}

\item
For small neutrino masses
($m_{i}\lesssim 10\,\mathrm{MeV}$) the nuclear matrix elements do not
depend on the neutrino masses \cite{Doi:1985dx}.

\item
A sensitivity of a few $10^{-2}$ eV for
$|\langle m \rangle|$ is planned to be reached in
future experiments on the search for neutrinoless double-$\beta$ decay
of \emph{different} nuclei.

\end{enumerate}

From Eq.~(\ref{107}) we have
\begin{equation}
R(A,Z/A',Z')
\equiv
\frac{T^{0\nu}_{1/2}(A,Z)}{T^{0\nu}_{1/2}(A',Z')}
=
\frac{|M(A',Z')|^{2}\,G^{0\nu}(E'_{0},Z')}{|M(A,Z)|^{2}\,G^{0\nu}(E_{0},Z)}
\,.
\label{109}
\end{equation}
Thus, if  the neutrinoless double $\beta$ -decay of \emph{different nuclei}
is observed, the calculated ratios of the corresponding
squared nuclear matrix elements
can be 
confronted with the experimental values. 
Table~\ref{bb-ratio} shows the
ratios of the lifetimes of $(\beta\beta)_{0\nu}$-decay
of several nuclei, calculated in six
different models,
using
the values of the lifetimes given in Ref.~\cite{Elliott:2002xe}. 
As it is seen from Table~\ref{bb-ratio}, 
the calculated ratios
vary
within about one order of magnitude.

\begin{table}[t]
\begin{center}
\begin{tabular}{|ccccccc|}
\hline
Lifetime Ratios
&
\protect\cite{Haxton:1984am}
&
\protect\cite{Caurier-PRL77-1996,Caurier-NPA654-1999}
&
\protect\cite{Engel:1988au}
&
\protect\cite{Staudt:1990qi}
&
\protect\cite{Toivanen:1995zi}
&
\protect\cite{Pantis:1996py,Simkovic:1999re,Simkovic:2001qf}
\\
\hline
$R(^{76}\mathrm{Ge}/^{130}\mathrm{Te})$
&
11.3
&
3
&
20
&
4.6
&
3.6
&
4.2
\\
$R(^{76}\mathrm{Ge}/^{136}\mathrm{Xe})$
&

&
1.5
&
4.2
&
1.1
&
0.6
&
2
\\
$R(^{76}\mathrm{Ge}/^{100}\mathrm{Mo})$
&

&

&
14
&
1.8
&
10.7
&
0.9
\\
\hline
\end{tabular}
\end{center}
\caption{ \label{bb-ratio}
The results of the calculation of the ratios of the lifetime of
$(\beta\beta)_{0\nu}$-decay of several nuclei in six different models.
The references
to the corresponding papers are given in brackets.
}
\end{table}

As one can see from Table~\ref{bb-ratio}, 
the values of the ratio $R(^{76}\mathrm{Ge}/^{130}\mathrm{Te})$
calculated in Ref.~\cite{Staudt:1990qi} and
Ref.~\cite{Toivanen:1995zi} are, correspondingly,
4.6 and 3.6.
It is clear that it will be difficult to distinguish
models \cite{Staudt:1990qi} and  \cite{Toivanen:1995zi} through the observation of the
neutrinoless double-$\beta$ decay of  
$^{76}\mathrm{Ge}$ and $^{130}\mathrm{Te}$.
However,
it will be possible to distinguish the corresponding models through the
observation of the
$(\beta\beta)_{0\nu}$-decay of $^{76}\mathrm{Ge}$ and $^{100}\mathrm{Mo}$
(the values of the corresponding ratio are 1.8 and 10.7, respectively).
This example illustrates the importance of the investigation of 
$(\beta\beta)_{0\nu}$-decay
of more than two nuclei.

The nuclear part of the matrix element of $(\beta\beta)_{0\nu}$-decay
is determined by the matrix element
of the $T$-product of two hadronic charged currents 
connected
by the propagator of a massless boson.
This matrix element cannot be connected with the matrix element of any observable
process.
The method proposed in Ref.~\cite{Bilenky:2002fi} is based only on the
smallness of neutrino masses and on the
factorization of the neutrino and nuclear parts of the matrix element of  
$(\beta\beta)_{0\nu}$-decay.
It requires observation of the 
$(\beta\beta)_{0\nu}$-decay of different nuclei.

Let us notice that,
if the ratio in Eq.~(\ref{109}), calculated in some model, is in agreement with
the experimental data, it could only mean that the model is correct
up to a possible factor, which does not depend on $A$ and $Z$ (and
drops out from the ratio (\ref{109})).
Such factor
was found and calculated in Ref.\cite{Simkovic:1999re},
where in addition to the usual axial and vector terms
in the nucleon matrix element pseudoscalar and weak magnetic form factors were taken into account.
It was shown that in the case of light Majorana neutrinos these
additional terms lead to
a universal reduction of the nuclear matrix elements
of $ (\beta \beta)_{0\,\nu}$-decay by about 30 \%.
This reduction, which practically does not depend on
the type of nucleus, causes a raise
of the value of the effective Majorana mass $|\langle m \rangle|$
that could be obtained
from the results of future experiments.

\section{Cosmology}
\label{Cosmology}

Perhaps the best example of the fruitful cross-fertilization of
high energy physics and cosmology is the momentous constraint by
Big-Bang Nucleosynthesis (BBN) \cite{Olive:2002qg} on the number
of light neutrino species. Indeed, the number of effective light
degrees of freedom affects the expansion rate of the Universe; the
larger this number, the larger is the expansion rate and hence the
higher the freeze out temperature of the weak interactions that
inter-convert neutrons and protons. Thus, the neutron to proton
ratio is correspondingly higher and so is the primordial helium
yield. These events took place when the temperature of the
universe was of the order of 1 MeV and therefore it is clear that
neutrino masses at the 1 eV scale or less play no significant role
in primordial light element formation. As a consequence, no
relevant information on the absolute value of light neutrino
masses from those early epochs of the history of the universe can
be gained. This does not mean, however, that cosmology cannot
supply interesting information on the neutrino mass issue.
Fortunately, we can learn about neutrino mass from various
cosmological and astrophysical instances as different as the
Cosmic Microwave Background radiation (CMB), the power spectrum in
large scale structure (LSS) surveys, and Lyman $\alpha$
(Ly$\alpha$) forest studies. We will address these issues in what
follows (see also the reviews in
Refs.~\cite{Pas:2001nd,Dolgov:2002wy,Dolgov:2002ad,Kainulainen:2002pu,%
Raffelt:2002ed,hep-ph/0210089}).

\subsection{The Gerstein-Zeldovich limit on neutrino masses}
\label{The Gerstein-Zeldovich limit on neutrino masses}

Before entering the issues mentioned explicitly above, let us
present the ``classical'' cosmological bound on the sum of the
masses of all neutrino species derived by Gerstein and Zeldovich
\cite{Gershtein:1966gg,Cowsik:1972gh}. Stable light neutrinos
(i.e. relativistic at neutrino decoupling) are present in the
Universe today with an abundance of about $100$ neutrinos and
antineutrinos per $\mathrm{cm}^3$. If they carry mass and this
mass is much larger than the present CMB temperature
(\textit{i.e.}
$m_{\nu} \gg k T_{\mathrm{CMB}} \sim 3 \times 10^{-4} \, \mathrm{eV}$, with
$T_{\mathrm{CMB}} \simeq 3 \, \mathrm{K}$), they contribute to the
known mass density $\Omega_{m}$ (relative to the critical density
$\rho_c = 3 H_0^2 / 8 \pi G_N$,
where $H_0$ is the Hubble constant
and
$G_N$ is the Newton gravitational constant)
associated to nonrelativistic matter (mainly dark). The energy
density\footnote{Neutrino masses relate directly to energy density
only if the chemical potential of relic neutrinos is negligible.
It has been shown in \cite{semikoz} that the neutrino chemical
potentials of the three species are very small. The cosmological
limits on neutrino mass that we will discuss in this review comply
with this fact.} associated to neutrino mass can be thus be
written as
\begin{equation}
\Omega_{\nu}h^2 =\sum_i \frac{ m_{i} }{ 93\, \eV }
\,,
\end{equation}
where, as usual, the Hubble constant is parameterized as $H_{0}=100h$ km/s/Mpc.

Since observationally $\Omega_{m}h^2 \approx 0.15$ and
$\Omega_{\nu}<\Omega_{m}$, it follows that $\sum_i m_{i}<14$ eV. For
mass degenerated neutrinos this bound implies that $m_i<5$ eV for
each species.

\subsection{Microwave Background Anisotropies }
\label{Microwave Background Anisotropies}

\begin{figure}[t]
\centering
\includegraphics*[bb=61 60 527 378, width=0.8\textwidth]{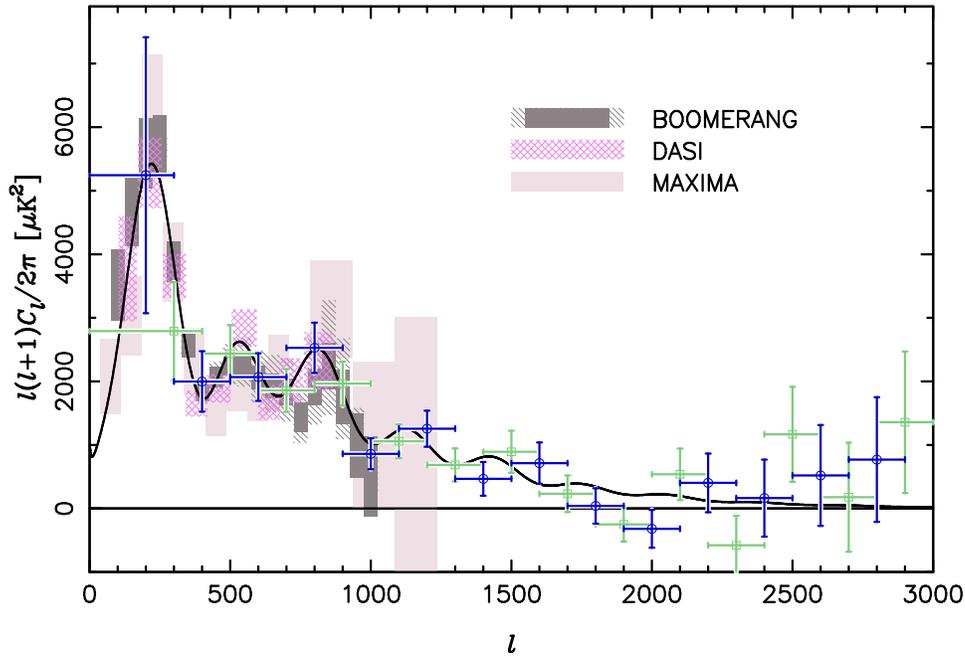}
\caption{
A recent plot of the anisotropy spectrum taken from Ref.~\cite{pearson}.
It shows BOOMERANG, DASI, MAXIMA, and CBI data
(squares and circles).
See Ref.~\cite{pearson} and references therein for details.
}
\label{figura1}
\end{figure}

The background radiation first detected by Penzias and Wilson in
the late fifties follows an almost perfect black-body spectrum
at the temperature $T_0 = 2.728 \pm 0.002 \, \mathrm{K}$. This
radiation is extremely isotropic so that this temperature on the sky is
direction independent to a precision of $10^{-5}$, once the Doppler effect
due to the peculiar velocity of the Solar System is removed.
However the Universe is highly inhomogeneous today and this means
that it should have been sufficiently inhomogeneous in the past
in order that structure could grow via gravitational instability.
Therefore, density inhomogeneities should give rise to
temperature inhomogeneities in the sky \cite{hu}. For many years such
temperature fluctuations in the cosmic background radiation have
been searched for until they have been finally established at the
aforementioned minute $10^{-5}$ level by COBE \cite{COBE}. It is customary to expand the
temperature fluctuations ${\Delta T / T_0}$ in spherical
harmonics
\begin{equation}
\frac{\Delta T}{T_0}=\sum_{l\ge 2} a_{lm}Y_{lm}(\theta, \varphi)
\,.
\end{equation}
The coefficients $a_{lm}$ are random variables with zero mean
$\langle a_{lm} \rangle=0$ and variance
$\langle a_{lm}^*a_{l'm'} \rangle=C_{l}\delta_{ll'}\delta_{mm'}$ as required by
the statistical isotropy of temperature fluctuations. The
$C_{l}$'s form the angular power spectrum and this angular power
spectrum is conventionally shown when presenting CMB results (see Fig.~\ref{figura1}).
The
CMB photons that we now record were last scattered at
recombination when the universe was about $300,000$ yr old and the
redshift was $z\sim 1100$. So, what we get from the temperature
fluctuation spectrum is essentially a snapshot of the density
inhomogeneities at recombination. It is not quite a $bona fide$
picture of the anisotropies at that time because in their way to
us the cosmic photons should have experienced gravitational
redshift by changing gravitational potentials, the so-called
Integrated Sachs-Wolfe effect (ISW) \cite{sachs}, and eventually,
rescattering by ionized gas in interposed clusters of galaxies
(Sunyaev-Zeldovich effect \cite{sunyaev}). The primary causes of
temperature anisotropy, those present at recombination, are
threefold. There is an intrinsic source associated to the fact
that denser spots are hotter and hence photons emerging from those
denser regions are bluer. But also, photons in denser regions will
be redshifted as they climb out of their potential wells
(Sachs-Wolfe effect (SW) \cite{sachs}). These are competing
effects and it depends on the scale under scrutiny that one or the
other dominates. And finally, the third source of temperature
anisotropy generation is associated to Doppler shifts arising from
the peculiar motion of matter in underdense regions being
attracted towards overdense regions and from which photons are
last scattered. We collect the three primary sources in the
formula:
\begin{equation}
\frac{\Delta T}{T_{0}}=\left(\frac{\Delta T}{T}\right)_{intr}+\phi -{\hat n} \cdot \vec v
\,,
\end{equation}
where $\phi$ is the gravitational potential well, $\vec v$ is the
peculiar velocity, and $\hat n$ is a unit vector pointing in the
direction $\theta,\varphi$.

To connect the observational CMB anisotropy data with the
underlying cosmological model and thus have a handle on the
different cosmological parameters one has to work out the
different pieces in the previous equation in terms of the matter
density inhomogeneities and peculiar velocity of the
photon-baryon fluid at recombination. These density fluctuations
and peculiar velocity, in turn, have to be obtained from the
general relativistic equations (and/or their newtonian
counterparts, when appropriate) to take into account their time
evolution from given initial (end of inflation) conditions (adiabatic)  up to recombination.
Since density perturbations are supposedly small over the whole period
of interest (up to recombination), one uses linear perturbation
theory to deal with the problem which then becomes easy to
solve. Indeed, a main feature of the linearized theory is that
by Fourier transforming from $r$ $space$ into $k$ $space$, the different
$k$ $modes$ become mutually independent and therefore the
corresponding spatial scales evolve independently during the
linear era of structure formation. Because each $k$ $mode$
corresponds to a different spatial size $\lambda$ ($\lambda ={2\pi/k}$),
a given mode enters the
horizon at a given epoch. But crossing the Hubble radius is
physically relevant since before crossing a scale evolves solely
under the rule of gravity and only after horizon crossing are also causal
effects operative. A sub-horizon sized fluctuation, therefore,
experiences both the gravitational pull and the pressure
gradients of the photon-baryon fluid. Too much gravity pull
cannot be counteracted by fluid pressure, hence there is a
critical size for a perturbation to stand gravity. Beyond that
size (so called Jeans size), collapse is unimpeded but below
the Jeans size collapse can be halted. This Jeans scale is set
by the sound speed in the primeval plasma because it determines
the distance over which a mechanical response of pressure forces
can propagate over a gravitational free-fall time and thus
restore hydrodynamical equilibrium in the fluid. For those
under-sized perturbations, acoustic oscillations set in:
compression is followed by rarefaction and back to compression
and so forth because the in-falling fluid bounces off
every time the pressure of the fluid rises to the point where it
can halt gravitational in-fall and reverse the process from contraction to
expansion. Since before recombination the pressure of the
baryon-photon plasma is dominated by photon pressure, the sound
velocity is roughly $c_{s}\sim  c/\sqrt 3$, close to the speed
of light. So, during the pre-recombination stages, the sound
horizon approximately matches the Hubble radius and thus scales
entering the horizon before recombination undergo acoustic oscillations from that
moment onwards. Later, when recombination takes place and photons
are freed from the baryons to which they were previously tightly
bound, the different modes (scales) are caught at different
phases of their oscillation with correspondingly different amplitudes
of their density perturbations. These compression and
rarefaction phases translate into peaks in the temperature power
spectrum that one observes. Odd peaks correspond to compression
maxima and even peaks correspond to compression minima (rarefaction peaks). The
first peak is associated to the scale that enters the horizon at
recombination and is thus caught in its first oscillation
height. The second peak corresponds to a scale that has already
gone through a complete oscillation cycle at recombination, etc. Because
the smaller the scale the sooner it entered the horizon, and therefore
will have got time for a longer period of oscillations before
photon decoupling, the corresponding peaks in the power spectrum
are progressively attenuated as compared to the first
compression maximum. The main source of damping, which is called Silk
damping \cite{silk}, is due to the fact that photons in the baryon-photon
fluid have a mean free path governed by the Thomson cross
section (photons are coupled to the electrons via Thomson scattering
and electrons in turn are tightly bound to protons via Coulomb
interactions) and so photons tend to leak out from overdense
regions to less dense regions whenever the photon mean free path
(which depends on the ionization history before recombination and on the baryon
content) exceeds the scale of the density fluctuation. In
addition to this there is also a limit to the pattern of peaks
supplied by the finite width of the last scattering surface.
Since recombination is not instantaneous but takes a finite
amount of time, observations of the cosmic background
temperature are actually an average over temperatures of photons that
reach us from a shell whose thickness is about one tenth the Hubble
distance at recombination \cite{jones}. Hence, scales that are of this order
of magnitude or less are completely washed out by the temperature averaging
process. For a flat universe this limit corresponds to angular
scales of about $0.1$ deg or to multipoles larger than about
$l\sim 2000$. We are prepared now to discuss what can be learned
from the peak structure of the power spectrum as far as
general cosmological parameters is
concerned (including the neutrino energy content, which is our main concern here).

The characteristics of the spectrum of CMB anisotropies and of its peaks ( i.e. their positions, heights,
and shapes) depend on the adopted cosmological model for the
initial anisotropies and on all the parameters that govern the evolution of the acoustic
oscillations before recombination and on the hindrances encountered by
photons in their paths from recombination onwards. Clearly, not
all parameters affect to the same extent the different aspects
of the power spectrum pattern. Let us remind here of the main
influences on the power spectrum features and let us leave for
later the neutrino mass related issues:

\begin{enumerate}

\item \label{if1}
At large angular scales (i.e., large compared to the horizon
scale at recombination), the SW effect dominates the
anisotropies. Those are the scales that, at recombination, have
had no microphysics processing yet and reflect directly the
character and strength of the primordial inhomogeneities.
Therefore, one can read from the observed spectrum at large angular scales (small $l$'s) the general
cosmological paradigm about the origin of inhomogeneities. For
instance, a flat spectrum (as it is indeed the case) in this low $l$ regime suggests an
adiabatic scale-invariant inflationary-like model of primordial
anisotropies.

\item \label{if2}
When a physical feature on the last scattering surface at
recombination is projected on the sky today, the projection
depends on the distance to last scattering and on the curvature
of space. The distance is fixed by $t_{rec}$ and hence depends
mainly on the Hubble constant and total energy density $\Omega$.
Curvature affects the projection because it is an obvious
geometric fact of curved spaces that objects seem to be smaller
(larger) in an open (closed) Universe then they would in a flat
Universe. Thus the angle $\theta$ subtended by a given structure
on the sky (its angular size) depends on curvature. Recall that the first peak corresponds to the
scale that enters the horizon at recombination. So, its location
in $l$ ($l\sim\theta^{-1}$) tells us about the curvature of the
Universe. For a zero curvature Universe the position of the first peak should be at $l_{1}\sim 200$,
which is actually the case of the observed spectrum (see Fig.~\ref{figura1}).

\item \label{if3}
Baryons contribute inertia to the oscillating photon-baryon
plasma in the dark matter potential wells and as a consequence
the bigger the baryon content is the larger the amplitudes of the acoustic vibrations are (i.e., the peaks are higher). Furthermore,
compression peaks (odd peaks) are enhanced relatively to
rarefaction peaks (even peaks) as a result of baryon drag.

\end{enumerate}

Unfortunately there is a substantial amount of degeneracy among the
cosmological parameters \cite{bond} that allow for a multiplicity of
different parameter choices giving an equally acceptable
spectrum. So it is very desirable to use alternative
measurements as complementary tools for determining
cosmological parameters and thus help break degeneracies. For
instance, for a flat Universe ($\Omega \sim 1$, as it is indeed the case;
see point~\ref{if2} above)
the position of the first peak is almost independent of the
relative weight of matter (baryonic plus dark) and dark energy
(cosmological constant) in $\Omega$.

Nonzero mass neutrinos affect the CMB anisotropy spectrum to a
much lesser extent than the previously stated effects \cite{dodelson}. Their
influence is twofold:
the position of the peaks is slightly modified and also their
amplitudes are enhanced. Although the position of the first peak
is mostly dictated by curvature, the peaks and the troughs move slightly
to lower $l's$ (to the left in Fig.~\ref{figura1}) due to massive neutrinos. This effect can
be traced back to the fact that neutrinos being massive, they
start being ultrarelativistic until their freeze out and beyond
and only later in the history of the Universe
the neutrinos become nonrelativistic. Compared to the massless
neutrino case where neutrinos are always relativistic degrees
of freedom, in the case under discussion, the expansion rate is
slightly modified since the competition between radiation
domination and matter domination is altered. While the propagation
of sound in the photon-baryon fluid depends only on the baryon
density and hence neutrino mass is not relevant here, the sound
horizon at decoupling is modified simply because decoupling is
slightly delayed due to the change in expansion rate. If the sound
horizon is larger, so it is the scale that enters the horizon at
last scattering. Therefore, larger angular scales corresponding to
lower $l's$, the pattern of acoustic peaks is shifted to the left.
The neutrino mass, on the other hand, has a larger impact on the
power spectrum than the shifting just mentioned; it leads to an
enhancement of the peaks. The origin of the effect is related to
the time variation of the gravitational potentials. In a radiation
dominated Universe potentials change with time whereas in a matter
dominated Universe gravitational potentials are constant. Since a
Universe with massive neutrinos implies that relativistic matter
turns into non-relativistic matter during relevant periods of its
evolution, the acoustic oscillations of the baryon-photon plasma
are being forced by time decaying potentials that differ from
those associated to a Universe with massless neutrinos only. As a
result the acoustic oscillations (mainly for $l\ge 300$) get an
extra boost in amplitude at last scattering (parametric
resonance). There is also a smaller effect associated to varying
potentials after last scattering that introduces a relative
difference between models with/out massive neutrinos (affecting
the ISW contribution to anisotropies) which is operative at
smaller $l's$.

\subsection{Galaxy Redshift Surveys}
\label{Galaxy Redshift Surveys}

In the previous section we gave a brief and general description of
the physics of the CMB angular power spectrum and noted that the
direct influence upon it of neutrino mass is only marginal. Yet,
the CMB power spectrum data is important in the determination of
neutrino mass because it can be used in combination with other
astrophysical sources -- where the neutrino mass plays a more
relevant role -- to help reduce the number of uncertainties in the
various cosmological parameters. One of these sources is large
scale structure. Neutrino mass affects large scale structure
formation and its effect can be studied via observation of the
distribution of galaxies. Since the distribution of galaxies
should trace the matter density of the Universe (related to each
other via a bias factor), large samples of galaxy redshifts in
surveys such as the 2 degree Field Galaxy Redshift Survey (2dFGRS)
\cite{colles} provide a tool to study the power spectrum of matter
fluctuations with very small random errors.

\begin{figure}[t]
\centering
\includegraphics*[bb=32.4481 298.7683 572.62 702.7, height=7cm]{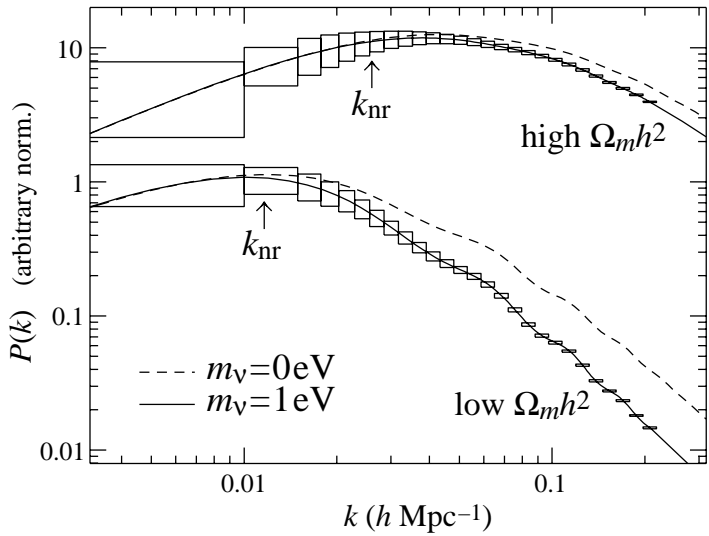}
\caption{
Effect of a 1 eV neutrino on $P(k)$ as found by the authors of Ref.~\cite{hu2}.
}
\label{figura2}
\end{figure}

Recall that in linear theory what one is dealing with is the
Fourier transformed density perturbations $\delta_{k}$. The
initial conditions for $\delta_{k}$  are set to reproduce a
property of inflation (and consistent with observations), i.e. a
flat or Harrison-Zeldovich spectrum is assumed. This implies that
the power spectrum behaves as $P(k)=\langle \delta_{k}^2 \rangle\sim k^n$ with
$n\simeq 1$. This initial spectrum has to be evolved from the very
high initial redshifts to the redshifts relevant for structure
formation surveys (the median redshift of 2dFGRS is $\sim 0.1$
\cite{colles}). This processing is dictated by the continuity,
Euler and Boltzmann equations that govern the physics of the
perturbations of the cosmic fluid. To be specific, what concerns
us here is the effect of massive neutrinos on the evolution of
perturbations, i.e on the power spectrum $P(k)$. Once decoupled very
early in the history of the Universe (at $T\sim 1$ MeV) neutrinos
free-stream at almost the speed of light. This is so until after
their momenta become on the order of their mass and less and hence
they enter a non-relativistic regime. During their relativistic
life-span they outflow from regions smaller than the horizon so
that these regions are being depleted and hence energy density
perturbations at those scales are effectively erased. This
phenomenon comes to an end when neutrinos cease to free-stream as
they become non-relativistic and can cluster with the cold
components of dark matter for all scales that are larger than the
Hubble radius at the time the neutrinos become non-relativistic.
This limiting scale is given by the formula \cite{hu2}:
\begin{equation}
k_{nr}\simeq 0.03(m_{\nu}/1\,\eV)^{1/2}\Omega_{m}^{1/2}h \mathrm{Mpc}^{-1}
\,.
\end{equation}
For all scales smaller than this (i.e. for $k\ge k_{nr}$) the
growth of perturbations is suppressed. So neutrino mass influences
the power spectrum of cosmological structure at small scales. The
loss of power on small scales can be approximated by \cite{hu2}
\begin{equation}
\frac{\Delta P}{P}\approx -8\frac{\Omega_{\nu}}{\Omega_{m}} \,.
\label{powerloss}
\end{equation}
This equation gives us a handle for extracting the bounds on
neutrino mass from the large samples of data in present and
upcoming galaxy distribution surveys (see Fig.~\ref{figura2}). The
analysis of the data has to be restricted to a band of scales for
which the data are precise enough and for which linear
perturbation theory holds. On the large scales side accuracy is
volume limited and on the small scales side linear theory is
jeopardized. The 2dFGRS data, for instance, are robust on scales
$0.02<k<0.15$ $h$ $\mathrm{Mpc}^{-1}$.

\subsection{Lyman $\alpha$ forests }
\label{Lyman alpha forests}

The last piece of information relevant to the neutrino mass issue
that we want to discuss is the Ly$\alpha$ forest \cite{peebles}
measurements of the power spectrum of mass fluctuations. Quasars
are of help in cosmology because for one thing they rank among the
oldest detected objects in the Universe and hence provide crucial
hints for structure formation studies. Furthermore, quasar spectra
are a means for studying the intergalactic medium. Atomic hydrogen
in the gas clouds in the vicinity of the quasar makes $2p \to 1s$
Ly$\alpha$ transitions that are seen redshifted today by a factor
$1+z_{quasar}$. But what is most important here is that quasar
spectra show a series of absorption lines associated to
intervening clouds that photons encounter in the way from the
quasar to us. Indeed, the spectrum displays a ``forest'' of
Ly$\alpha$ absorption lines to the left of the Ly$\alpha$ emission
line of the quasar, i.e. blueshifted relatively to the emitter,
that correspond to the resonant absorption of those photons -- with
wavelengths $\lambda$ stretched by cosmic expansion in the
proportion $(1+z_{quasar})/(1+z_{cloud})$ -- that exactly match the
Ly$\alpha$ transition. The forest of absorption lines in the
quasar spectrum refers therefore to a sequence of clouds at
various redshifts along our line-of-sight towards the quasar that
absorb radiation from the quasar at specific wavelengths given by
the redshift of each cloud relative to the quasar's own redshift.
The distribution of such clouds thus can provide astronomers with
important clues about structure formation. In particular, since
light neutrinos do not cluster on small scales as was explained in
the previous section, the measurement of cluster formation on
small scales extracted from Ly$\alpha$ forest observations can
lead to definite predictions for neutrino mass in the eV range.

\subsection{Neutrino mass bounds }
\label{Neutrino mass bounds}

Now that we discussed the different cosmological and astrophysical
sources of information on neutrino mass, we can summarize the
constraints on neutrino masses that follow from these sources \cite{hannestad}.

 Absolute neutrino masses cannot be measured in neutrino
oscillation experiments. Only mass-squared differences have been
established so far. If the three neutrino mass spectrum is
hierarchical then
$m_{1}\simeq 0$,
$m_{2} \simeq \sqrt{\Delta{m}^2_{\mathrm{sol}}}$, and
$m_{3} \simeq \sqrt{\Delta{m}^2_{\mathrm{atm}}}$.
Should, on the other hand, the
mass hierarchy be inverted, then
$m_{1} \simeq 0$,
$m_{2} \simeq \sqrt{\Delta{m}^2_{\mathrm{atm}}}$, and
$m_{3} \simeq \sqrt{\Delta{m}^2_{\mathrm{atm}}}$.
The third possibility is mass degeneracy, i.e.
$m_{1} \simeq  m_{2} \simeq m_{3}$.
In this case the three masses would
be much bigger than $\sqrt{\Delta{m}^2_{\mathrm{atm}}}$ and any hint as to
their absolute mass scale is lost in neutrino oscillations. In this instance
astrophysical/cosmological tests come to the rescue.

The Ly$\alpha$ forest seen in quasar spectra, as mentioned in the
previous section, can be used to study mass distribution
fluctuations. There is a well-understood theory of Ly$\alpha$
forest formation embodied in the standard cosmology which
establishes a rather simple local connection between the absorbed
flux in a quasar spectrum and the underlying matter density. From
this relationship, the power spectrum $P(k)$ can be extracted, at
least over a limited range of scales \cite{katz}.

Indeed, in the usual cosmological scenario where structure
formation proceeds via gravitational instabilities, the behavior
of gas plus a background of UV ionizing radiation leads naturally
to quasar absorption effects. The physics of the gas is driven by
the competition of two phenomena, namely adiabatic cooling due to
Hubble flow, and heating by the photo-ionizing UV background.
Hydrodynamic simulations show that the Ly$\alpha$ forest arises in
gas of moderate overdensity \cite{miralda}. This density field can
be then locally related to the Ly$\alpha$ optical depth
\cite{katz2}, and consequently to the observed transmitted flux in
a quasar spectrum.

Light neutrinos delay the growth of perturbations at small scales
and therefore a constraint on their mass is made possible through
the recovery of the power spectrum $P(k)$ from an observational
measurement of the Ly$\alpha$ forest. The authors of
Ref.~\cite{croft} used the measurement in \cite{katz3} to restrict
light neutrino masses. Their general strategy consisted of two
steps. In a first step they tested hydrodynamic Ly$\alpha$ forest
simulations in the context of a cosmological model with non-zero
mass neutrinos. Specifically, what they wanted to check is their
ability to recover, in a given cosmological setting, the power
spectrum $P(k)$ from a hydrodynamic simulation of the forest. The
method had been previously used in models that do not include ``hot
particles'', i.e. models without massive neutrinos. The assumed
underlying cosmological scenario is an adiabatic, cold dark matter
dominated Universe with gaussian initial fluctuations. Such models
contain six free parameters, namely: the matter density
$\Omega_{m}$, the Hubble parameter $h$, the baryon density
$\Omega_{b}$, the neutrino density $\Omega_{\nu}$, and the
amplitude and tilt of the initial power spectrum of density
perturbations $Ak^n$. A flat spatial geometry is also assumed
throughout as the CMB anisotropy and supernova measurements seem
to corroborate. The particular model used in the tests has
$\Omega_{m}=1$, $h=0.5$, $\Omega_{b}=0.075$, and
$\Omega_{\nu}=0.2$. The hydrodynamic simulation \cite{hernquist}
allows then to follow the evolution of structure in this model and
to know the physical condition and distribution of the gas at
$z=2.5$ (i.e. at the redshift of the actual measurements). From
this, they generated artificial Ly$\alpha$ spectra for 1200 random
lines of sight through the simulation volume. The recovery
procedure was applied then to these spectra to see if the
recovered $P(k)$ agrees with the linear theory prediction of the
model under scrutiny. The authors obtained a fair amount of
consistency over the whole observational band of scales and
concluded from their simulation tests that their recovered $P(k)$
was systematically too low in amplitude and had a somewhat flatter
slope than the linear theory prediction (however, the authors
point out that correcting for this underestimation would lead only
to tighter neutrino mass limits). Having proved that the method
works for a specific model, it was assumed that it would also work
for all cosmological models with massive neutrinos available in
parameter space. The final step involves using the observational
results on the Ly$\alpha$ forest to explore the six dimensional
parameter space and place an upper limit on the neutrino mass.

The loss of power in the Ly$\alpha$ forest induced by a neutrino
mass is given by equation (\ref{powerloss}) but other parameters
could produce similar power suppression effects. To avoid as much
as possible undesired degeneracies, the authors use additional
cosmological constraints. They use the Hubble constant measurement
\cite{madore}, the COBE detection of large scale anisotropies
\cite{cobe2}, the present abundance of galaxy clusters
\cite{viana}, galaxy surveys \cite{dodds}, nucleosynthesis limits
on baryon abundance \cite{burles2}, and the age of the oldest
globular clusters to set a lower limit to the age of the Universe
\cite{carretta}. In their analysis the authors reject every model
that violates the 95\% C.L. on any of the aforementioned
constraints. With such reduced parameter space, the analytic
approximations of \cite{eisenstein} were used to find the model
that maximizes $\Omega_{\nu}$ as a function of $\Omega_m$. The
result is
\begin{equation}
\sum_{i}m_{i}<5.5\, \eV
\qquad
(95\% \, \mathrm{C.L.})
\,,
\end{equation}
independently of the value of $\Omega_{m}$.

If $\Omega_{m}$ is restricted to lie in the range $0.2-0.5$ as
favored by observation,
then the authors parameterize their bounds
as\footnote{Here the authors give the parameterization for
a single massive neutrino species of mass $m_{\nu}$; in the
degenerate three neutrino case, one should interpret
$m_{\nu}\equiv \sum_{i}m_{i}$.}
\begin{equation}
m_{\nu} \leq 2.4 \left( \frac{\Omega_{m}}{0.17} -1 \right) \eV
\qquad
(95\% \, \mathrm{C.L.})
\,.
\end{equation}

There have been several analysis that
put constraints on neutrino masses from LSS redshift surveys. The
most recent ones make use of the 2dF galaxy survey \cite{colles}.
The 2dFGRS is a sample of over 220,000 galaxy redshifts that
permits the measurement of large-scale structure statistics with
very small random errors. Ref.~\cite{elgaroy} computes the matter
power spectrum from linear theory for a multiplicity of
cosmological models described in terms of the components: baryons,
cold dark matter, cosmological constant, and of course hot dark
matter (i.e. neutrinos with non-zero mass). The calculated matter
power spectrum and the measured galaxy power spectrum can be put in
correspondence through a bias parameter, i.e. $b^2\equiv
P_{g}(k)/P_{m}(k)$,
where
$P_{g}(k)$ is the measured galaxy distribution power spectrum
and
$P_{m}(k)$ is the matter power spectrum.
Although $b$ is in principle scale dependent,
there are good reasons to believe that $b$ is a constant on the
scales considered in the analysis \cite{benson}. The authors of
this study absorbed this constant in the amplitude $A$ of the
power spectrum of density fluctuations taken as a free parameter.
There is a vast parameter space available for the analysis, and
again it helps to take other cosmological inputs into
consideration. In this way the implications for neutrino mass will
be less uncertain.

From primordial nucleosynthesis one has the constraint
$\Omega_{b}h^2=0.020 \pm 0.002$ on the density of baryons
\cite{burles3}. The Hubble parameter $h$ as measured by the Hubble
Space Telescope (HST) key project \cite{freedman} is $h=0.70 \pm
0.07$. Another prior is the total matter density $\Omega_{m}$. As
stated before when we discussed the peak structure of the CMB
anisotropy spectrum, there is strong evidence for a spatially flat
Universe \cite{efstathiou}. This means
$\Omega_{m}+\Omega_{\Lambda}=1$. This last relation, used together
with the results from surveys of high redshift Type Ia supernovae
(SNIa) \cite{perlmutter,riess}, leads to the constraint
$\Omega_{m}=0.28 \pm 0.14$. On the other hand, independent studies
give a wider spread of values. For instance, mass-to-light ratio
studies of galaxy clusters render typically lower values for
$\Omega_{m}$ ($\sim 0.15$) \cite{bahcall} whereas cluster
abundance studies deliver $\Omega_{m}\approx 0.3-0.9$
\cite{pierpaoli}. Given these facts that make $\Omega_{m}$ the
most poorly known parameter, the authors of the present analysis
employed two kinds of priors on $\Omega_{m}$. One was a Gaussian
at $\Omega_{m}=0.28$ and standard deviation 0.14 as required by
supernova and CMB results and the other was a uniform prior in the
range $0.1<\Omega_{m}<0.5$. The latter upper limit
($\Omega_{m}<0.5)$ is dictated by the values of $h$ used which
would imply, for $\Omega_{m}>0.5$, an age of the Universe shorter
than 12 Gyr. Although the value $n=1$ for the spectral index is
the usual theoretical choice, $n=1 \pm 0.1$ is also acceptable and
consistent with the CMB data. Therefore, the authors of
Reference~\cite{elgaroy} considered the cases $n=0.9$ and $n=1.1$
and they ran a grid of models with $n$ as an added parameter
restricted to the values $n=1 \pm 0.1$ (Gaussian prior).

Their results can be summarized as
\begin{equation}
\sum_{i}m_{i}<1.8\, \eV
\qquad
(95\% \, \mathrm{C.L.})
\,,
\end{equation}
for $\Omega_{m}h^2=0.15$ (for the central
values of $\Omega_{m}$ and $h=0.7$; actually, almost identical
results follow from the two distinct priors on $\Omega_m$) and
spectral index $n=1$. If this latter condition is relaxed to
$n=1.0\pm 0.1$, then a somewhat looser bound is obtained:
\begin{equation}
\sum_{i}m_{i}<2.2\, \eV
\qquad
(95\% \, \mathrm{C.L.})
\,.
\end{equation}
Other groups \cite{hannestad2,lewis} obtain results from LSS that
are fully compatible with the previous upper bounds on neutrino
masses. Hannestad in Ref.~\cite{hannestad2} performs a full
numerical likelihood analysis in a cosmological parameter space
with the following free parameters (other than $\Omega_{\nu}$):
$\Omega_m$, $\Omega_b$, $h$, $n$, the normalization of the CMB
power spectrum $Q$ and the optical depth to reionization $\tau$.
Also, he restricts the study to flat models, i.e. with zero
curvature. This is by no means a drawback since there is little
degeneracy between neutrino mass and curvature. The analysis is
presented in three stages, depending on the data sets used and on
the corresponding priors for the cosmological parameters other
than neutrino mass. In the first stage LSS \cite{peacock} data and
CMB \cite{COBE,netterfield} data alone are used. The resulting
bound is, in this instance,
\begin{equation}
\sum_{i} m_{i} < 2.96 \, \eV
\qquad
(95\% \, \mathrm{C.L.})
\,.
\end{equation}

The second stage incorporates Big-Bang Nucleosynthesis \cite{burles} and
Hubble Space Telescope \cite{freedman} data in the analysis (in
addition to the previous data sets). This entails the Big-Bang
Nucleosynthesis prior on the baryon density $\Omega_b h^2=0.020\pm
0.002$ and the HST key project prior on the Hubble parameter
already given above. This leads to a slight improvement of the
bound:
\begin{equation}
\sum_{i} m_{i} < 2.65 \, \eV
\qquad
(95\% \, \mathrm{C.L.})
\,.
\end{equation}

Finally, including the data from high redshift Type Ia supernova
surveys \cite{perlmutter} Hannestad obtains:
\begin{equation}
\sum_{i} m_{i} < 2.47 \, \eV
\qquad
(95\% \, \mathrm{C.L.})
\,.
\end{equation}
Here, the result $\Omega_m=0.28\pm 0.14$ (valid for a
flat Universe) has been used. (In all three cases above, $Q$ and
$b$ are allowed to vary freely, $\tau=0-1$ and $n=0.66-1.34$. In
the first and second stages, $\Omega_m=0.1-1$. In stage one,
$\Omega_b h^2=0.008-0.040$ and $h=0.4-1.0$.)

Lewis and Bridle \cite{lewis} use the sets of observational data
on the CMB, LSS, BBN, HST and SNIa, that we are already familiar
with, to explore the consequences of a non-zero neutrino mass
under somewhat less restricted assumptions than in the analysis
just discussed \cite{hannestad2}. In particular, they consider
nine parameter model universes that include parameters that
account for a ``quintessential'' equation of state and tensor
contributions to the power spectrum (allowed in inflationary
models). Perhaps the most distinctive feature of the present
analysis is the use of powerful Markov Chain Monte-Carlo
techniques to perform a fast and efficient exploration of a high
dimensional cosmological parameter space. As a result of these
methods, reference~\cite{lewis} reports
\begin{equation}
\sum_{i} m_{i} < 1.5 \, \eV
\qquad
(95\% \, \mathrm{C.L.})
\,.
\end{equation}

For the sake of completeness we should include here the result
obtained in Ref.~\cite{fukugita}, namely (for the case of mass
degeneracy):
\begin{equation}
m_{i}<0.9\,\eV
\end{equation}
for $h\leq 0.8$, $\Omega_{m}\leq 0.4$, and an age for the Universe
in excess of $11.5$ Gyr. However, the approach of this work is
different from what has been discussed here so far. It is based on
the matching condition for the cosmic mass density fluctuation
power at the COBE scale and the matter clustering power at the
cluster scale. As the authors put it, the advantage of using the
cluster abundance information is that it refers to the mass
function which is not affected by any biasing uncertainties (i.e.
to what extent galaxies trace the mass distribution).

To summarize all these findings, we can say that the bound
\begin{equation}
\sum_{i}m_{i} \lesssim 3 \,\eV
\end{equation}
should be a reliable upper limit on the sum of neutrino masses
which implies that (again, for three almost degenerate neutrinos)
\begin{equation}
m_{i} \lesssim 1 \,\eV
\end{equation}
for each of the three masses.

As to future prospects, MAP/PLANCK CMB data in conjunction with
high precision galaxy surveys such as the Sloan Digital Sky Survey
\cite{sdss} could render \cite{hu2}
\begin{equation}
\sum_{i}m_{i}<0.3\,\eV
\end{equation}
or \cite{hannestad3}
\begin{equation}
\sum_{i}m_{i}<0.12\,\eV
\end{equation}
at 95\% C.L.,
or even go down to an ultimate sensitivity of about $0.04$ eV when
 weak lensing of galaxies by large scale structure is also taken
into account \cite{hu3}.

To conclude, perhaps we should mention here a result
\cite{buchmuller} that has an extra theoretical input, namely
leptogenesis as the origin of matter-antimatter asymmetry. In such
scenario, neutrinos are Majorana particles, and for the whole
picture to work, such neutrinos should weigh less than $0.2$ eV.

\section{Cosmic Rays}
\label{Cosmic Rays}

Since the sixties \cite{Greisen:1966jv,Zatsepin:1966jv} it is well-known that the universe 
is opaque to protons (and other nuclei) on cosmological distances. An 
ultra high energy (UHE) 
proton with energy $E$ exceeding the Greisen-Zatsepin-Kuzmin (GZK) 
energy
\begin{equation}
E_{\mathrm{GZK}} \sim 5 \times 10^{19}\ \eV
\label{GZK}
\end{equation}
interacts with the photons of the cosmic background producing pions through 
the $\Delta^*$ resonance, 
$p+\gamma_{CB}\rightarrow \Delta^* \rightarrow N\pi$. 
In this way, the initial proton 
energy is degraded with an attenuation length of about 50 Mpc 
\cite{Greisen:1966jv,Zatsepin:1966jv}. The UHE photons have even shorter 
absorption lengths ($\sim 10 \, \mathrm{Mpc}$ for $E\sim 10^{20}$ eV 
\cite{Halzen:1994gy}) 
due to their interactions with cosmic background photons 
\cite{Protheroe:1996si}. Since plausible
astrophysical sources for UHE particles (like AGNs) are located at 
distances larger than 50-100 Mpc, one expects the so-called GZK cutoff in the 
cosmic ray flux at the energy given by (\ref{GZK}).
\begin{figure}[t]
\begin{center}
\includegraphics*[bb=0 16 519 515, height=7cm]{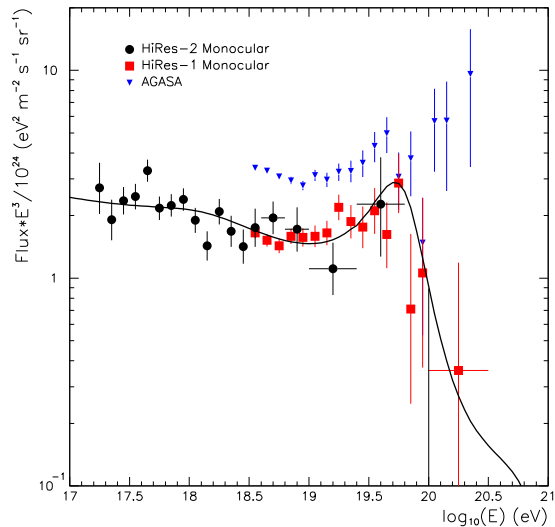}
\end{center}
\caption{ \label{hires-0208301-fig_14_je3}
UHE Cosmic Ray Flux times E$^3$.
Results from the
HiRes-I and HiRes-II detectors, and the AGASA experiment are
shown.  
Also shown (solid curve) is a fit to the data assuming a model
with two sources of cosmic rays,
galactic and extragalactic,
which includes the GZK cutoff.
Figure taken from Ref.~\cite{Abu-Zayyad:2002sf}.
}
\end{figure}

However, in the cosmic ray data \cite{Efimov:1990rk,Lawrence:1991cc,Bird:1993yi,Bird:1994wp,Bird-1995ApJ...441..144B,Takeda:1998ps,CosmicRay-1999,Ave:2000nd,Takeda:2002at,Abu-Zayyad:2002sf},
there are about twenty cosmic ray events with energies just above the 
GZK energy (see Fig.~\ref{hires-0208301-fig_14_je3}).
Yet, the whole observational status in the UHE regime is controversial.
While the HiRes collaboration claim \cite{Abu-Zayyad:2002sf} that they see
the expected event reduction, a recent reevaluation of AGASA data seems 
to confirm the violation of the GZK cutoff \cite{Takeda:2002at}. 
Indeed, some apparent
inconsistencies among data have been pointed out \cite{Bahcall:2002wi}.
The observational status is not settled, but it is clear that
if the GZK violation is confirmed, the origin of the super-GZK particles 
constitutes one of the most pressing puzzles in modern 
high-energy astrophysics (for a recent review, see Ref.~\cite{Sigl:2002ky}).

Several hypothetical explanations have been put forward to account for
this phenomenon. For example, there are scenarios where the UHE cosmic 
rays are decay products of exotic super-massive particles or relic topological 
defects. Also, a way to solve the problem might be to postulate a
violation of Lorentz symmetry, or introduce new particles or/and
new interactions (for a recent review, see Ref.~\cite{Sigl:2002ky}).
We are concerned here with a possible explanation that is based 
on the so-called $Z$-bursts \cite{Fargion:1997ft,Weiler:1997sh}.
If the GZK cutoff is violated and the Z-burst mechanism
is indeed the solution to the GZK puzzle, it may be used to determine 
the absolute value of neutrino masses and, in fact, it would be an indirect 
proof of the existence of the relic cosmic background neutrinos.

The main hypothesis of the Z-burst explanation of the GZK puzzle is the 
existence of a very high flux of UHE neutrinos. And the main
criticism to it is that standard 
astrophysical objects cannot produce such fluxes. Thus, the Z-burst
hypothesis requires new sources producing UHE neutrinos copiously.
(If the flux is so large, it may be measured by the next generation of
neutrino telescopes; see Ref.~{\cite{Kalashev:2002kx}.)
However, compared to the other models quoted before, the Z-burst scenario 
does not need new physics
beyond the Standard Model of particle physics with neutrino masses.

The crucial observation is that neutrinos, contrary 
to protons and photons, propagate over cosmological distances with negligible 
opacity. The most important interaction they have is with cosmic
background neutrinos. Although this represents a extremely small opacity it
might be enough to generate the UHE cosmic rays.

Cosmic background neutrinos are the relics of neutrino decoupling in
the early Universe,
that happened when the Universe was about 1 second old and had a temperature
of about 1 MeV. Their number density 
today can be calculated easily in terms of the observed  
number density of the relic microwave background  $n_\gamma$ 
(see, for example, Ref.~\cite{Kolb:vq}). 
In the case of no net leptonic number it is given by
\begin{equation}
n_{\nu_i}=n_{\bar \nu_i}= \frac{3}{22} n_\gamma= 56 \ {\rm cm}^{-3}
\,,
\label{n_nu0} 
\end{equation}
where $\nu_i$ is a neutrino species. The neutrino cosmic 
background is in many aspects similar to the photon cosmic 
background.
The photons have been detected but the neutrinos not yet,
because the photons have much stronger interactions than the neutrinos.
In any case, the neutrino background is a firm prediction of 
cosmology. These neutrinos decoupled when relativistic, but today they are 
non-relativistic (if the neutrino mass $m_i>10^{-4}$ eV.)

The interaction of UHE neutrinos with background neutrinos is 
strongly enhanced when
it proceeds through the $Z$-resonance, $\nu\bar \nu \rightarrow Z 
\rightarrow$ hadrons. To be exactly on top of the resonance,
the UHE neutrino has to have an energy $E^R$ such that
$M_Z^2= (E^R+m_i)^2-(E^R)^2 \simeq 2 m_i E^R$, so that
\begin{equation}
E^R_{\nu_i} \simeq \frac{M_Z^2}{2 m_i}\simeq 4.2\,
\left( \frac{\eV}{m_i}\right)\, \times 10^{21}\ \eV
\,.
\label{ER}
\end{equation}
We see there is a resonant energy for each neutrino mass $m_i$,
and for this reason we have added the subscript $\nu_i$ to $E^R$.
The lowest value of $E^R_{\nu_i}$ corresponds to the highest
neutrino mass. 

To have the $Z$-resonance enhancement,
the neutrino energy $E$ must be in the range 
\begin{equation}
E \simeq E^R_{\nu_i} \pm \Gamma_Z
\,,
\label{band}
\end{equation}
where $\Gamma_Z$ is the $Z$ width.

The idea of the $Z$-burst mechanism
\cite{Fargion:1997ft,Weiler:1997sh} is that when a UHE neutrino
with energy $E$ in the band (\ref{band})
scatters off a relic neutrino, about 70\% of the times gives
\begin{equation}
\nu\bar \nu \rightarrow Z \rightarrow {\rm hadrons}
\,.
\label{nunuZ} 
\end{equation}
The hadrons form a highly collimated final state, since in the
relic neutrino rest frame the $Z$ particle
has a Lorentz factor of $\gamma \sim 10^{11}$.
Provided the scattering takes place at a distance from Earth of less
than 50 Mpc, attenuation is small and the produced particles may induce 
air showers in
the Earth atmosphere giving rise to the observed super-GZK events.

The properties of the process 
$e^-e^+ \rightarrow Z \rightarrow {\rm hadrons}$,
have been studied with
huge statistics at LEP (see Ref.~\cite{Hagiwara:pw} for a review),
and can be used to determine the properties of the reaction 
(\ref{nunuZ}).
The average multiplicity is $\sim 30$, with final states having an 
average of about 2 nucleons, 10
neutral pions and 17 charged pions. The nucleons will then have
an average energy
\begin{equation}
\langle E_p \rangle \, \sim  \frac{ E^R_{\nu_i}}{30}
\simeq 1.4\,
\left( \frac{\eV}{m_i}\right)\, \times 10^{20}\ \eV
\,.
\label{avegnucleon}
\end{equation}
Neutral pions originate photons, $\pi^0 \rightarrow \gamma\gamma$, with
average energy
\begin{equation}
\langle E_\gamma \rangle \, \sim  \frac{ E^R_{\nu_i}}{60}
\simeq 0.7\,
\left( \frac{\eV}{m_i}\right)\, \times 10^{20}\ \eV
\,.
\label{aveggamma}
\end{equation}
In the $Z$-burst scenario these protons and/or photons are the cosmic
ray primaries. We see that one then needs a mass $m_i \sim 0.1-1$ eV to be 
just above the GZK cutoff (\ref{GZK}). 

Relevant for the purposes of this review is the determination of the neutrino
mass if the $Z$-burst mechanism turns out to be the solution to 
the GZK puzzle, as discussed in Ref.~\cite{Pas:2001nd}
(see also Ref.\cite{Fargion:2000pv}). 
A detailed analysis has been done in Ref.~\cite{Fodor:2001qy}, where 
the observed UHE cosmic ray spectrum is compared with the
predictions of the $Z$-burst model. In the analysis
one uses collider data to derive the spectra of the final state, and
finally one determines the energy losses in the propagation of
particles until reaching the Earth atmosphere. A maximum
likelihood analysis gives the interval 
\begin{equation}
0.08\, {\eV} < m_i < 1.3\, {\eV}
\qquad
(68\% \, \mathrm{C.L.})
\,,
\end{equation}
assuming that the super-GZK events are originated by 
protons produced in $Z$-bursts outside our galaxy. 
The authors of Ref.~\cite{Fodor:2001qy} claim that
this neutrino mass determination is fairly robust 
against variations of presently unknown quantities.

Another interesting study concerns the $Z$-burst model in presence of a 
leptonic asymmetry \cite{Gelmini:1999qa}, 
where one has $n_{\nu_i} \neq n_{\bar \nu_i}$ and (\ref{n_nu0}) 
is no longer valid. The authors of Ref.~\cite{Gelmini:1999qa} conclude that 
a neutrino mass $m_i \sim$ 0.07
eV, consistent with Super-Kamiokande data, explains the cosmic ray events
in this leptonic asymmetric case.
Also, we would like to mention that the possibility that $Z$-bursts may
account for events just below the GZK energy and above the ankle
of the cosmic ray spectrum is considered in Ref.~\cite{Gelmini:2002xy},
again for a mass $m_i \sim$ 0.07 eV.

Many other aspects of the $Z$-burst scenario have been treated 
in the literature \cite{others}.
For example, to what extent may neutrino clustering 
(which is quite likely) enhance the signal,
which other observations may put constrains to the model, and which are
the distinctive features that may help us in discriminating between
$Z$-bursts and other explanations of the UHE cosmic ray puzzle.

The future experimental projects aimed at the detection of cosmic rays 
to probe the UHE regime will be crucial to shed more light on this subject
(for a recent review, see for example Ref.~\cite{Sigl:2002yk}).

\section{Supernova Neutrinos}
\label{Supernova Neutrinos}

Supernovae are extremely powerful explosions
that terminate the life of some stars.
Typically
some solar masses are ejected in the interstellar space
with a kinetic energy
of the order of $10^{51} \, \mathrm{ergs}$.
The turbulence produced in the stellar medium can help
the formation of new stars.
The ejecta contain heavy elements
that are important for the chemical evolution of galaxies, stars, planets and life.
Some supernovae produce a compact remnant,
a neutron star or a black hole,
that may be observed.
Reviews and extensive bibliographies on the physics of supernovae
can be found in Refs.~\cite{Wheeler:2002gw,Woosley-Heger-Weaver-RMP74-2002,Neutrino-Unbound-note}.
A fairly updated general introduction and review of supernova neutrino physics
is given in Refs.~\cite{Raffelt:2002tu,hep-ex/0202043}.

Several supernovae that have exploded in parts of our galaxy
not obscured by dust
have been observed with naked eye during the last 2000 years.
Most famous is the 1054 supernova that
produced the Crab nebula and the Crab pulsar.
The 1006 supernova is the brightest supernova of all times.
The last galactic supernovae have been observed by naked eye
in 1572 (Tycho Brahe)
and 1604 (Joannes Kepler).
In the last centuries
many supernovae occurring in other galaxies
have been observed with telescopes because their luminosity
is comparable to that of an entire galaxy.
Supernova SN1987A,
which occurred on 23 February 1987
in the Large Magellanic Cloud,
is the best studied of all supernovae and
it is the only one which has been detected also through its neutrino burst.
As we will see in the following,
this first historical observation
of neutrinos produced out of the solar system
(and even out of our galaxy)
is important not only for the study of supernova dynamics,
but also for the study of neutrino properties,
and in particular neutrino mass.

\subsection{Supernova Types and Rates}
\label{Supernova Types and Rates}

For historical reasons,
supernovae are divided in the four different types
listed in Table~\ref{ch:Supernova-ta:suptyp},
characterized by their
spectroscopic characteristics near maximum luminosity,
which depend on the composition of the envelope of the supernova
progenitor star.
The two wide categories
called Type I and Type II
are characterized by the absence or presence of
hydrogen.
However,
the most important physical
characteristic is the mechanism that generate the supernova,
that distinguishes supernovae of Type Ia
from supernovae of Type Ib, Ic and II, as shown in
Table~\ref{ch:Supernova-ta:suptyp}.
This difference becomes noticeable from the
electromagnetic spectrum some months
after maximum luminosity,
when the ejecta become optically thin
and the innermost regions become visible.

Typically the optical emission of both Type I and II
supernovae start with a rise in luminosity during a week or two,
due to the expansion of the luminous surface.
Type I supernovae have typically a narrow luminosity peak,
whereas Type II have broad peaks,
of the order of 100 days.
After the peak the luminosity decreases during about one year.

Type Ia supernovae are thought to be generated by
carbon-oxygen white dwarfs
that have a close companion star from which the white dwarf can
accrete mass.
When the mass of the white dwarf reaches the Chandrasekhar limit\footnote{
White dwarfs are the evolutionary product of
stars that have finished thermonuclear fuel burning.
They weight about one solar mass,
they have a radius of about 5000 km
and a density of the order of
$ 10^6 \, \mathrm{g} \, \mathrm{cm}^{-3} $.
The pressure of degenerate electrons
support white dwarfs against the inward pull of gravity
(see, for example, Ref.~\cite{Shapiro:1983du}).
In 1931 Chandrasekhar discovered that
white dwarfs have a maximum mass of about $ 1.4 \, M_{\odot} $,
above which the star collapses.
}
of about $ 1.4 \, M_{\odot} $,
the star becomes unstable,
because the pressure of the degenerate electron gas
cannot sustain any more the gravitational weight.
The white dwarf begins to collapse,
triggering the fusion of carbon and oxygen to heavy nuclei,
that liberate an enormous quantity of energy
causing the explosion of the star
(see Ref.~\cite{Woosley:1986ta}).
This explosion disrupts the progenitor white dwarf
and generates an expanding nebula without a central compact object.

Since supernovae of Type Ia are all generated under similar
physical circumstances,
they have almost identical characteristics,
the most important being the total luminosity
and the ``light curve''
(luminosity as a function of time).
An empirical relation
between the duration of the peak phase of the light curve
and the luminosity of Type Ia supernovae
has been discovered by Phillips in 1993 \cite{Phillips-1993ApJ...413L.105P}
from the catalog of observed Type Ia supernovae
in nearby galaxies with known distance.
This width-luminosity relation (``broader is brighter'')
allows to use
Type Ia supernovae as standard candles
for the measurement of the distance of galaxies
as far as 100 Mpc or more
(see Ref.~\cite{Signore:2000mg} and references therein).

The observation by the Hubble Space Telescope
of supernovae of Type Ia in
galaxies at cosmological distances
have recently been used for the measurement of the Hubble parameter
and the deceleration constant.
Contrary to the expectations,
it has been found that the rate of expansion of the Universe is accelerating
\cite{Riess:1998cb,Perlmutter:1998np}.
This surprising behavior
can be explained
in the framework of the standard Friedmann-Robertson-Walker cosmology
(see Ref.~\cite{astro-ph/0207347})
through the presence of a relatively large vacuum energy
(``dark energy'' or a cosmological constant).

\begin{table}[t]
\begin{center}
\begin{tabular}{|c|c|c|c|c|c|c|c|}
\hline
\vphantom{\bigg|}
&
\multicolumn{3}{c|}{near maximum}
&
\multicolumn{2}{c|}{months later}
&
&
\\
\cline{2-6}
\vphantom{\bigg|}
\raisebox{1em}[0pt][0pt]{
Type
}
&
H
&
He
&
Si
&
Fe
&
O and C
&
\vphantom{\bigg|}
\raisebox{1em}[0pt][0pt]{
Mechanism
}
&
\vphantom{\bigg|}
\raisebox{1em}[0pt][0pt]{
Remnant
}
\\
\hline
\vphantom{\bigg|}
Ia
&
&
No
&
Yes
&
Yes
&
No
&
\begin{tabular}{c}
Mass\\Accretion
\end{tabular}
&
None
\\
\cline{1-1}
\cline{3-8}
\vphantom{\bigg|}
Ib
&
No
&
Yes
&
No
&
&
&
&
\\
\cline{1-1}
\cline{3-4}
\vphantom{\bigg|}
Ic
&
&
No
&
No
&
No
&
Yes
&
\raisebox{0pt}[0pt][0pt]{
\begin{tabular}{c}
Core\\Collapse
\end{tabular}
}
&
\raisebox{0pt}[0pt][0pt]{
\begin{tabular}{c}
Neutron Star\\or\\Black Hole
\end{tabular}
}
\\
\cline{1-4}
\vphantom{\bigg|}
II
&
Yes
&
?
&
?
&
&
&
&
\\
\hline
\end{tabular}
\end{center}
\caption{ \label{ch:Supernova-ta:suptyp}
Main characteristics of supernova types.
}
\end{table}

From the point of view of neutrino physics,
Type Ib, Ic and II supernovae are much more interesting than
Type Ia supernovae,
because they produce a huge flux of neutrinos of all types.
This is due to the fact that these supernovae
originate from the collapse of the core of massive stars
($M \gtrsim 8 \, M_\odot$)
that leaves a compact remnant.
During the few seconds following the collapse,
the compact remnant is very hot and neutrinos of all types
are copiously produced.
Since the remnant and the surrounding envelope are optically thick,
about 99\% of the gravitational binding energy liberated by the collapse
(about $3 \times 10^{53} \, \mathrm{ergs}$)
is carried away by neutrinos.
The average energy of the emitted neutrinos and antineutrinos is
of the order of 10 MeV,
and their number
is about
$10^{58}$,
about one order of magnitude
larger than the lepton number of the collapsed core.

Type II supernovae are though to be generated by the core collapse of
red (or blue as SN1987A) giant stars with a mass between about 8 and 60 solar masses.
Since the size and mass of the hydrogen envelope can be very different
from star to star,
even if they have the same initial mass,
the visible effects of the supernova explosion
have a wide range of variability,
leading to a further classification
of Type II supernovae
as Type IIL if the decrease of the luminosity is approximately linear in time,
as Type IIP if the time evolution of the luminosity shows a plateau,
as Type IIF if the supernova is faint,
as Type IIb if helium dominates over hydrogen,
as Type IIn if the spectrum shows narrow line emissions,
as Type IIpec if the supernova has peculiar characteristics
(see Refs.~\cite{Cappellaro:2000ez,Supernova-Taxonomy};
subclasses determined by spectral properties are denoted by lower-case letters
and
subclasses determined by properties of the light curve are denoted by upper-case letters).
It is believed that
if the exploding star does not have a hydrogen envelope
the supernova is of Type Ib,
and if also the helium shell is missing
the supernova is of Type Ic.
All these classes are not clear-cut and intermediate cases exist.

\begin{table}[t]
\begin{center}
\begin{tabular}{|c|c|c|c|c|}
\hline
\vphantom{\bigg|}
galaxy
&
\multicolumn{4}{|c|}{supernova type}
\\
\cline{2-5}
\vphantom{\bigg|}
type
&
Ia
&
Ib, Ic
&
II
&
All
\\
\hline
\vphantom{\bigg|}
E -- S0
&
$ 0.32 \pm 0.11 $
&
$ < 0.02 $
&
$ < 0.04 $
&
$ 0.32 \pm 0.11 $
\\
\hline
\vphantom{\bigg|}
S0a -- Sb
&
$ 0.32 \pm 0.12 $
&
$ 0.20 \pm 0.11 $
&
$ 0.75 \pm 0.34 $
&
$ 1.28 \pm 0.37 $
\\
\hline
\vphantom{\bigg|}
Sbc -- Sd
&
$ 0.37 \pm 0.14 $
&
$ 0.25 \pm 0.12 $
&
$ 1.53 \pm 0.62 $
&
$ 2.15 \pm 0.66 $
\\
\hline
\vphantom{\bigg|}
All
&
$ 0.36 \pm 0.11 $
&
$ 0.14 \pm 0.07 $
&
$ 0.71 \pm 0.34 $
&
$ 1.21 \pm 0.36 $
\\
\hline
\end{tabular}
\end{center}
\caption{ \label{suprat}
Supernova rates in units of $h^2 \, \mathrm{SNu}$ from
Refs.~\cite{Cappellaro-Evans-Turatto-AA351-P459-1999,Cappellaro:2000ez}.
One supernova unit (SNu) is defined as one supernova per
$10^{10} \, L_{\odot,B}$
per
$100 \, \mathrm{yr}$,
where $L_{\odot,B}$ is the solar luminosity in the blue spectral band.
}
\end{table}

Supernova SN1987A was an extreme case of Type IIP,
since the luminosity increased for about 3 months after collapse
and
the supernova was rather faint.
Therefore,
sometimes SN1987A is classified as IIP \cite{Cappellaro:2000ez,Supernova-Taxonomy},
sometimes as IIF \cite{Asiago-Supernova-Catalogue}
and sometimes as IIpec \cite{Supernova-Taxonomy}.
It is believed that its faintness is due to the compactness of the progenitor
(a radius of about $10^{12} \, \mathrm{cm}$).
In this case
much of the available energy
is used in the expansion and the luminosity
increases for some time because of the growing contribution
of radioactive decay of heavy elements
in inner shells,
that become more visible as the envelope expands.

A very important problem is the estimation of supernova rates.
Table~\ref{suprat}
shows the recent estimates of supernova rates
presented in
Refs.~\cite{Cappellaro-Evans-Turatto-AA351-P459-1999,Cappellaro:2000ez},
that have been obtained from the Asiago Supernova Catalog
\cite{Asiago-Supernova-Catalogue-AA139-1999,Asiago-Supernova-Catalogue}.
Some of these rates are significantly smaller than previously thought
\cite{vandenBergh-Tammann-ARAA29-1991}.
One can see that the rate of core-collapse supernovae
of Type Ib, Ic and II depends rather strongly
on the galaxy type,
being very small in elliptical galaxies.
These galaxies are very old
and have little star formation
that could produce short-lived massive stars that end their life
with a core-collapse supernova explosion.
Instead, the rate of Type Ia supernovae is almost independent
from the galaxy type, because mass accretion can occur also
in old population II star.

One of the most crucial questions for supernova neutrino astronomy
is the rate of core-collapse supernovae in our galaxy,
that could produce an observable neutrino burst with high statistics
in neutrino telescopes.
The morphological type of the Milky Way is thought to be
Sb--Sbc
and the luminosity is
$2.3 \times 10^{10} \, L_{\odot,B}$.
From Table~\ref{suprat},
using a Hubble parameter $h \simeq 0.7$,
the rate of core-collapse supernovae in the Milky Way
is about $2 \pm 1$ per century.
This rate is about a factor of two smaller than previous estimates
derived from counts of historical supernovae
and of supernovae remnants
\cite{vandenBergh-Tammann-ARAA29-1991},
but the large uncertainties do not allow to claim a disagreement
and leave the problem open to further study.
The lack of observation of neutrinos from
core-collapse supernova in our galaxy
since the Baksan Underground Scintillator Telescope
began observations in June 1980
is consistent with the estimated rate
and implies that the true rate cannot be much higher
\cite{Alekseev:1993dy}.

\begin{table}[t]
\begin{center}
\begin{tabular}{|c|ccc|ccc|}
\hline
&
\multicolumn{3}{|c|}{$1 \, M_{\odot}$}
&
\multicolumn{3}{|c|}{$25 \, M_{\odot}$}
\\
\cline{2-7}
phase
&
\begin{tabular}{c}
$T_c$
\\
(keV)
\end{tabular}
&
\begin{tabular}{c}
$\rho_c$
\\
(g cm$^{-3}$)
\end{tabular}
&
\begin{tabular}{c}
$\Delta{t}$
\\
(yr)
\end{tabular}
&
\begin{tabular}{c}
$T_c$
\\
(keV)
\end{tabular}
&
\begin{tabular}{c}
$\rho_c$
\\
(g cm$^{-3}$)
\end{tabular}
&
\begin{tabular}{c}
$\Delta{t}$
\\
(yr)
\end{tabular}
\\
\hline
H burning
&
$1.3$
&
$153$
&
$1.1 \times 10^{10}$
&
$3.3$
&
$3.8$
&
$6.7 \times 10^{6}$
\\
He burning
&
$11$
&
$2.0 \times 10^4$
&
$1.1 \times 10^{8}$
&
$17$
&
$762$
&
$8.4 \times 10^{5}$
\\
C burning
&
&
&
&
$72$
&
$1.3 \times 10^{5}$
&
$522$
\\
Ne burning
&
&
&
&
$135$
&
$4.0 \times 10^{6}$
&
$0.89$
\\
O burning
&
&
&
&
$180$
&
$3.6 \times 10^{6}$
&
$0.40$
\\
Si burning
&
&
&
&
$314$
&
$3.0 \times 10^{7}$
&
$2.0 \times 10^{-3}$
% \\
% Collapse
% &
% &
% &
% &
% $600$
% &
% $3 \times 10^{9}$
% &
% $10^{-6}$
\\
\hline
\end{tabular}
\end{center}
\caption{ \label{StarPhases}
Central temperature $T_c$,
central density $\rho_c$
and time scale $\Delta{t}$
of the evolutionary phases of
Population I stars with initial masses
$1 \, M_{\odot}$
and
$25 \, M_{\odot}$
(values taken from Ref.~\cite{Woosley-Heger-Weaver-RMP74-2002}).
}
\end{table}

\subsection{Core-Collapse Supernova Dynamics}
\label{Core-Collapse Supernova Dynamics}

Since only supernovae produced by the collapse of the core
of massive stars produce large fluxes of neutrinos
that could be detected on Earth,
here we present a short description of the current standard theory
of the dynamics of core-collapse supernovae
and the resulting neutrino flux
(see Refs.~\cite{Shapiro:1983du,%
Burrows:1992kf,%
astro-ph/0008432,%
Loredo:2001rx,%
Wheeler:2002gw,%
Woosley-Heger-Weaver-RMP74-2002,%
Neutrino-Unbound-note}
and references therein).
As explained in the previous subsection,
core-collapse supernovae are classified as
of Types II, Ib or Ic
depending on their spectroscopic characteristics at maximum luminosity.
However,
these characteristics depend only on the composition of the envelope,
which play no role in the collapse of the core and neutrino production.
Hence,
the following theory applies equally well to all Types II, Ib, Ic
core-collapse supernovae.

It is believed that core-collapse supernovae are the final stage of
the evolution of
stars with mass between about 8 and 60 solar masses.
Lighter stars end their life as white dwarfs
(but may explode as Type Ia supernovae if they belong to a multiple system),
whereas heavier stars are unstable and probably collapse into black holes
without a supernova explosion.
Stars with mass in excess of 12 solar masses
are thought to undergo all the stages of nuclear fusion
of hydrogen, helium, carbon, neon, oxygen, silicon
(see Table~\ref{StarPhases} and Ref.~\cite{Woosley-Heger-Weaver-RMP74-2002}), until the star
has an onion-like structure shown in Fig.~\ref{onion},
with an iron core
surrounded by shells composed by elements with decreasing atomic mass.
At this point the iron core has a mass of about 1 solar mass,
a radius of a few thousand km,
a central density of about $10^{10} \, \mathrm{g} \, \mathrm{cm}^{-3}$,
a central temperature of about 1 MeV,
and its weight is sustained by the pressure of degenerate relativistic electrons.
Since iron is the most bound nucleus,
there is not any more thermonuclear fuel to burn:
the iron core is endothermic;
it can only absorb energy by breaking into lighter nuclei or
creating heavier elements.
The equilibrium between
the inward pull of gravity
and
the electron pressure
that sustain the core
is destabilized shortly before the core has reached the standard Chandrasekhar mass
of about $ 1.4 \, M_{\odot} $,
because
the core contracts and the increased temperature
causes photodissociation of iron
through the process
\begin{equation}
\gamma + {}^{56}\mathrm{Fe} \to 13 \, \alpha + 4 \, n
\,.
\label{SN151}
\end{equation}
This reaction absorbs about 124 MeV of energy
and reduces the kinetic energy and pressure of the electrons.
Electron capture of nuclei,
\begin{equation}
e^- + \mathcal{N}(Z,A) \to \mathcal{N}(Z-1,A) + \nu_e
\,,
\label{SN152}
\end{equation}
and free protons,
\begin{equation}
e^- + p \to n + \nu_e
\,,
\label{SN153}
\end{equation}
favored by the high electron Fermi energy,
additionally reduces the number and pressure of the electrons.
At the onset of collapse,
when the density of the iron core is not too high,
the electron neutrinos produced by electron capture
leave the core
carrying away most of the kinetic energy of the captured electrons.
The combined effect of iron photodissociation and electron capture
lowers the electron pressure,
decreasing the value of the Chandrasekhar mass,
until the Chandrasekhar mass becomes smaller than the core mass.
At this moment
the pressure of degenerate relativistic electrons
cannot sustain the weight of the core any more
and
collapse commences.
As the density and temperature increase
the processes (\ref{SN151})--(\ref{SN153})
proceed faster,
lowering further the electron pressure and favoring the collapse,
which accelerates.

\begin{figure}[t]
\begin{center}
\includegraphics*[bb=82 425 418 752, height=7cm]{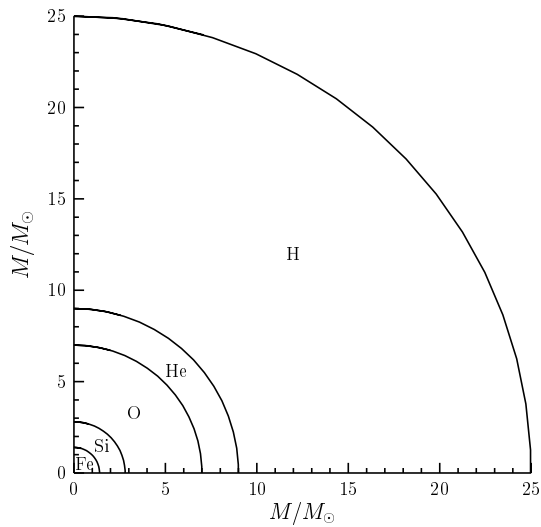}
\end{center}
\caption{ \label{onion}
Onion-like interior structure
of a Population I star of $25 \, M_{\odot}$
just before the onset of collapse
(see Ref.~\cite{Trimble-RMP54-1183-1982}).
Fe represents assorted iron-peak elements:
$^{48}\mathrm{Ca}$,
$^{50}\mathrm{Ti}$,
$^{54}\mathrm{Fe}$,
$^{56}\mathrm{Fe}$,
$^{58}\mathrm{Fe}$,
$^{66}\mathrm{Ni}$.
The Si shell contains less abundant amounts of S, O, Ar, Ca,
the O shell contains less abundant amounts of Ne, C, Mg, Si,
the He shell contains less abundant amounts of C, Ne, O,
and
the H shell contains less abundant amounts of He, Ne, O, N, C.
}
\end{figure}

According to theory
(see Ref.~\cite{Wheeler:2002gw} and references therein),
stars with mass between about 8 and 12 solar masses
burn hydrogen, helium, carbon,
but the core does not get hot enough to burn oxygen.
However,
the core contains neon and magnesium at high density,
which can undergo electron capture,
reducing the electron pressure that sustains the core against gravity.
As a result,
the core collapses and during the collapse
oxygen, neon and magnesium are converted to iron.
Therefore, also in this case the supernova explosion
is produced by the gravitational
energy released by the collapse of an iron core.

The collapse of the core produces a neutron star
and the huge liberated gravitational energy
is released mainly as a flux of neutrinos,
with a small fraction as electromagnetic radiation
and kinetic energy of the ejecta,
which constitute the visible explosion.
The liberated gravitational energy is about
\begin{equation}
\Delta{E}_{\mathrm{B}}
\sim
\frac{ G \, M_{\mathrm{core}}^2 }{R}
=
3 \times 10^{53}
\left( \frac{ M_{\mathrm{core}} }{ M_{\odot} } \right)^2
\left( \frac{ R }{ 10 \, \mathrm{km} } \right)^{-1}
\mathrm{ergs}
\,,
\label{SN241}
\end{equation}
where $G$
is Newton gravitational constant,
$M_{\mathrm{core}}$ is the mass of the core
and
$R$ is the radius of the neutron star.
Since
$M_{\mathrm{core}} \sim 1 M_{\odot}$
and
$R \sim 10 \, \mathrm{km}$,
the released gravitational energy is of the order of
$ 3 \times 10^{53} \, \mathrm{ergs}$.
Only
about 0.01\% of this energy is transformed into electromagnetic radiation
and
about 1\% is transformed into kinetic energy of the ejecta.

Let us examine in more detail the mechanism of formation of the neutron star,
of neutrino production and of supernova explosion.

The electron neutrinos produced by the electron capture processes
(\ref{SN152}) and (\ref{SN153})
initially
leave freely the core,
carrying away energy and lepton number,
since their mean free path is longer than the radius of the core.
In this so-called ``capture'' phase,
electron neutrinos have a
non-thermal spectrum and average energy that grows from about 12 to about 16 MeV
(see Ref.~\cite{Burrows:1992kf}).
The luminosity reaches about
$ 10^{53} \, \mathrm{ergs} \, \mathrm{sec}^{-1} $,
but in total only about $10^{51} \, \mathrm{ergs} $
are released before core bounce,
because the capture phase is too short (less than about 10 ms).

When the density of
the inner part of the core
(about $ 0.8 \, M_{\odot} $)
exceeds about
$3 \times 10^{11} \, \mathrm{g} \, \mathrm{cm}^{-3}$
neutrinos are trapped in the collapsing material
leading to an adiabatic collapse with constant lepton number.
During this stage,
the inner part of the core
collapses homologously,
\textit{i.e.} with subsonic velocity proportional to radius.
The outer part of the core collapses with supersonic free-fall velocity.

After about one second from the start of instability,
the density of the inner core reaches
the density of nuclear matter, about
$10^{14} \, \mathrm{g} \, \mathrm{cm}^{-3}$,
and the pressure of degenerate non-relativistic nucleons
abruptly stops the collapse.
The inner core settles into hydrostatic equilibrium,
forming a proto-neutron star with a radius of about 10 km,
while a supersonic shock wave
caused by the halting and rebound of the inner core
forms at its surface.
The shock propagates outward through the outer iron core,
which is still collapsing,
with an initial velocity of the order of
$ 100 \, \mathrm{km} \, \mathrm{msec}^{-1} $.
The gas that is infalling at a velocity near free-fall
is abruptly decelerated within the shock.
Below the shock it falls much more slowly
on the surface of the proto-neutron star,
accreting it.
Therefore,
the proto-neutron star develops an unshocked core with nuclear density,
of the order of $10^{14} \, \mathrm{g} \, \mathrm{cm}^{-3}$
and radius of the order of 10 km,
and a shocked mantle
with decreasing density,
down to about $10^{9} \, \mathrm{g} \, \mathrm{cm}^{-3}$
and a radius of about 100 km,
up to the surface of the proto-neutron star,
where the density has a steep decrease
of several orders of magnitude.

As the shock propagates
through the infalling dense matter of the outer core,
its energy is dissipated by the photodissociation of nuclei into
protons and neutrons.
Thus, the material behind the shock wave
is mainly composed of free nucleons.
Free protons have a high electron capture rate,
leading to the transformation of most protons in neutrons,
with huge production of electron neutrinos.
These neutrinos pile up behind the shock,
which is dense and opaque to them,
until the shock reaches a zone with density about
$10^{11} \, \mathrm{g} \, \mathrm{cm}^{-3}$
(``shock breakout'')
a few milliseconds after bounce
and the electron neutrinos behind the shock are released
in a few milliseconds.
This neutrino emission is usually called
``prompt electron neutrino burst''
or
``neutronization burst'',
to be distinguished from the thermal production of all neutrino flavors.
The neutronization burst has a luminosity of about
$ 6 \times 10^{53} \, \mathrm{ergs} \, \mathrm{sec}^{-1} $
and carries away an energy of the order of
$ 10^{51} \, \mathrm{ergs} $
in a few milliseconds.
In spite of his name,
the neutronization burst is too short to carry away a significant part of the electron lepton number
of the core,
which remains trapped.
Only the low-density periphery of the proto-neutron star
is neutronized.

The energy lost by photodissociation of nuclei and neutrino emission
weakens the shock
(about $ 1.5 \times 10^{51} \, \mathrm{ergs} $ are dissipated
for each 0.1 solar masses of photodissociated material).
In the so-called ``prompt'' supernova
explosion scenario, the shock,
although somewhat weakened,
is able to expel the envelope of the star
generating the supernova explosion
on a time scale of the order of 100 msec.
If the star weights more than about 10 solar masses,
the shock is weakened and stalls
about 100 ms after bounce, at a radius of about 200-300 km,
with insufficient energy to reach the outer layers of the star.
Matter continues to fall through the stalled shock
and be photodissociated.
If too much matter lands on the proto-neutron star,
the pressure of degenerate nucleons is not sufficient to maintain stability
and
the core collapses into a black hole,
presumably
without a supernova explosion.

The conditions that lead to a prompt supernova
explosion,
without a stalling shock,
are controversial
and are thought to depend on the mass of the progenitor star and on the
equation of state of nuclear matter,
which determines the energy transferred to the shock wave by the bounce.
It is widely believed that
in order to obtain a supernova explosion if the shock stalls,
the shock must be revived by some mechanism
that is able to renew its energy.
The mechanism which is currently thought to
be able to revive the shock
is the energy deposition by the huge neutrino flux produced thermally
in the proto-neutron star
\cite{Colgate-White-ApJ143-1966,Bethe:1985ux}.
In this case,
a so-called ``delayed'' supernova
explosion is produced on a time scale of the order of 0.5 sec after bounce.

Neutrinos of all flavors are produced in the hot core of the proto-neutron star
(see Refs.~\cite{Shapiro:1983du}),
which has a temperature of about 40 MeV,
through
electron-positron pair annihilation,
\begin{equation}
e^- + e^+ \to \nu + \bar\nu
\,,
\label{SN154}
\end{equation}
electron-nucleon bremsstrahlung,
\begin{equation}
e^{\pm} + N \to e^{\pm} + N + \nu + \bar\nu
\,,
\label{SN155}
\end{equation}
nucleon-nucleon bremsstrahlung,
\begin{equation}
N + N \to N + N + \nu + \bar\nu
\,,
\label{SN156}
\end{equation}
plasmon decay
\begin{equation}
\gamma \to \nu + \bar\nu
\,,
\label{SN157}
\end{equation}
and
photoannihilation
\begin{equation}
\gamma + e^{\pm} \to e^{\pm} + \nu + \bar\nu
\,.
\label{SN158}
\end{equation}
Electron neutrinos are also produced by the electron capture process (\ref{SN153}),
and electron antineutrinos are produced by positron capture on neutrons
($ e^+ + n \to p + \bar\nu_e $).
In spite of their weak interactions,
these neutrinos are trapped in a supernova core
because of the very high matter density.
Neutrino can free-stream out of the mantle proto-neutron star
only at a distance from the center where
the matter density is low enough
(of the order of $10^{11} \, \mathrm{g} \, \mathrm{cm}^{-3}$)
that the mean neutrino free path
is larger than the radius of the core.
The sphere from which neutrinos free stream
is called \emph{neutrinosphere},
and it lies within the mantle of the proto-neutron star.
Since neutrino interactions depend on flavor and energy
(see Ref.~\cite{astro-ph/0211404}),
there are different energy-dependent
neutrinospheres for different flavor neutrinos.
More precisely,
since the medium is composed by protons, neutrons and electrons,
and the neutrino energy does not allow creation of muons and taus,
the flavor neutrinos $\nu_e$ and $\bar\nu_e$ can interact
with the medium through both charged-current and neutral-current weak
processes,
whereas the neutrinos
$\nu_\mu$, $\bar\nu_\mu$, $\nu_\tau$, $\bar\nu_\tau$
can interact
only through neutral-current weak
processes,
which are flavor-independent
(small differences between neutrino and antineutrino interactions can be neglected).
Therefore,
there are three energy-dependent neutrinospheres:
one for $\nu_e$,
one for $\bar\nu_e$
and
one for $\nu_\mu$, $\bar\nu_\mu$, $\nu_\tau$, $\bar\nu_\tau$.
From now on we will denote
$\nu_\mu$, $\bar\nu_\mu$, $\nu_\tau$, $\bar\nu_\tau$
collectively as
$\nu_x$,
as usually done in the literature.
Each energy-dependent
neutrinosphere emits a black-body thermal flux of neutrinos
at the considered energy.
The estimated radii of the neutrinospheres lie between about 50 and 100 km.
As we have seen above,
when the shock
passes through the electron neutrino neutrinosphere (shock breakout)
a few milliseconds after bounce,
a large flux of electron neutrinos is released
in a few milliseconds in the neutronization burst.
After shock breakout
each neutrinosphere produces a thermal flux of the corresponding neutrino flavor.

The opacities of $\nu_e$ and $\bar\nu_e$
are dominated, respectively,
by the charged-current weak interaction processes
\begin{align}
\null & \null
\nu_e + n \to p + e^-
\,,
\label{SN001}
\\
\null & \null
\bar\nu_e + p \to n + e^+
\,.
\label{SN002}
\end{align}
These reactions allow exchange of energy and lepton number
between the neutrinos and the medium
(which is composed by electrons, positrons, nucleons and photons).
For example,
in the process (\ref{SN001})
the neutrino energy is mainly transferred to the final electron\footnote{
The recoil kinetic energy of the final proton is negligible.
Indeed,
momentum conservation implies that the momentum $p_p$ of the final proton
is of the order of the momentum $p_{\nu_e}$ of the initial neutrino,
which is practically equal to the neutrino energy,
because of the smallness of neutrino masses.
Since the neutrino energy is smaller than a few tens of MeV,
the recoil kinetic energy of the proton,
$p_p^2/2m_p$,
is suppressed by the large mass $m_p$ of the proton.
}
whose creation increases by one unit the lepton number of the medium.

Since the mantle of the proto-neutron star is neutron-rich,
the opacity of $\nu_e$ of a given energy
is larger than the opacity of $\bar\nu_e$
with the same energy,
and the corresponding $\nu_e$ neutrinosphere
has larger radius than the
$\bar\nu_e$ neutrinosphere.
Therefore,
for a fixed neutrino energy
$\bar\nu_e$'s are emitted by a deeper and hotter layer
than $\nu_e$'s,
leading to a $\bar\nu_e$ mean energy larger than the
$\nu_e$ mean energy.
Moreover,
the spectra do not have a perfect black-body shape,
but are ``pinched'',
\emph{i.e.}
both the low- and high-energy tail are suppressed with respect to the tails of
a black-body thermal spectrum with the same mean energy.
Figure~\ref{totani-9710203-fig01}
shows the time evolution of neutrino luminosity and average energy
obtained with the numerical supernova model in Ref.~\cite{Totani:1998vj}.
Other similar estimations of the neutrino luminosity and average energy
have been obtained with the numerical simulations in
Refs.~\cite{Liebendorfer:2000cq,astro-ph/0205006}.
A rough estimate of the time-integrated average energies is
\begin{equation}
\langle E_{\nu_e} \rangle \approx 10 \, \mathrm{MeV}
\,,
\quad
\langle E_{\bar\nu_e} \rangle \approx 15 \, \mathrm{MeV}
\,,
\quad
\langle E_{\nu_x} \rangle \approx 20 \, \mathrm{MeV}
\,.
\label{SN159}
\end{equation}

\begin{figure}[t]
\begin{center}
\includegraphics*[bb=18 144 592 718, width=0.7\linewidth]{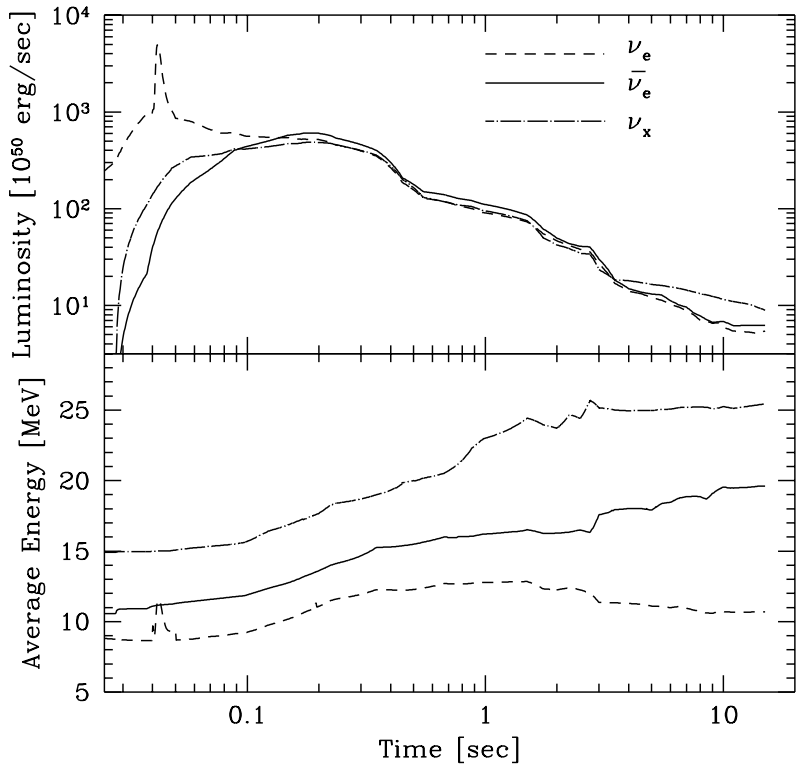}
\end{center}
\caption{ \label{totani-9710203-fig01}
Time evolution of neutrino luminosity and average energy
obtained with the numerical supernova model in Ref.~\protect\cite{Totani:1998vj}.
The dashed
lines are for $\nu_e$,
the solid lines for $\bar\nu_e$,
and the dot-dashed lines for
$\nu_\mu, \bar\nu_\mu, \nu_\tau, \bar\nu_\tau$,
which are collectively denoted by $\nu_x$.
The neutronization burst is visible as a
peak of luminosity and energy of electron neutrinos,
which happens 3-4 msec after bounce.
}
\end{figure}

In the delayed supernova explosion scenario the stalled shock lies at
a radius of about 100--300 km,
well outside of the neutrinosphere.
The post-shock temperature is about 1.5 MeV
and the density of the order of $ 10^8 \, \mathrm{g} \, \mathrm{cm}^{-3} $.
The capture of a small fraction,
about 5--10\% \cite{astro-ph/0210212},
of the thermal flux of neutrinos emitted from the neutrinosphere
could revive the shock,
leading to the explosion.
The largest energy deposition is due to electron neutrinos and antineutrinos,
which have a charged-current cross section
on the free nucleons behind the shock
that is larger than the neutral-current cross section
of all neutrino types
and is able to deposit more energy.

If enough energy is deposited behind the shock,
about half second after bounce the shock is revived and starts to sweep the outer layers of the star
generating the explosion.
Unfortunately,
most one-dimensional (\textit{i.e.} spherically symmetric) computer simulations
\cite{Rampp:2000ws,Liebendorfer:2000cq,Burrows:2000tj}
did not obtain a successful explosion,
which was recently obtained only by the Livermore group
\cite{Totani:1998vj}
(they used the so-called ``neutron finger convection''
in the proto-neutron star to enhance the early neutrino luminosity
which leads to a large
energy deposition behind the shock).
In recent years
several groups have performed two-dimensional simulations
(\textit{i.e.} cylindrically symmetric)
with unsatisfactory results
(see Ref.~\cite{Mezzacappa:2000jj,astro-ph/0210212})
and recently a successful three-dimensional simulation of explosion
has been presented in Ref.~\cite{Fryer-Warren-2002ApJ...574L..65F}.
The multi-dimensionality of the simulations is important in order
to take into account convection effects
that enhance the efficiency of the neutrino energy deposition
behind the shock.

While the shock is stalled,
matter continues to accrete on the proto-neutron star passing through the shock.
During this so-called ``accretion phase''
the shocked hot material behind the shock,
composed mainly by free nucleons, electrons and photons,
is heated by the accretion and
produces neutrinos and antineutrinos of all flavors
through the processes
(\ref{SN154})--(\ref{SN158}).
Since the stalled shock is out of the neutrinosphere,
these neutrinos can free-stream out of the star
and cause the so-called ``hump'' in the neutrino luminosity curve
shown in Fig.~\ref{totani-9710203-fig01}.
The average neutrino energy is low during the hump
because the dense matter in the shock
is opaque to high-energy neutrinos.
As the shock gradually revives,
the matter density decreases and the average neutrino energy increases.

Summarizing,
in the prompt explosion scenario
there are two phases of the neutrino flux:
first a brief and intense burst of prompt electron neutrinos
from shock breakout,
with a degenerate spectrum of high energy,
which is however so brief that little energy
(about $10^{51} \, \mathrm{ergs} $)
and lepton number are carried away.
Then there is a less intense
thermal emission of neutrinos of all flavors which last for a few seconds
and carries away most of the binding energy of the neutron star
(about $3 \times 10^{53} \, \mathrm{ergs} $).
The total number of emitted neutrinos and antineutrinos
exceeds by an order of magnitude the original
lepton number of the collapsed core.

In the delayed explosion scenario,
in addition to the prompt electron neutrino burst
and the thermal emission of neutrinos of all flavors
one expects an accretion phase which prolongs the peak of the thermal
neutrino luminosity over a time scale of about half second.

The delayed explosion scenario constitutes a sort of standard model
of core-collapse supernova explosion.
However,
the possibility of shock revival through neutrino heating is still under study
(see Ref.~\cite{Mezzacappa:2000jj,astro-ph/0210212}).

\subsection{SN1987A}
\label{SN1987A}

On 24 February 1987
a new very bright Type II supernova,
SN1987A,
was discovered in the Large Magellanic Cloud,
which is a satellite galaxy of the Milky Way,
at a distance of about 50 kpc from the solar system
(see Refs.~\cite{Trimble-RMP60-859-1988,Wheeler:2002gw}).
At that time four large underground neutrino detectors
potentially sensitive to supernova neutrinos
were in operation:
Kamiokande-II \cite{Hirata:1987hu,Hirata:1988ad},
IMB \cite{Bionta:1987qt,Bratton:1988ww,VanDerVelde:1988hh},
Baksan \cite{Alekseev:1987ej,Alekseev:1988gp,Chudakov:1987my}
and LSD \cite{Dadykin:1987ek}.
These detectors saw an unusual number of events
with energy of the order of 10 MeV
within a time window of the order of 10 sec
in the hours before the optical discovery of SN1987A.
The events observed in the
Kamiokande-II, IMB and Baksan
happened at the same time
(within uncertainties of the absolute time calibration of the detectors
and the random occurrence of the events),
whereas
the LSD events
have been recorded about five hours before those
of the other detectors,
at a time when the other detectors did not see any signal.
Therefore,
there is a controversy on the origin of the LSD events
(see Refs.~\cite{DeRujula:1987pg,Schramm:1987ra})
and
usually the LSD events are not included in the analysis
of SN1987A data.
In the following sub-subsections
we describe briefly the data of the
Kamiokande-II, IMB and Baksan
detectors,
which are used to set limits on neutrino properties.

Supernova SN1987A is the best studied of all supernovae not only
because of the detection of its neutrinos but also because
it was the first supernova visible to the naked eye after
the Kepler in 1604
and because it is the only supernova for which the
progenitor star is known:
it was a blue supergiant B3 I star
named Sanduleak - $69^{\circ}202$
\cite{Kirshner-ApJ323-1987}.

The evolution of the remnant of SN1987A
has been deeply studied in all spectral bands
(see references in Refs.~\cite{Trimble-RMP60-859-1988,Wheeler:2002gw,Neutrino-Unbound-note}).
Although no compact remnant
has been identified with certainty so far,
there is some indication of the presence
of a 2.14 ms optical pulsar \cite{Middleditch:2000it}.

The observation of SN1987A neutrinos
marked the beginning of
\emph{extra solar system neutrino astronomy}\footnote{
Solar neutrino astronomy was started by Raymond Davis Jr.,
co-winner of the 2002 Nobel Prize in Physics,
with the Homestake Chlorine experiment.
}.
It has been one of the great achievements of the Kamiokande detector,
which was designed by Masatoshi Koshiba and earned him
the 2002 Nobel Prize in Physics.

\subsubsection{Kamiokande-II}
\label{Kamiokande-II}

The Kamiokande detector
(see \cite{Koshiba:1992yb,Totsuka:1992dm})
was a water Cherenkov detector
with a fiducial volume containing 2140 tons of water
surrounded by 948 photomultiplier tubes of 50 cm diameter,
covering about 20\% of the surface area.
It was located
in the Kamioka mine in Gifu prefecture, Japan,
with a 2400 m.w.e. overhead shielding.

The Kamiokande
detector was built in 1983 for the search of nucleon decay
(Kamiokande is the acronym of
Kamioka Nucleon Decay Experiment),
although the possible detection of supernova neutrinos
was mentioned in the original proposal
(see Ref.~\cite{Koshiba:1992yb}).
In 1986 the Kamiokande detector was upgraded to
Kamiokande-II
for the detection of solar $^{8}\mathrm{B}$ neutrinos
with a threshold of about 6 MeV.
In 1990 the detector was upgraded to
Kamiokande-III
and continued operation
until 1995.
Besides the search for nucleon decay
\cite{Hirata:1989kn,Suzuki:1993zp}
and the observation of solar $^{8}\mathrm{B}$ neutrinos
\cite{Fukuda:1996sz},
the Kamiokande detector obtained two unexpected important results
during the
Kamiokande-II phase:
the observation of SN1987A neutrinos
\cite{Hirata:1987hu,Hirata:1988ad}
and the discovery of the atmospheric neutrino anomaly
\cite{Hirata:1988uy,Hatakeyama:1998ea}.
In 1996 the Kamiokande detector
was replaced by the Super-Kamiokande detector
located in the same mine,
which has a fiducial volume of 22.5 kt
(see Ref.~\cite{Malek:2002ns}).

After the optical discovery of
supernova SN1987A
the Kamiokande-II collaboration
examined carefully 
their data looking for
a significant
number of events above background
in a time interval of the order of 10 sec
and energy of the order of 10 MeV.
They found such a collection of events
at 7:35:35 UT of 23 February 1987.
Unfortunately,
before the discovery of supernova SN1987A
the Kamiokande-II collaboration
did not think that an accurate time measurement
was necessary and the clock of the experiment
was set by hand.
As explained in Ref.~\cite{Hirata:1988ad},
``it would be straightforward after SN1987A to have made an absolute calibration
of the clock ..., but and abrupt power outage took place in the Kamioka mine
on 25 February 1987,
and precluded that alternative measure''.
Therefore there is an uncertainty of about one minute in the
Kamiokande-II
determination of the time in which the
SN1987A neutrino burst passed the Earth.

Electron antineutrinos with energy larger than 1.8 MeV
can be detected with the ``inverse $\beta$-decay'' reaction
\begin{equation}
\bar\nu_e + p \to n + e^+
\,.
\label{SN160}
\end{equation}
with cross section
\begin{equation}
\sigma_{\bar\nu_e + p \to n + e^+}
\simeq
8.5 \times 10^{-44}  \left( \frac{E_{e^+}}{\mathrm{MeV}} \right)^2 \mathrm{cm}^2
\,.
\label{SN161}
\end{equation}
The produced positron is emitted almost isotropically
and can be observed in water Cherenkov detectors,
as well as in scintillator detectors.
The energy of the incident $\bar\nu_e$
is given by
\begin{equation}
E_{\bar\nu_e}
=
E_{e^+}
+
m_n - m_p
=
E_{e^+}
+
1.3 \, \mathrm{MeV}
\,.
\label{SN1611}
\end{equation}

The Kamiokande-II detector observed also neutrinos
through the elastic scattering reaction
\begin{equation}
\nu + e^- \to \nu + e^-
\,,
\label{SN162}
\end{equation}
which is mainly sensitive to electron neutrinos
(indeed, it has been used to observe solar electron neutrinos):
\begin{equation}
\sigma^{\mathrm{ES}}_{\nu_e e}
\simeq
3
\,
\sigma^{\mathrm{ES}}_{\bar\nu_e e}
\simeq
6
\,
\sigma^{\mathrm{ES}}_{\nu_x e}
\,,
\label{SN163}
\end{equation}
with
\begin{equation}
\sigma^{\mathrm{ES}}_{\nu_e e}
\simeq
9.2 \times 10^{-45} \, \frac{E_{\nu_e}}{\mathrm{MeV}} \, \mathrm{cm}^2
\,.
\label{SN164}
\end{equation}

\begin{table}[t]
\begin{center}
\begin{tabular}{|c|c|c|c|}
\hline
\multicolumn{4}{|c|}{Kamiokande II}
\\
\hline
Event
&
\begin{tabular}{c}
Time $t$
\\
(sec)
\end{tabular}
&
\begin{tabular}{c}
Energy $E_e$
\\
(MeV)
\end{tabular}
&
\begin{tabular}{c}
Angle $\theta_e$
\\
(degrees)
\end{tabular}
\\
\hline
$ 1$ & $ 0    $ & $20.0 \pm 2.9$ & $ 18 \pm 18$ \\
$ 2$ & $ 0.107$ & $13.5 \pm 3.2$ & $ 40 \pm 27$ \\
$ 3$ & $ 0.303$ & $ 7.5 \pm 2.0$ & $108 \pm 32$ \\
$ 4$ & $ 0.324$ & $ 9.2 \pm 2.7$ & $ 70 \pm 30$ \\
$ 5$ & $ 0.507$ & $12.8 \pm 2.9$ & $135 \pm 23$ \\
$ 6$ & $ 0.686$ & $ 6.3 \pm 1.7$ & $ 68 \pm 77$ \\
$ 7$ & $ 1.541$ & $35.4 \pm 8.0$ & $ 32 \pm 16$ \\
$ 8$ & $ 1.728$ & $21.0 \pm 4.2$ & $ 30 \pm 18$ \\
$ 9$ & $ 1.915$ & $19.8 \pm 3.2$ & $ 38 \pm 22$ \\
$10$ & $ 9.219$ & $ 8.6 \pm 2.7$ & $122 \pm 30$ \\
$11$ & $10.433$ & $13.0 \pm 2.6$ & $ 49 \pm 26$ \\
$12$ & $12.439$ & $ 8.9 \pm 1.9$ & $ 91 \pm 39$ \\
\hline
\end{tabular}
\end{center}
\caption{ \label{KII}
Relative time $t$, energy $E_e$ and angle $\theta_e$
of the observed charged lepton
with respect to the direction opposite to SN1987A
of the
Kamiokande II events \cite{Hirata:1988ad}.
The event number 6 is likely to be due to background
because it has a low number of hit photomultipliers.
}
\end{table}

Table~\ref{KII}
shows the relative time $t$, the energy $E_e$ and the angle $\theta_e$
of the observed charged lepton
with respect to the direction opposite to SN1987A
of 12 events measured in the Kamiokande-II detector
during the supernova SN1987A neutrino burst.
The event number 6 is reported in the Kamiokande II
original publication \cite{Hirata:1988ad},
but is excluded in their signal analysis
because its low number of hit photomultipliers
indicates that it is likely to be due to background.

Most authors agree that
it is most likely that all Kamiokande events
have been generated through the inverse $\beta$-decay reaction
(\ref{SN160})
\cite{Sato:1987rd},
because of the dominance of its cross section.
Nevertheless,
some authors
\cite{Hirata:1988ad,Rosen:1988yh}
have speculated on the fact that the first event
point almost in the opposite direction of the LMC\footnote{
In the first Kamiokande II publication, Ref.~\protect\cite{Hirata:1987hu},
the angle of the second event was reported to be
$ 15^{} \pm 27^{} $,
pointing almost backward from the direction of the Large Magellanic Cloud.
This angle was corrected
in the second Kamiokande II publication, Ref.~\protect\cite{Hirata:1988ad}.
},
which could be an indication that it is due to an
electron neutrino interacting in the detector trough the elastic scattering
process (\ref{SN162}).

\subsubsection{IMB}
\label{IMB}

The IMB water Cherenkov detector was located
in a salt mine near Fairport, Ohio, USA,
at a depth of 1570 m.w.e..
It consisted of a rectangular tank filled with purified water
with an active volume of about 6800 tons
viewed by 2048 8-inch photomultipliers
arranged on an approximate 1 m grid.

On 23 February 1987
the IMB detector recorded eight neutrino-produced events with energies
between 20 and 40 MeV
in a time interval of 6 sec
starting from 7:35:41.37 UT
(the clock had an absolute uncertainty of 50 msec
and a relative uncertainty of 0.5 msec).
The background rate is negligible,
about 2 per day in the range 20--2000 MeV.

The important characteristics of the eight IMB events are listed
in Table~\ref{ch:Supernova-ta:IMB}.
Since these events are most likely due to the inverse $\beta$-decay process
(\ref{SN160}),
the neutrino energy is given by Eq.~(\ref{SN1611}).

Taking into account the trigger efficiency and about 13\% dead time
of the detector,
the IMB collaboration estimated that $35\pm15$
neutrino events with energy above 20 MeV occurred in the detector
\cite{Bratton:1988ww}.

\begin{table}[t]
\begin{center}
\begin{tabular}{|c|c|c|c|}
\hline
\multicolumn{4}{|c|}{IMB}
\\
\hline
Event
&
\begin{tabular}{c}
Time $t$
\\
(sec)
\end{tabular}
&
\begin{tabular}{c}
Energy $E_e$
\\
(MeV)
\end{tabular}
&
\begin{tabular}{c}
Angle $\theta_e$
\\
(degrees)
\end{tabular}
\\
\hline
$1$ & $0    $ & $38 \pm 7$ & $ 80 \pm 10$ \\
$2$ & $0.412$ & $37 \pm 7$ & $ 44 \pm 15$ \\
$3$ & $0.650$ & $28 \pm 6$ & $ 56 \pm 20$ \\
$4$ & $1.141$ & $39 \pm 7$ & $ 65 \pm 20$ \\
$5$ & $1.562$ & $36 \pm 9$ & $ 33 \pm 15$ \\
$6$ & $2.684$ & $36 \pm 6$ & $ 52 \pm 10$ \\
$7$ & $5.010$ & $19 \pm 5$ & $ 42 \pm 20$ \\
$8$ & $5.582$ & $22 \pm 5$ & $104 \pm 20$ \\
\hline
\end{tabular}
\end{center}
\caption{ \label{ch:Supernova-ta:IMB}
IMB supernova SN1987A events
from Ref.~\protect\cite{Bratton:1988ww}.
The time of each event is relative to the first one,
which occurred at 7:35:41.37 UT of 23 February 1987.
There is an additional systematic uncertainty in the energy scale estimated
to be about 10\%.
The background rate is negligible (about 2 events per day).
}
\end{table}

\subsubsection{Baksan}
\label{Baksan}

The Baksan Underground Scintillation Telescope
\cite{Alekseev:1987ej,Alekseev:1988gp,Chudakov:1987my}
is located in the Baksan neutrino Observatory at a depth of 850 m.w.e.
in the Baksan Valley in North Caucasus, Russia.
The telescope consists of 3150 parallelepipedal tanks filled with oil-based liquid scintillator
viewed by a 15 cm photomultiplier.
The energy threshold for supernova neutrinos is about 10 MeV.
The total target mass is about 330 tons.
The background,
mainly caused by
cosmic ray muons and discharges in the photomultipliers,
is relatively large.
Therefore,
only 1200 inner tanks with lower background and a mass of about 130 tons
are used as signal triggers,
and the inner tanks plus part of the external tanks
are used as fiducial volume, with a mass of about 200 tons.

As water Cherenkov detectors,
the Baksan Underground Scintillation Telescope
is mostly sensitive to electron antineutrinos
which interact with protons
through the inverse $\beta$-decay reaction (\ref{SN160}).

At the time of SN1987A
the Baksan Underground Scintillation Telescope
had been in operation for about six years.
During this period of time,
including 23 February 1987,
it never happened that more than 7 events were observed in an interval of 20 sec.
The Baksan Collaboration were expecting about 35 antineutrino events
in the trigger mass and about 54 events in the fiducial mass
for a supernova at a distance of 10 kpc
(\textit{i.e.} within the Milky Way).
In the period from 1 to 23 February 1987
the Baksan Underground Scintillation Telescope
did not measure pulse clusters that differ significantly from the background.
Therefore,
the Baksan Collaboration
could not claim an independent observation of SN1987A neutrinos.
However,
when supplemented by the information
of the Kamiokande-II and IMB observations,
the Baksan Collaboration
identified 5 events in a 10 sec interval that may overlap
with the Kamiokande-II and IMB,
taking into account an uncertainty
of ${}^{+2}_{-54}$ sec in the absolute Baksan clock measurement.
The Baksan clock had a relative accuracy of about one millisecond and a nominal
absolute accuracy of about 2 sec,
but on 11 March 1987 it was found that the clock had
developed a forward shift in time of 54 sec that could have happened
in one step or gradually since 17 February 1987.
Since the Baksan clock time of the
five Baksan events is about 30 sec after the IMB
events
(which were measured with absolute time uncertainty of about 50 msec),
the simultaneous occurrence of Baksan and IMB events is possible.

Table~\ref{Bak}
shows the relative time $t$ and the energy $E_e$
of the five
Baksan events \cite{Alekseev:1988gp}.
However,
since the background rate in the Baksan detector is rather high,
it is impossible to know which, if any, of the events
is due to SN1987A neutrinos.
Therefore,
most authors did not include
the Baksan data in the analysis of SN1987A neutrino events.

\begin{table}[t]
\begin{center}
\begin{tabular}{|c|c|c|}
\hline
\multicolumn{3}{|c|}{Baksan}
\\
\hline
Event
&
\begin{tabular}{c}
Time $t$
\\
(sec)
\end{tabular}
&
\begin{tabular}{c}
Energy $E_e$
\\
(MeV)
\end{tabular}
\\
\hline
$ 1$ & $ 0    $ & $12   \pm 2.4$ \\
$ 2$ & $ 0.435$ & $18   \pm 3.6$ \\
$ 3$ & $ 1.710$ & $23.3 \pm 4.7$ \\
$ 4$ & $ 7.687$ & $17   \pm 3.4$ \\
$ 5$ & $ 9.099$ & $20.1 \pm 4.0$ \\
\hline
\end{tabular}
\end{center}
\caption{ \label{Bak}
Relative time $t$ and energy $E_e$
of the
Baksan SN1987A events \cite{Alekseev:1988gp}.
}
\end{table}

\subsubsection{Comparison with Theory}
\label{Comparison with Theory}

The neutrino events have been compared with theoretical predictions
in many papers
\cite{Bahcall:1987ua,%
Spergel:1987ch,%
Sato:1987rd,%
Sato:1987yi,%
Schramm:1987ra,%
Krauss:1987re,%
Suzuki:1988qi,%
Loredo:2001rx}.
Although only about two dozens of the estimated $10^{28}$
neutrinos that passed through the Earth were detected,
these few events delivered us precious information about
the physics of core-collapse supernovae.
Most authors agree that the detected neutrino events are compatible with
the general features of the standard core-collapse supernova scenario
described in section~\ref{Core-Collapse Supernova Dynamics}.

The most accurate analysis
of SN1987A neutrino data
has been performed recently
by Loredo and Lamb \cite{Loredo:2001rx},
which, for the first time,
took into account the background in the Kamiokande-II
and Baksan detectors.
This is important,
because it is impossible
to know with certainty which events have been really produced by neutrinos
coming from SN1987A
and which events are due to background.

Table~\ref{Loredo-KII}
shows the relative time $t$, the energy $E_e$,
the estimated background rate $B(E_e)$,
and the probabilities
$P_{B}$(prompt)
and
$P_{B}$(delayed)
that the event is due to background
according to the best fit prompt and delayed supernova explosion models
(see section~\ref{Core-Collapse Supernova Dynamics})
of the 16 Kamiokande-II events
taken into account in the analysis of Loredo and Lamb \cite{Loredo:2001rx}.
The events number 13--16
have not been considered as SN1987A events
by the Kamiokande II Collaboration
(see Table~\ref{KII}),
although they can be seen in Fig.~9 of Ref.~\cite{Hirata:1988ad}.
Indeed,
from the last two columns of Table~\ref{Loredo-KII} one can see that,
according to the calculation in Ref.~\cite{Loredo:2001rx},
these events have a high probability to be due to
background.
However,
the probability that at least one of them is a signal event
is not negligible
and it is correct to include them in the data analysis,
as done in Ref.~\cite{Loredo:2001rx}.

From Table~\ref{Loredo-KII}
one can also see that
the event number 6,
which was excluded
from the Kamiokande-II signal analysis \cite{Hirata:1988ad}
has indeed a non-negligible probability to be a
background event
according to the best fit prompt and delayed supernova explosion models
(see section~\ref{Core-Collapse Supernova Dynamics})
calculated in Ref.~\cite{Loredo:2001rx}.

\begin{table}[t]
\begin{center}
\begin{tabular}{|c|c|c|c|c|c|}
\hline
\multicolumn{6}{|c|}{Kamiokande II}
\\
\hline
Event
&
\begin{tabular}{c}
Time $t$
\\
(sec)
\end{tabular}
&
\begin{tabular}{c}
Energy $E_e$
\\
(MeV)
\end{tabular}
&
\begin{tabular}{c}
$B(E_e)$ \protect\cite{Loredo:2001rx}
\\
(s$^{-1}$)
\end{tabular}
&
\begin{tabular}{c}
$P_{B}$ \protect\cite{Loredo:2001rx}
\\
(prompt)
\end{tabular}
&
\begin{tabular}{c}
$P_{B}$ \protect\cite{Loredo:2001rx}
\\
(delayed)
\end{tabular}
\\
\hline
$ 1$ & $ 0    $ & $20.0 \pm 2.9$ & $1.6\times10^{-5}$ & $5.8\times10^{-5}$ & $2.4\times10^{-5}$ \\
$ 2$ & $ 0.107$ & $13.5 \pm 3.2$ & $1.9\times10^{-3}$ & $6.3\times10^{-3}$ & $1.9\times10^{-3}$ \\
$ 3$ & $ 0.303$ & $ 7.5 \pm 2.0$ & $2.9\times10^{-2}$ & $0.16            $ & $4.7\times10^{-2}$ \\
$ 4$ & $ 0.324$ & $ 9.2 \pm 2.7$ & $1.2\times10^{-2}$ & $5.4\times10^{-2}$ & $1.7\times10^{-2}$ \\
$ 5$ & $ 0.507$ & $12.8 \pm 2.9$ & $2.1\times10^{-3}$ & $7.6\times10^{-3}$ & $3.2\times10^{-3}$ \\
$ 6$ & $ 0.686$ & $ 6.3 \pm 1.7$ & $3.7\times10^{-2}$ & $0.25            $ & $0.15            $ \\
$ 7$ & $ 1.541$ & $35.4 \pm 8.0$ & $4.5\times10^{-5}$ & $1.2\times10^{-3}$ & $1.5\times10^{-3}$ \\
$ 8$ & $ 1.728$ & $21.0 \pm 4.2$ & $8.2\times10^{-5}$ & $5.7\times10^{-4}$ & $1.0\times10^{-3}$ \\
$ 9$ & $ 1.915$ & $19.8 \pm 3.2$ & $1.5\times10^{-5}$ & $9.9\times10^{-5}$ & $1.9\times10^{-4}$ \\
$10$ & $ 9.219$ & $ 8.6 \pm 2.7$ & $1.5\times10^{-2}$ & $0.33            $ & $0.49            $ \\
$11$ & $10.433$ & $13.0 \pm 2.6$ & $1.9\times10^{-3}$ & $0.11            $ & $0.12            $ \\
$12$ & $12.439$ & $ 8.9 \pm 1.9$ & $1.6\times10^{-2}$ & $0.54            $ & $0.60            $ \\
$13$ & $17.641$ & $ 6.5 \pm 1.6$ & $3.8\times10^{-2}$ & $0.92            $ & $0.89            $ \\
$14$ & $20.257$ & $ 5.4 \pm 1.4$ & $2.9\times10^{-2}$ & $0.97            $ & $0.94            $ \\
$15$ & $21.355$ & $ 4.6 \pm 1.3$ & $2.8\times10^{-2}$ & $0.97            $ & $0.93            $ \\
$16$ & $23.814$ & $ 6.5 \pm 1.6$ & $3.8\times10^{-2}$ & $0.99            $ & $0.94            $ \\
\hline
\end{tabular}
\end{center}
\caption{ \label{Loredo-KII}
Relative time $t$, energy $E_e$,
event background rate $B(E_e)$,
and probabilities
$P_{B}$(prompt)
and
$P_{B}$(delayed)
that each event is due to background
in the best fit prompt and delayed supernova explosion models
of the Kamiokande-II events taken into account
in Ref.~\cite{Loredo:2001rx}.
}
\end{table}

The ability to take into account background events
of the Loredo and Lamb method \cite{Loredo:2001rx}
is mostly useful for the inclusion in the analysis
of SN1987A of the Baksan data.
Table~\ref{Loredo-Bak} shows the
relative time $t$, the energy $E_e$,
the event background rate $B(E_e)$,
and probabilities
$P_{B}$(prompt)
and
$P_{B}$(delayed)
that each event is due to background
in the best fit prompt and delayed supernova explosion models
in Ref.~\cite{Loredo:2001rx}
of the Baksan events.
One can see that the background rate
in the Baksan detector is rather high.
For this reason most authors did not include
the Baksan data in the analysis of SN1987A neutrino events.
However,
Loredo and Lamb \cite{Loredo:2001rx}
properly took into account the background rate and
proved that the Baksan events are compatible
with a supernova signal.
The probabilities
$P_{B}$(prompt)
and
$P_{B}$(delayed)
show that some of the Baksan events could be
due to supernova electron antineutrinos.

\begin{table}[t]
\begin{center}
\begin{tabular}{|c|c|c|c|c|c|}
\hline
\multicolumn{6}{|c|}{Baksan}
\\
\hline
Event
&
\begin{tabular}{c}
Time $t$
\\
(sec)
\end{tabular}
&
\begin{tabular}{c}
Energy $E_e$
\\
(MeV)
\end{tabular}
&
\begin{tabular}{c}
$B(E_e)$ \protect\cite{Loredo:2001rx}
\\
(s$^{-1}$)
\end{tabular}
&
\begin{tabular}{c}
$P_{B}$ \protect\cite{Loredo:2001rx}
\\
(prompt)
\end{tabular}
&
\begin{tabular}{c}
$P_{B}$ \protect\cite{Loredo:2001rx}
\\
(delayed)
\end{tabular}
\\
\hline
$ 1$ & $ 0    $ & $12.0 \pm 2.4$ & $8.4\times10^{-4}$ & $2.1\times10^{-2}$ & $4.9\times10^{-3}$ \\
$ 2$ & $ 0.435$ & $17.9 \pm 3.6$ & $1.3\times10^{-3}$ & $3.6\times10^{-2}$ & $1.9\times10^{-2}$ \\
$ 3$ & $ 1.710$ & $23.5 \pm 4.7$ & $1.2\times10^{-3}$ & $7.4\times10^{-2}$ & $0.12            $ \\
$ 4$ & $ 7.687$ & $17.6 \pm 3.5$ & $1.3\times10^{-3}$ & $0.30            $ & $0.35            $ \\
$ 5$ & $ 9.099$ & $20.3 \pm 4.1$ & $1.3\times10^{-3}$ & $0.55            $ & $0.52            $ \\
\hline
\end{tabular}
\end{center}
\caption{ \label{Loredo-Bak}
Relative time $t$, energy $E_e$,
event background rate $B(E_e)$
and probabilities
$P_{B}$(prompt)
and
$P_{B}$(delayed)
that each event is due to background
in the best fit prompt and delayed supernova explosion models
of the Baksan events taken into account
in Ref.~\cite{Loredo:2001rx}.
}
\end{table}

Loredo and Lamb \cite{Loredo:2001rx}
found that models of supernova explosion with the delayed mechanism
explained in section~\ref{Core-Collapse Supernova Dynamics}
are about 100 times more probable than prompt explosion models.
The electron antineutrino average energy is about 15 MeV,
as expected from the cooling of the proto-neutron star
(see Eq.~(\ref{SN159})).
The cooling time scale is about 4 sec,
and the time scale of the accretion component is about 0.7 sec,
in agreement with numerical calculations.
The total inferred number of electron antineutrinos
emitted is about $3 \times 10^{57}$,
implying a binding energy of the neutron star of about
$3 \times 10^{53} \, \mathrm{ergs}$,
as expected from simple estimations
(see section~\ref{Core-Collapse Supernova Dynamics}).
Unfortunately,
as explained in Ref.~\cite{Loredo:2001rx},
the SN1987A neutrino data are too sparse to
obtain more detailed information on the supernova mechanism.

\subsection{Neutrino Mass}
\label{Neutrino Mass}

The basic idea of constraining neutrino masses from the
observation of supernova neutrinos
was proposed many years ago
\cite{Zatsepin:1968,Pakvasa:1972gz,Cabibbo:1980,Piran:1981zz,Fargion:1981xc}.

An extremely relativistic neutrino with mass $m \ll E$
propagates with a group velocity
\begin{equation}
v
=
\frac{p}{E}
=
\sqrt{1 - \frac{m^2}{E^2} }
\simeq
1
-
\frac{m^2}{2E^2}
\,.
\label{SN101}
\end{equation}
If a neutrino flux is emitted by a source at a distance $D$,
the time-of-flight delay of a massive neutrino with respect to
a massless particle (as a photon or a graviton)
emitted by the same source is
\begin{equation}
\Delta{t}
=
\frac{D}{v} - D
\simeq
\frac{m^2}{2E^2} \, D
=
2.57
\left( \frac{m}{\mathrm{eV}} \right)^2
\left( \frac{E}{\mathrm{MeV}} \right)^{-2}
\frac{D}{50\mathrm{kpc}}
\,
\mathrm{sec}
\,.
\label{SN102}
\end{equation}
If neutrinos are emitted in a burst with intrinsic duration
$\Delta{T}_0$,
the observation of events separated by a time interval
larger than
$\Delta{T}_0$
would provide a direct measurement of the neutrino mass
(assuming $D$ known and $E$ measurable).
If the neutrino energy spectrum has mean value $E$ and width $\Delta{E}$,
neutrinos produced at the same time with different energies
would reach
a detector at a distance $D$
in the time interval
\begin{equation}
\Delta{T}
\simeq
\frac{m^2}{E^2} \, D \, \frac{\Delta{E}}{E}
\,.
\label{SN103}
\end{equation}
The model-independent sensitivity to the neutrino mass is found by requiring
this time interval to be smaller than the intrinsic
duration of the neutrino burst:
\begin{equation}
\Delta{T}
<
\Delta{T}_0
\leq
\Delta{T}_{\mathrm{obs}}
\,,
\label{SN104}
\end{equation}
where
$\Delta{T}_{\mathrm{obs}}$
is the observed time interval of the neutrino burst.
The inequalities
(\ref{SN104})
imply the upper bound
\begin{equation}
m
\lesssim
E
\sqrt{ \frac{E}{\Delta{E}} \, \frac{\Delta{T}_{\mathrm{obs}}}{D} }
\simeq
14 \, \mathrm{eV}
\left( \frac{ E }{ 10 \, \mathrm{MeV} } \right)
\sqrt{ \frac{E}{\Delta{E}} }
\left( \frac{ \Delta{T}_{\mathrm{obs}} }{ 10 \, \mathrm{sec} } \right)^{1/2}
\left( \frac{ 50 \, \mathrm{kpc} }{ D } \right)^{1/2}
\,.
\label{SN105}
\end{equation}
It is clear that a
large distance,
a quick neutrino burst,
a low neutrino energy,
and a wide energy range
are advantageous for the measurement of
an effect due to the neutrino mass.
Unfortunately,
increasing the distance decreases the neutrino flux at the detector
in proportion to $D^{-2}$
and decreasing the energy decreases the detection event rate.
In practice, the energy of neutrinos coming from a supernova
is of the order of 10 MeV
and the existing detectors allow to
observe only neutrinos produced by supernovae in our galaxy
or in its satellites (the Small and Large Magellanic Clouds).

Supernova SN1987A occurred in the Large Magellanic Cloud,
at a distance of about 50 kpc from the Solar System.
Since the measured neutrino burst had an average energy
$ E \simeq 15 \, \mathrm{MeV} $,
a width
$ \Delta{E} \sim 15 \, \mathrm{MeV} $,
and an estimated original time duration
$ \Delta{T}_{\mathrm{obs}} \sim 10 \, \mathrm{sec} $,
from Eq.~(\ref{SN105})
one can see that the observation of the neutrinos SN1987A
allows a model-independent sensitivity to a neutrino mass
$ m \gtrsim 20 \, \mathrm{eV} $.
Since the time duration of the neutrino signal
is compatible with theoretical predictions,
the SN1987A data allow only to obtain
an upper limit on the neutrino mass
of the order of
$ 20 \, \mathrm{eV} $.
Indeed,
Schramm \cite{Schramm:1987ra}
argued that
\emph{without making specific model assumptions,
all that can be safely said is}
\begin{equation}
m_{\bar\nu_e} \lesssim 30 \, \mathrm{eV}
\,.
\label{SN211}
\end{equation}

Many authors have calculated an upper bound on the electron antineutrino mass
from the SN1987A neutrino data
with some specific assumptions,
often well-motivated, about the intrinsic spread
of the neutrino burst,
obtaining upper bounds for $m_{\bar\nu_e}$
lying in the $ 5 - 30 \, \mathrm{eV} $ range
\cite{%
Bahcall:1987nx,%
Arnett:1987iz,%
Sato:1987rd,%
Schramm:1987ra,%
Krauss:1987re,%
Spergel:1988ex,%
Abbott:1988bm,%
Loredo:2001rx},
as shown in Table~\ref{SN1987A-mUB}.
These bounds were also obtained with different 
statistical techniques for the analysis of the few available events.
However,
Loredo and Lamb \cite{Loredo:2001rx}
noticed that not all of these statistical procedures
are appropriate.

\begin{table}[t]
\begin{center}
\begin{tabular}{|c|c|}
\hline
Authors
&
Upper Bound
\\
\hline
Bahcall and Glashow (1987) \protect\cite{Bahcall:1987nx}
&
$m_{\bar\nu_e} < 11 \, \mathrm{eV}$
\\
Arnett and Rosner (1987) \protect\cite{Arnett:1987iz}
&
$m_{\bar\nu_e} < 12 \, \mathrm{eV}$
\\
Sato and Suzuki (1987) \protect\cite{Sato:1987rd}
&
$m_{\bar\nu_e} < 26 \, \mathrm{eV}$
\\
Schramm (1987) \protect\cite{Schramm:1987ra}
&
$m_{\bar\nu_e} \lesssim 30 \, \mathrm{eV}$
\\
Krauss (1987) \protect\cite{Krauss:1987re}
&
$m_{\bar\nu_e} \lesssim 8 - 15 \, \mathrm{eV}$
\\
Spergel and Bahcall (1988) \protect\cite{Spergel:1988ex}
&
$m_{\bar\nu_e} < 16 \, \mathrm{eV}$ (95\% C.L.)
\\
Abbott, De~Rujula and Walker (1988) \protect\cite{Abbott:1988bm}
&
$m_{\bar\nu_e} < 11 \, \mathrm{eV}$ (95\% C.L.)
\\
Loredo and Lamb (2001) \protect\cite{Loredo:2001rx}
&
$m_{\bar\nu_e} < 5.7 \, \mathrm{eV}$ (95\% P.)
\\
\hline
\end{tabular}
\end{center}
\caption{ \label{SN1987A-mUB}
Upper bounds for $m_{\bar\nu_e}$
obtained by several authors
from the analysis of SN1987A neutrino data.
C.L. = Confidence Level in Frequentist analyses
and
P. = Probability in Bayesian analyses.
}
\end{table}

In their accurate
recent analysis of SN1987A neutrino data,
Loredo and Lamb \cite{Loredo:2001rx}
applied the Bayesian method,
which is rather easily implemented in a correct way
and leads to results with a clear meaning
(see also Ref.~\cite{Loredo-90}).
Their upper limit on the electron antineutrino mass is
\begin{equation}
m_{\bar\nu_e} < 5.7 \, \mathrm{eV}
\,,
\label{SN201}
\end{equation}
with 95\% probability.
This limit is comparable with the other most stringent existing limits
on the electron neutrino mass.
It is significantly more stringent than Schramm's model-independent limit (\ref{SN211})
because it follows from a detailed fit of the data
in terms of a delayed explosion model,
which is favored by the data as explained in section~\ref{Comparison with Theory}.

\subsubsection{Neutrino Mixing}
\label{Neutrino Mixing}

If there is neutrino mixing
(see Section~\ref{Status of neutrino oscillations}),
an electron antineutrino does not have a definite mass,
since it is a superposition of different massive neutrinos.
In this case the kinematical upper limit (\ref{SN201})
applies to all the massive neutrinos that have a substantial mixing with $\bar\nu_e$.

Current experimental data on solar and atmospheric neutrinos
indicate the existence of three-neutrino mixing
(see Section~\ref{Status of neutrino oscillations})
with
\begin{equation}
\Delta{m}^2_{\mathrm{sol}} \approx 5 \times 10^{-5} \, \mathrm{eV}^2
\,,
\quad
\Delta{m}^2_{\mathrm{atm}} \approx 2.5 \times 10^{-3} \, \mathrm{eV}^2
\,,
\label{SN202}
\end{equation}
where
$\Delta{m}^2_{\mathrm{sol}}$
and
$\Delta{m}^2_{\mathrm{atm}}$
are,
respectively,
the effective neutrino squared-mass differences
in two-neutrino analyses of solar and atmospheric data.
Since the squared-mass differences in Eq.~(\ref{SN202}) are very small,
the kinematical
upper limit (\ref{SN201}) applies to
all the three neutrino masses $m_1$, $m_2$ and $m_3$.

Since the Kamiokande-II SN1987A events appear to be clustered in time
in two groups separated by an interval of about 10 sec,
some authors
\cite{Cowsik:1988xr,Huzita:1987cg}
have claimed that
there is an evidence of two mass groupings at about
4 eV and 22 eV
\cite{Cowsik:1988xr}.
However,
these authors had to assume that electron antineutrinos
are emitted from the supernova in a very short time,
of the order of 0.1 sec.
This assumption is contrary to our understanding of the core-collapse
supernova mechanism,
according to which electron antineutrinos
are emitted during the cooling phase of the proto-neutron star
on a time scale of about 10 sec
(see section~\ref{Core-Collapse Supernova Dynamics}).
Moreover,
the existence of neutrinos with masses of about
4 eV and 22 eV
which have large mixing with the electron antineutrino is
excluded by the Tritium upper bound on the effective electron antineutrino mass
(see Section~\ref{Neutrino mass from beta-decay experiments}).

Other information on neutrino mixing
has been obtained from SN1987A data
considering the effect of vacuum oscillations
or MSW \cite{Wolfenstein:1978ue,Mikheev:1985gs}
resonant transitions
on the fluxes of different flavors.
Large
$\bar\nu_x \leftrightarrows \bar\nu_e$
transitions are disfavored,
because they would imply a harder spectrum of $\bar\nu_e$'s
on Earth than observed
(see Refs.~\cite{Lunardini:2000sw,Minakata:2000rx,Kachelriess:2001sg} and references therein).

\subsubsection{Other Neutrino Properties}
\label{Other Neutrino Properties}

For the sake of completeness,
let us briefly list some of the other neutrino properties
that have been constrained using SN1987A neutrino data.

Since electron antineutrinos arrived at the Earth from a distance of about 50 kpc,
their lifetime $\tau_{\bar\nu_e}$
is constrained by
\cite{Hirata:1988ad,Schramm:1987ra}
\begin{equation}
\tau_{\bar\nu_e}
\gtrsim
1.6 \times 10^{5} \left( m_{\bar\nu_e} / E_{\bar\nu_e} \right) \, \mathrm{yr}
\,.
\label{SN221}
\end{equation}

The total amount of emitted energy inferred from
the measured $\bar\nu_e$ flux is compatible with
the binding energy of a neutron star only if the number
$N_\nu$ of neutrino flavors is limited by
\cite{Ellis:1987pk,Schramm:1987ra,Krauss:1987re}
\begin{equation}
N_\nu \lesssim 6
\,.
\label{SN222}
\end{equation}

The cooling of the proto-neutron star constrains
the Dirac masses of $\nu_\mu$ and $\nu_\tau$ by
\cite{Raffelt:1988yt,Grifols:1990jn,Gandhi:1990bq,Turner:1992ax,Burrows:1992ec}
\begin{align}
\null & \null
m_{\nu_\mu}^{\mathrm{Dirac}}
\lesssim
14 \, \mathrm{keV}
\nonumber
\\
\null & \null
m_{\nu_\tau}^{\mathrm{Dirac}}
\lesssim
14 \, \mathrm{keV}
\quad
\text{or}
\quad
m_{\nu_\mu}^{\mathrm{Dirac}}
\gtrsim
34 \, \mathrm{MeV}
\,.
\label{SN223}
\end{align}

The absence of
$\gamma$ emission
accompanying the SN1987A neutrino burst
implies a lower bound between about
$10^6$ and $10^{10}$ yr
for the lifetime of a heavy massive neutrino
with mass $ 2 m_e < m_h \lesssim 100 \, \mathrm{MeV}$
which as a substantial mixing with the active light flavor neutrinos
and decays via
$ \nu_h \to \nu_k + e^+ + e^- $
\cite{Bahcall:1987fz,Dar:1987nq,Takahara:1987gb}.

The observed 10 sec timescale of cooling of the proto-neutron star
implies an upper bound
\cite{Goldman:1988fg,Lattimer:1988mf,Barbieri:1988nh,Barbieri:1988xw}.
\begin{equation}
\mu_{\nu_e} \lesssim 10^{-12} \, \mu_B
\label{SN224}
\end{equation}
for the electron neutrino magnetic moment
which could flip neutrino helicity through scattering with electrons and nucleons
or through interactions with the strong magnetic field,
generating sterile right-handed neutrinos that escape freely, cooling the
proto-neutron star in less than 1 sec.

The absence of a similar cooling by right-handed neutrino emission
constrains also the
charge radius of right-handed neutrinos by
\cite{Grifols:1989vi}
\begin{equation}
\langle r^2 \rangle_R
\lesssim
2 \times 10^{-33} \, \mathrm{cm}^2
\,.
\label{SN228}
\end{equation}

The electric charge of the electron neutrino is bounded by
\cite{Barbiallini-Cocconi-1987}
\begin{equation}
q_{\nu_e} \lesssim 10^{-17} \, e
\,,
\label{SN225}
\end{equation}
otherwise the galactic magnetic field would
have lengthened the neutrino path
and neutrinos of different energy
could not have arrived on Earth within a few seconds.

\subsection{Future}
\label{Future}

Several detectors sensitive to supernova neutrinos
are currently in operation
(Super-Kamiokande \cite{Malek:2002ns},
SNO \cite{Virtue:2001mz},
LVD \cite{Aglietta:2001jf},
KamLAND \cite{hep-ex/0212021},
AMANDA \cite{Ahrens:2001tz},
MiniBooNE \cite{Sharp:2002as})
or under preparation or study
(see Ref.~\cite{hep-ex/0202043} and references therein).
Many authors have studied future possibilities of supernova neutrino detection
and its potential sensitivity to neutrino masses
(see Refs.~\cite{%
Burrows:1992kf,%
Fiorentini:1997hi,%
Totani:1998nf,%
Beacom:1998yb,%
Beacom:1999bn,%
Beacom:2000ng,%
Beacom:2000qy}
and references therein).
Current and future
supernova neutrino detectors are much larger than the detectors in operation during 1987
and the order of magnitude of the total number of events expected
when the next galactic supernova will explode 
is $10^4$.
Such impressive statistics will be precious in order to
test our understanding of supernova physics and
improve our knowledge of neutrino properties.

There is a general agreement among workers in the field that
future supernova neutrino detections
cannot be sensitive to an effective\footnote{
In this context we use the adjective ``effective''
for the masses of flavor neutrinos
in order to keep in mind that in reality
what are measured are the masses of the massive neutrinos
which have a large mixing with the flavor neutrino under consideration.
}
electron neutrino mass smaller than a few eV,
because of the intrinsic spread in time of the neutrino burst.
Totani \cite{Totani:1998nf}
has shown that using the correlation between neutrino energy and arrival time
implied by Eq.~(\ref{SN102}),
it is possible to reach a sensitivity of about 3 eV for
the effective electron neutrino mass.
Beacom, Boyd and Mezzacappa \cite{Beacom:2000ng,Beacom:2000qy}
have shown that an abrupt termination of the neutrino signal
due to black-hole formation may allow
the Super-Kamiokande detector to be sensitive to an electron neutrino mass
as low as 1.8 eV.
Another interesting possibility is the measurement
of the time delay between gravitational waves
generated by core collapse
and
the neutronization neutrino burst
\cite{Pakvasa:1972gz,Fargion:1981xc,Fargion:2000iz},
which may allow a sensitivity to the neutrino mass of about 1 eV
\cite{Arnaud:2001gt}.

However,
since the current
upper limit for the effective
electron neutrino mass is already a few eV
(see Section~\ref{Neutrino mass from beta-decay experiments})
and the future KATRIN experiment \cite{Osipowicz:2001sq}
will be able to push the limit down to about 0.3 eV,
a supernova limit on $m_{\nu_e}$ will not be extremely exciting.
Therefore,
several authors have concentrated on the possibility to constrain the
effective masses
of $\nu_\mu$ and $\nu_\tau$
\cite{%
Burrows:1992kf,%
Fiorentini:1997hi,%
Beacom:1998yb,%
Beacom:1999bn,%
Beacom:2000ng,%
Beacom:2000qy},
whose laboratory limits are well above the eV scale
(see Section~\ref{Muon and tau neutrino mass measurements}).

The flux of supernova
$\nu_\mu$, $\bar\nu_\mu$, $\nu_\tau$ and $\bar\nu_\tau$
is of the same order as that of $\nu_e$ and $\bar\nu_e$,
but the problem is to distinguish them,
because they can be observed only through neutral-current interactions,
which are flavor blind
(the energy is too low to produce $\mu$ or $\tau$
in charged-current reactions).
Therefore,
the $\nu_\mu$, $\bar\nu_\mu$, $\nu_\tau$, $\bar\nu_\tau$
signal
can be only extracted on a statistical basis by subtracting the
$\nu_e$ and $\bar\nu_e$
contributions from the measured neutral-current signal.
The $\nu_e$ and $\bar\nu_e$
contributions
are estimated from the $\nu_e$ and $\bar\nu_e$
charged-current signals.

Unfortunately,
in usual neutral-current neutrino interactions,
as that in SNO,
\begin{equation}
\nu + d \to p + n + \nu
\,,
\label{SN226}
\end{equation}
the energy of the neutrino is not determined.
Therefore,
it is not possible to use the correlation between neutrino energy and arrival time
implied by Eq.~(\ref{SN102})
for the measurement of neutrino masses,
and the upper limit on the effective masses
of $\nu_\mu$ and $\nu_\tau$
cannot be pushed below about 30 eV
\cite{%
Burrows:1992kf,%
Fiorentini:1997hi,%
Beacom:1998yb}.
An interesting exception is the
abrupt termination of the neutrino signal
due to black-hole formation,
which may allow a sensitivity to
$\nu_\mu$ and $\nu_\tau$
effective masses
as low as about 6 eV \cite{Beacom:2000ng,Beacom:2000qy}.
Another promising technique
which has been recently proposed by Beacom, Farr and Vogel \cite{Beacom:2002hs}
is the measurement of the recoil proton kinetic energy
in neutral-current neutrino-proton elastic scattering,
\begin{equation}
\nu + p \to \nu + p
\,.
\label{SN227}
\end{equation}
Since the recoil protons have a kinetic energy of the order of 1 MeV,
they are non-relativistic and cannot be seen in water Cherenkov detectors,
but they can be observed
in liquid scintillator detectors as
KamLAND \cite{hep-ex/0212021}
and
BOREXINO \cite{Tartaglia:2001sh}.
Unfortunately,
the proton direction cannot be measured in scintillator detectors,
denying the possibility to reconstruct the neutrino energy from simple kinematics
on a event-by-event basis.
However,
Beacom, Farr and Vogel \cite{Beacom:2002hs}
have shown that a fit of the proton kinetic energy distribution
could allow to measure the neutrino temperature
and the total neutrino energy
with an accuracy of about 10\%.
As far as we know,
the possibility to obtain information on the
effective masses of $\nu_\mu$ and $\nu_\tau$
with this method has not been explored so far.

Of course a major problem in supernova neutrino physics is the
actual occurrence of a supernova at a galactic scale distance.
As we have seen in section~\ref{Supernova Types and Rates},
the estimated rate of core-collapse supernovae in the Milky Way is about $2\pm1$ per century.
Such low rate is just at the border of the patience of
very patient scientists.
Since most scientists are not so patient,
there is an active research to study the feasibility
of huge detectors
that could observe a few tents of events
produced by a supernova in the local group of galaxies
(see Ref.~\cite{hep-ex/0202043} and references therein).

\section{Conclusions}
\label{Conclusions}

During many years there were indications in favor of the disappearance
of solar $\nu_{e}$'s (the so-called ``solar neutrino problem''),
obtained in the
Homestake \cite{Cleveland:1998nv}, 
Kamiokande \cite{Fukuda:1996sz},
GALLEX \cite{Hampel:1998xg}
and
SAGE \cite{Abdurashitov:1999zd}
solar neutrino experiments,
and indications in favor of the disappearance of
atmospheric $\nu_{\mu}$'s (the so-called ``atmospheric neutrino anomaly''),
found in the
Kamiokande \cite{Hirata:1992ku,Fukuda:1994mc,Hatakeyama:1998ea},
IMB \cite{Becker-Szendy:1992hq},
Soudan 2 \cite{Allison:1999ms}
and
MACRO \cite{Ambrosio:2000qy}
atmospheric neutrino experiments.
Neutrino oscillations were considered to be the
natural interpretation of these data, in spite of the fact that
other possibilities were not excluded at that time
(as
large anomalous magnetic moment of neutrino,
neutrino decay,
etc.;
see Refs.~\cite{Bahcall-book-89,Pulido:1992fb,CWKim-book-93,Mohapatra-book-98}).

In the latest years with the impressive results
of the atmospheric and solar neutrino experiments
Super-Kamiokande and SNO
and the long-baseline reactor experiment KamLAND,
the status of neutrino
masses, mixing and oscillations drastically changed. 
The up-down asymmetry of the atmospheric multi-GeV  muon 
events discovered in the Super-Kamiokande experiment
\cite{Fukuda:1998mi},
the evidence of transitions of solar $\nu_e$ into $\nu_\mu$ and $\nu_\tau$
obtained in the SNO experiment
\cite{Ahmad:2002jz}
from the observation of solar
neutrinos through the detection of CC and NC reactions,
and
the evidence of disappearance of reactor $\bar\nu_e$
found in the long-baseline KamLAND experiment
\cite{hep-ex/0212021}
imply that neutrino oscillations driven by small neutrino masses and neutrino mixing
is the only viable explanation of the experimental data.

The generation of neutrino masses,
many orders of magnitude smaller than the masses of their family
partner leptons and quarks,
requires a new mechanism beyond the Standard Model.
Several possibilities for new mechanisms which could generate
small neutrino masses are open today. They are based on the 
see-saw type violation of the
total lepton number at a very large scale, large extra dimensions, etc.
(see Refs.~\cite{Altarelli:1999gu,Fritzsch:1999ee,Masina:2001pp,hep-ph/0206077}).
It is obvious that the understanding of the true mechanism of the generation
of neutrino masses and mixing requires new experimental data.

First of all, we need to know how many massive neutrino exist in nature.
The minimal number of massive neutrinos is equal to the number
of flavor neutrinos (three). If, however, the LSND
indication in favor of
short-baseline $\bar\nu_\mu\to\bar\nu_e$ oscillations
is confirmed,
at least four massive neutrinos are needed.
In spite of the fact that such scenario is disfavored by the analysis of the existing
data,
in order to reach a conclusion it is needed
to wait the decisive check of the LSND result
which is under way at Fermilab with the MiniBooNE experiment.

The problem of the nature of the massive neutrinos (Dirac or Majorana?)
is crucial for the understanding of the origin of neutrino masses.
This problem can be solved by the experiments searching for neutrinoless
double-$\beta$ decay of some even-even nuclei. From the existing 
data for the effective Majorana mass $|\langle m \rangle|$ the bound
$|\langle m \rangle| \leq (0.3-1.2) \, \mathrm{eV}$
has been obtained. Several new experiments
with a sensitivity to $|\langle m \rangle|$
improved by about one order of magnitude
with respect to present experiments
are under preparation.

Neutrino oscillation experiments allow to determine the
neutrino mass-squared differences $\Delta m^{2}$.
Since neutrino oscillations is an interference phenomenon,
neutrino oscillation experiments are sensitive to 
very small values of $\Delta m^{2}$.
On the other hand,
the problem of the
\emph{absolute values of neutrino masses}
is apparently the most difficult one in the experimental
investigation of the
physics of massive and mixed neutrinos.
In this paper we presented a review of our present knowledge
of the absolute values of neutrino masses
and
the prospects for the future
(see also Ref.~\cite{Pas:2001nd}).

At present,
the bound $m_{\beta} < 2.2 \, \mathrm{eV}$
at 95\% C.L. has been obtained
from $\beta$-decay experiments
\cite{hep-ex/0210050,Lobashev:2001uu}.
About one order of the magnitude improvement is expected in the
future.

With enormous progress in cosmology in the last years 
the possibility to obtain information about the values of neutrino masses from
cosmological data has strongly increased.
The existing data allowed to obtain
the bound
$\sum_{i}m_{i} \lesssim 2 \, \mathrm{eV}$
for the total mass of neutrinos
\cite{elgaroy,lewis,hannestad2}.
About one order of magnitude improvement is expected with the future data.

A possible explanation of the observation
\cite{Efimov:1990rk,Lawrence:1991cc,Bird:1993yi,Bird:1994wp,Bird-1995ApJ...441..144B,
Takeda:1998ps,CosmicRay-1999,Ave:2000nd,Takeda:2002at}
of very high-energy cosmic rays beyond the GZK cutoff
\cite{Greisen:1966jv,Zatsepin:1966jv}
is the existence of $Z$-bursts
\cite{Fargion:1997ft,Weiler:1997sh},
which are possible if there is at least one neutrino mass
in the
$0.08 - 1.3 \, \mathrm{eV}$
range at 68\% C.L..

The observation of supernova 1987A neutrinos
by the
Kamiokande-II \cite{Hirata:1988ad},
IMB \cite{Bratton:1988ww} and
Baksan \cite{Alekseev:1988gp}
detectors
allowed to obtain important information
on supernova physics and neutrino properties.
The absence of an energy-dependent spread of the neutrino burst
constrains the absolute value of the effective electron neutrino mass
below 5.7 eV with 95\% probability
\cite{Loredo:2001rx}.

In the framework of the minimal scheme with three massive neutrinos
the values of the neutrino masses are determined
by the minimal neutrino mass $m_{1}$ and the neutrino mass-squared differences
$\Delta m^{2}_{21}$ and $\Delta m^{2}_{32}$, which will be measured
with high precision in solar, atmospheric, long-baseline reactor and accelerator
neutrino oscillations experiments.
Thus, the problem of the absolute values of neutrino masses is the problem of determination of
the minimal mass $m_{1}$.
If $m_{1}$ is much smaller than $\sqrt{\Delta m^{2}_{\mathrm{sol}}}\simeq 7\cdot 
10^{-3}\,\mathrm{eV}$ (hierarchical mass spectrum),
it will be very difficult to reach the
absolute values of neutrino masses in terrestrial experiments.
Future cosmological measurements could be a possibility in this case
(see Ref.~\cite{hannestad3}),
although the inferred information is somewhat model dependent.
If  $m_{1}$ is much smaller than $\sqrt{\Delta m^{2}_{\mathrm{atm}}}
\simeq 5\cdot 
10^{-2}\,\mathrm{eV}$ and the neutrino mass spectrum is characterized by
an inverted hierarchy,
information about the
absolute values of neutrino masses can be obtained from the future experiments
searching for neutrinoless double-$\beta$ decay.
If $m_{1}$ is much larger than $\sqrt{\Delta m^{2}_{\mathrm{atm}}}$
(practically degenerate neutrino mass spectrum), there is a possibility
that $m_{1}$ will be measured in future $\beta$-decay experiments.
Of course, progress in experimental techniques could bring some unexpected
surprises for the solution of the exciting problem of the absolute neutrino masses.

\section{Note added in proof}

After completion and submission of this review, the Wilkinson
Microwave Anisotropy Probe (WMAP) Science Team issued a most
impressive set of data on the cosmic temperature angular power
spectrum and the temperature-polarization angular power spectrum
\cite{wmap}. Analysis of the data by the WMAP team has
strengthened the notion that a Standard Model in Cosmology is
emerging. Indeed, a flat universe filled with vacuum energy, dark
matter, baryons and structure arisen from a nearly scale-invariant
spectrum of primordial fluctuations seems to fit not only the WMAP
high precision data but also smaller scale CMB data, large scale
structure data, and Supernova Ia data. Furthermore, it is fully
consistent with a much wider set of astronomical data such as the
baryon to photon ratio derived from observations of D/H in distant
quasars, the Hubble parameter measurement by the HST Key Project,
the size of mass fluctuations obtained from galaxy clusters
studies, and the inferred ages of stars \cite{spergel}.

As was stated in Section 6, neutrino mass has an impact on
cosmological and astrophysical observables, notably on the
suppression of power at low scales of the matter density
perturbation spectrum (see Section 6.3). This is in fact the key
ingredient in the determination of neutrino mass bounds from
cosmology, as neutrinos with masses of the order of $1$ eV are
easily mistaken at recombination for cold dark matter and thus
have small influence on the CMB. Hence the new WMAP data
\cite{wmap,spergel} permits an improvement of these bounds only
indirectly, as the parameters that define the overall cosmological
picture become more precise. So, in order to constrain neutrino
masses beyond what was reached in previous analysis (see Section
6.5) the WMAP team undertook a study of their data combined with
other CMB data that explore smaller angular scales (ACBAR
\cite{kuo} and CBI \cite{pearson}) and, most importantly, with the
large scale structure 2dFGRS data, and with the power spectrum
recovered from Lyman $\alpha$ forest measurements. As a result of
their analysis they find:
\begin {equation}
\sum_i m_{i}<0.71\,\mathrm{eV} \qquad (95\% \, \mathrm{C.L.}) \, ,
\end {equation}
which implies, for 3 degenerate neutrino species, $m_{i}<0.23$ eV
. This is about a factor three better than the bounds (6.8) and
(6.9) (which, moreover, had to assume strong priors on the Hubble
constant and the matter density). The precise role played by
priors in lifting degeneracies among cosmological parameters when
using WMAP data to constrain neutrino masses has been analyzed in
\cite {1} and \cite {2}. Two priors turn out to be particularly
important: the prior on the matter density $\Omega_{m}$ (not
surprisingly, because the suppression of growth of matter
fluctuations at small scales depends on the ratio
$\Omega_{\nu}/\Omega_{m}$) and the prior on the Hubble parameter
$h$. Hence, the WMAP results (where only flat models and $h>0.5$
were allowed) $\Omega_{m}h^2=0.14\pm 0.02$ and $h=0.72\pm 0.05$
 \cite{spergel}  are crucial in setting the neutrino mass limit given above. In
their study, the authors in \cite {1} include an analysis of the
2dFGRS data complemented with simple gaussian priors from WMAP
\cite{spergel} that leads to the following bound:
\begin {equation}
\sum_i m_{i}<1.1\,\mathrm{eV} \qquad (95\% \, \mathrm{C.L.}) \, .
\end {equation}

In \cite {2} Hannestad performs a search for neutrino mass bounds
in three stages. In his most conservative analysis, where only
WMAP data and 2dFGRS are being used, he obtains
\begin {equation}
\sum_i m_{i}<2.12\,\mathrm{eV} \qquad (95\% \, \mathrm{C.L.}) \, .
\end {equation}
In a second stage, he includes extra CMB data from the compilation
in \cite {3} that incorporates data at large $l$'s. He finds the
considerably more restrictive limit:
\begin {equation}
\sum_i m_{i}<1.20\,\mathrm{eV} \qquad (95\% \, \mathrm{C.L.}) \, .
\end {equation}
The reason for improvement can be traced to the fact that the
scales probed by the CMB data at high $l$'s \cite {3} overlap
significantly with the scales under scrutiny in the 2dFGRS survey.
As a consequence a normalization of the 2dFGRS power spectrum from
the CMB data is possible. In the final analysis Hannestad adds the
constraints on $h$ from the Hubble Space Telescope key project HST
\cite{freedman} and the constraints on matter density from the
SNIa project \cite{perlmutter}. As a result, the bound on the
total neutrino mass is,
\begin {equation}
\sum_i m_{i}<1.01\,\mathrm{eV} \qquad (95\% \, \mathrm{C.L.}) \, .
\end {equation}

We see from these numbers that neither \cite {1} nor \cite {2} can
match the limit obtained by the WMAP team, a fact that the authors
in these latter studies attribute mainly to not using Ly$\alpha$
data whereas the WMAP team does. Given that the extraction of the
matter power spectrum from the Ly$\alpha$ forest involves complex
numerical simulations, both \cite {1} and \cite {2} argue that one
is on a safer position if these data are not used.

\section*{Acknowledgments}
S.M.B. acknowledges the support of the
``Programa de Profesores Visitantes de IBERDROLA de Ciencia y Tecnologia''.
Two of us (J.A.G and E.M) are partially supported by the CICYT Research 
Projects AEN99-0766 and FPA2002-00648,  
by the EU network on Supersymmetry and the Early Universe 
(HPRN-CT-2000-00152), and by the
\textit{Departament d'Universitats, Recerca i Societat de la Informaci\'o},
Project 2001SGR00188.

\end{document}